\documentclass[a4paper,11pt]{article}

\usepackage{jinstpub} 

\usepackage{comment}
\usepackage{subcaption}
\usepackage{lineno}
\usepackage{placeins}
\usepackage[page,toc,titletoc,title]{appendix}

\title{\boldmath Test-Beam and Simulation Studies Towards RPWELL-based DHCAL }

\author[a,1]{D. Shaked-Renous,\note{Corresponding author.}}
\author[b]{F.D. Amaro}
\author[a,2]{P. Bhattacharya\note{Now at School of Basic and Applied Sciences, Adams University, Barasat-Barrackpore Road, Kolkata, 700126 India.}}
\author[a]{A. Breskin}
\author[c]{M. Chefdeville}
\author[c]{C. Drancourt}
\author[d]{T. Geralis}
\author[c]{Y. Karyotakis}
\author[a]{L. Moleri}
\author[a]{A. Tesi}
\author[e]{M. Titov}
\author[f]{J. Veloso}
\author[c]{G. Vouters}
\author[a]{S. Bressler}

\affiliation[a]{Weizmann Institute of Science,\\76100 Rehovot, Israel}
\affiliation[b]{LIBPhys, Department of Physics, University of Coimbra,\\RuaLarga, PT3004-516 Coimbra, Portugal}
\affiliation[c]{Univ. Grenoble Alpes, Univ. Savoie Mont Blanc, CNRS, IN2P3-LAPP,\\9 Chemin de Bellevue - BP 110, 74941 Annecy-le-Vieux Cedex, France}
\affiliation[d]{NCSR Demokritos/INPP,\\Ag. Paraskevi Attikis, 15310 Athens, Greece}
\affiliation[e]{CEA Saclay / Irfu,\\Route Nationale, Gif-sur-Yvette, 91191 France}
\affiliation[f]{I3N, Physics Department, University of Aveiro,\\3810-193 Aveiro, Portugal}

\emailAdd{dan.shaked@cern.ch}

\abstract{Digital Hadronic Calorimeters (DHCAL) were suggested for future Colliders as part of the particle-flow concept. Though studied mainly with Resistive Plate Chambers (RPC), studies focusing on MPGD-based sampling elements have shown the potential advantages; they can be operated with environmental friendly gases and reach similar detection efficiency at lower average pad multiplicity. We summarize here the experimental test-beam results of a small-size DHCAL prototype, incorporating six Micromegas and two Resistive-Plate WELL (RPWELL) sampling elements, interlaced with  steel-absorber plates. It was investigated with 2--6 GeV pion beams at the CERN/PS beam facility. The data permitted  validating a GEANT4 simulation framework of a DHCAL, and evaluating the expected pion energy resolution of a full-scale RPWELL-based calorimeter. The pion energy resolution of $\frac{\sigma}{E[GeV]}=\frac{50.8\%}{\sqrt{E[GeV]}} \oplus 10.3\%$ derived for the RPWELL concept is competitive to that of glass RPC and MM sampling techniques.}


\arxivnumber{2208.12846} 

\begin{document}
\maketitle
\flushbottom

\section{Introduction}
\label{sec:intro}

Particle-flow \cite{1} is a leading approach towards reaching the challenging jet energy resolution ($\frac{\sigma}{E}\leq 30\%/\sqrt{E[\text{GeV}]}$) required in future lepton collider experiments \cite{2,3,4,5}. It is based on the observation that, on average, over 60\% of the particles in a jet are charged hadrons. Hence, their energy can be measured by the tracking system with higher precision than a traditional measurement based on the Hadronic Calorimeter (HCAL). In this approach, only the energy of the neutral hadrons ($\sim$10\% of the jet energy) is measured in the HCAL. Particle-flow calorimeters are designed to allow associating the energy deposits with individual particles, ignoring the ones deposited by charged particles. This requires high transverse and longitudinal granularity --- thus many readout channels. In this respect, Digital and Semi-Digital Hadronic Calorimeters ((S)DHCAL) \cite{dhcal} with a 1-2 bit Analog to Digital Converter (ADC) readout are appealing; they offer a cost-effective solution for reading out a large number of channels.

A typical configuration consists of alternating layers of absorbers, where the hadronic shower develops, and sampling elements with pad-readout. The absorber's thickness (a few cm) and the pad's size (a few cm$^2$) define the longitudinal and transverse granularity, respectively. The energy of a single hadron is reconstructed from the number and the pattern of all fired pads (hits), under the assumption of a one-to-one relation between the energy and these hits' features. The performance of a sampling element is characterized in terms of a minimum-ionizing particle (MIP) detection efficiency and the average pad multiplicity --- the number of pads firing per impinging MIP. Low detection efficiency reduces the total number of hits, and large average pad multiplicity introduces two effects. On the one hand, large average pad multiplicity increases the probability for overcounting: e.g., attributing two-particle hits to two neighboring pads fired by a single particle. On the other hand, it can be thought of as larger pads and thus increase the probability for overlapping or undercounting: e.g., detecting the clusters of hits from two nearby particles as a single one. Both effects may degrade the energy resolution.

Sampling elements of different technologies have been studied: 1$\times$1 m$^2$ glass-Resistive Plate Chambers (RPC) \cite{9}, 1$\times$1 m$^2$ Micromegas (MM) \cite{1sqmMM, mmMaxAddition}, 30$\times$30 cm$^2$ double Gaseous Electron Multipliers (GEM) \cite{doubleGEM}, 50$\times$100 cm$^2$ Resistive WELL (RWELL) \cite{RWELL talk}
, and 30$\times$30 cm$^2$ Resistive-Plate WELL (RPWELL) \cite{mediumSizeRPWELL}. Their measured performances are summarized in Table \ref{tab:i}. Although RPC is the most studied technology for DHCAL, Micro-Pattern Gaseous Detectors (MPGD) sampling elements have some advantages; they demonstrate lower average pad multiplicity for a similar MIP detection efficiency and are operated in environment-friendly gas mixtures.

\begin{table}[htbp]
\centering
\caption{\label{tab:i} A summary of the average pad multiplicity and MIP detection efficiency measured with sampling elements of different technologies.}
\smallskip
\begin{tabular}{lcc}
\hline
& \begin{tabular}{c}
     Average  \\
     Pad Multiplicity 
\end{tabular}&\begin{tabular}{c}
     MIP detection  \\
     Efficiency 
\end{tabular}\\
\hline
Glass RPC \cite{9} & 1.6 & 98\%\\
MM \cite{1sqmMM} & 1.1 & 98\%\\
Resistive MM \cite{newMM2021} & $\sim$1.1 & 95\%\\
Double GEM \cite{doubleGEM} & $\sim$1.2 & 98\%\\
RWELL \cite{RWELL talk} & Not Reported & $\sim$96\%\\
RPWELL \cite{mediumSizeRPWELL} & 1.2 & 98\%\\
\hline
\end{tabular}
\end{table}

In November 2018, we operated the small DHCAL prototype having six MM- and two RPWELL-based sampling elements at the CERN/PS test-beam facility \cite{VCI19,MPGD19}. The data collected was used to validate a simulation framework for studying MPGD-DHCALs, based on which, the expected performance of a full-scale (50 layers) RPWELL-based DHCAL was evaluated. For more details regarding the material presented in this article see \cite{DanThesis}.
 
This article is organized as follows: Section \ref{sec:simFram} describes the test-beam campaign and the validation of the simulation framework, while Section \ref{sec:50layers}  presents the simulation studies of a full-scale RPWELL-based DHCAL. A summary and a discussion are provided in Section \ref{subsec:simDis}. 

\section{Simulation Framework: Validation with test-beam data}
A simulation framework for MPGD-DHCAL was developed based on the GEANT4 toolkit (version 10.06.p01 \cite{g4}). It was validated by comparing its outcome to the data collected with a small, eight layers, MPGD-DHCAL prototype at the CERN/PS T10 beamline. Only low energy (2--6 GeV) pions were used to mitigate the effect of shower fragments escaping our setup.
\label{sec:simFram}
\subsection{Methodology}
\label{sec:method}
\subsubsection{Experimental Setup}
\label{subsec:expSetup}
\begin{figure}[htbp]
\centering 
\includegraphics[width=0.95\textwidth]{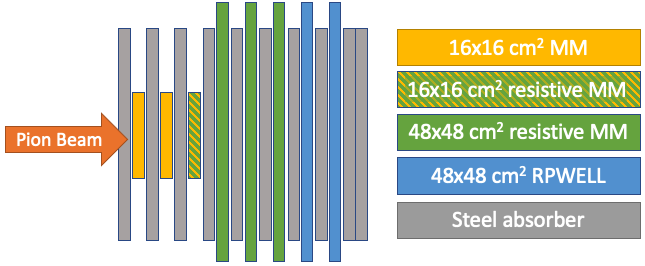}
\qquad

\caption{\label{fig:DHCAL} Schematic description of a small-DHCAL prototype comprising three 16$\times$16 cm$^2$ (two non-resistive and one resistive) and three 48$\times$48 cm$^2$ resistive MM followed by two 48$\times$48 cm$^2$ RPWELL sampling elements. 2 cm thick steel absorber plates were inserted between neighboring sampling elements.}
\end{figure}

A schematic description of the DHCAL experimental setup is shown in Figure \ref{fig:DHCAL}. The prototype comprised eight alternating layers of 2 cm thick steel absorber plates and sampling elements \cite{VCI19,MPGD19}. The total thickness of $\sim$16 cm steel corresponds to 0.8 pion interaction lengths ($\lambda_\pi$) and $\sim$8.9 radiation lengths ($X_0$), yielding a probability of 45\% and 99.9\% for a pion and an electron shower to start within the calorimeter, respectively. The sampling elements consisted of three 16$\times$16 cm$^2$ bulk MM (two non-resistive and one resistive), three 48$\times$48 cm$^2$ resistive MM, and two 48$\times$48 cm$^2$ RPWELL. All eight sampling elements were equipped with semi-digital readout electronics based on the MICROROC chip \cite{MICROROC}. The small bulk MMs had a square-geometry pad-matrix, while the large MM and the RPWELL detectors were equipped with a circular geometry pad-matrix (48 cm diameter) to reduce the total number of channels. All the readout pads were 1$\times$1 cm$^2$ in size. The chambers were operated in a digital mode, where only the lowest threshold of the MICROROC (set to 0.8 fC) was used. All the chambers were read out with a common DAQ system.

\begin{figure}[htbp]
    \centering
    \begin{subfigure}[b]{0.45\textwidth}
        \includegraphics[width=\textwidth]{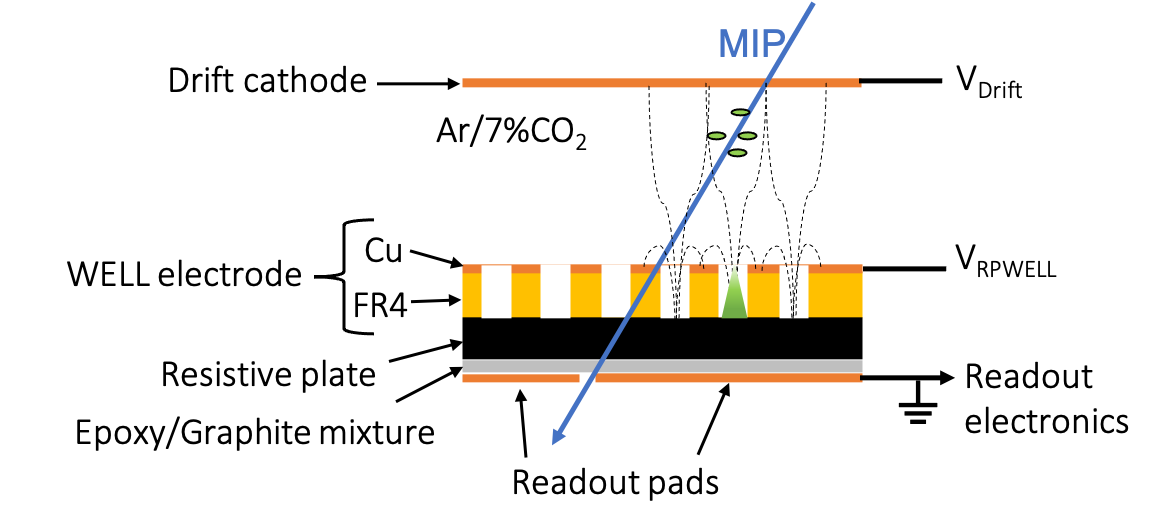}
        \centering
        \caption{RPWELL}
        \label{fig:RPWELL}
    \end{subfigure}
    \centering
    \begin{subfigure}[b]{0.45\textwidth}
        \includegraphics[width=\textwidth]{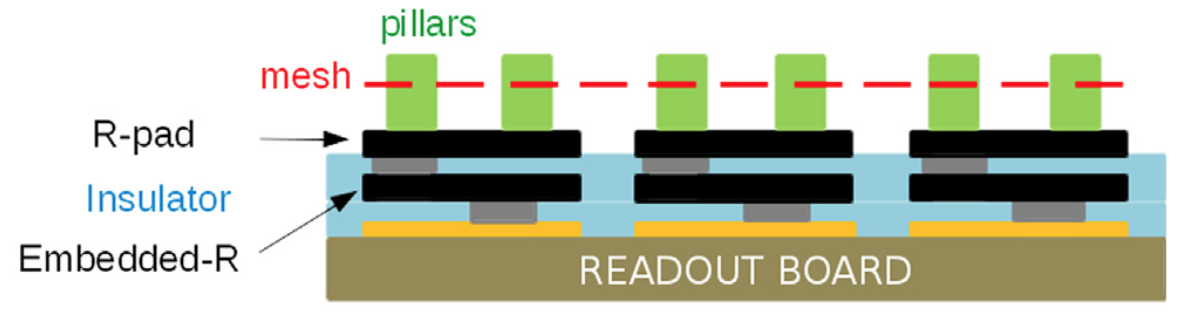}
        \centering
        \caption{Resistive MM}
        \label{fig:rMM}
    \end{subfigure}
\caption{\label{fig:detSchem} Schematic descriptions of (a) the RPWELL detector and (b) resistive MM with embedded resistors.}
\end{figure}

The RPWELL detector \cite{adamRPWELL} is shown schematically in Figure \ref{fig:RPWELL}. It consists of a single-sided WELL electrode (a THick Gaseous Electron-Multiplier (THGEM) \cite{THGEM1st} with copper-clad on its top side only) coupled to the readout anode through a layer of high bulk and surface resistivity. Radiation-induced ionization electrons drift into the WELL holes, where charge avalanche-multiplication occurs under a high electric field. Signals are induced on the readout anode by the motion of the avalanche charges. Due to electrical instabilities, the RPWELL sampling elements were operated below the efficiency plateau. The voltage applied to the 0.8 mm thick WELL electrode ($V_{\text{RPWELL}}$) and the drift cathode ($V_\text{drift}$) were 1525 V and 1675 V, respectively, corresponding to an efficiency at the order of 50\% \cite{DanThesis}. The MM sampling elements are described in details in \cite{newMM2021}, and their schematic description is shown in Figure \ref{fig:rMM}. The charge avalanche-multiplication region is defined by two parallel electrodes, a micro-mesh and a readout padded-anode located $\sim$0.1 mm away from each other. In the resistive configuration, the pads are coupled to the readout boards via embedded resistors. The current MM prototypes were operated close to the efficiency plateau; the voltage applied to the mesh ($V_\text{mesh}$) and the drift cathode were 480 V and 550 V, respectively, corresponding to an efficiency at the order of 90\% \cite{newMM2021}. The eight sampling elements were operated in Ar:CO$_2$ 93:7.

The pion beam covered an area of $\sim$40$\times$80 cm$^2$. In order to reduce the probability of shower fragments escaping the detection region, only showers at the center of the calorimeter were selected. The trigger region of 1$\times$1 cm$^2$ at the center of the transverse plane was defined using a coincidence of three scintillators. 

The DHCAL prototype was located downstream on the beamline, such that pion showers could have been initiated by interaction with other material before reaching the first layer of the DHCAL setup. These were vetoed in the analysis (see Section \ref{subsec:valid}). At low energies, the beam had a non-negligible electron component. Possible contamination from these electrons was mitigated using Cherenkov counters as electron-veto. 

\subsubsection{The Performance of a Single Sampling Element}
\label{subsec:measureMIP}

For each sampling element, we measured the MIP detection efficiency and the distribution of pad multiplicity using MIP tracks --- pions that did not initiate a shower. For a given sampling element (layer $i$), the MIP tracks were reconstructed from hits recorded by the other seven layers (reference layers) using Hough transform \cite{Hough}. To ensure high-quality MIP tracks, we selected only those parallel to the beam axis with exactly one hit, less than 1 cm away from the reconstructed track, in each reference layer. A group of neighboring hits in layer $i$ formed a cluster. Its position was defined by the average position of the hits in the cluster. An event was considered efficient if the cluster position was less than 1 cm away from the expected intersection point of the MIP track with layer $i$. The measured MIP detection efficiency was defined as the ratio of the number of efficient events over the total number of MIP tracks. For each efficient event, the measured pad multiplicity was defined by the number of hits in the cluster.

Since a MIP can induce signals on a cluster of pads, an inefficient pad does not necessarily degrades the event's detection efficiency. In other words, the MIP detection efficiency and the probability to measure a signal-above-threshold, the single hit detection efficiency (HDE), could be different. Further, the measured cluster's size (measured pad multiplicity) could be smaller than the cluster induced by the MIP (underlying pad multiplicity). Using binomial statistics as detailed in Appendix \ref{appendix:HDE}, the HDE and the underlying pad multiplicity distribution were estimated from the measured MIP detection efficiency and pad multiplicity. For detectors of low efficiency, the MIP detection efficiency is larger than the HDE. A difference at the level of 40\% was observed for the RPWELL detectors used.

\subsubsection{Simulation Framework}
\label{subsec:sim}
The simulation framework consisted of two parts. In the first one, the interaction of the impinging particle with the DHCAL under study was modeled using the GEANT4 simulation toolkit with three different physics lists: QGSP\_BERT, QGSP\_BERT\_EMZ, and FTFP\_BERT\_EMZ \cite{g4phylist, g4em}. The GEANT4 output consists of a set of energies deposited in predefined sensitive regions along with their positions. In a second part, the energy deposits are translated into a set of hits using a dedicated parametric model of the detector's response.

The DHCAL geometric model was formed by concatenating alternating absorber and sampling element modules. The RPWELL module was defined by a cover supporting the ASU (epoxy), an Active Sensor Unit (ASU, defined by PCB), a resistive-plate (SiO$_2$), electrode-plate material (FR4), electrodes (Cu), and gas (Ar:CO$_2$ 93:7). The MM layers consisted of PCB, gas (Ar:CO$_2$ 93:7), and steel covers. Only the gas volumes were defined as sensitive media. A detailed description of the material used in the simulation, along with their thickness, is given in Table \ref{tab:model}.

\begin{table}[htbp]
\centering
\caption{\label{tab:model} Details of the layers constituting the  simulated DHCAL modules.}
\smallskip
\begin{tabular}{lccc}
\hline
Module & Layer & Material & Thickness [mm]\\
\hline
Absorber & \begin{tabular}{@{}c@{}}
            air gap \\ steel plate\\ air gap
        \end{tabular}& \begin{tabular}{@{}c@{}}
            air \\ steel \\ air 
        \end{tabular} & \begin{tabular}{@{}c@{}}
            1 \\ 20\\ 1
        \end{tabular}\\ \hline
RPWELL & \begin{tabular}{@{}c@{}}
            supporting cover \\ ASU\\ 
            resistive-plate\\ WELL electrode \\
            gas gap \\ cathode
        \end{tabular} & 
        \begin{tabular}{@{}c@{}}
            epoxy \\ PCB\\ 
            SiO$_2$\\ 
            FR4 + Cu
            \\
            Ar:CO$_2$ 93:7 \\
                Cu + FR4 + Cu
        \end{tabular} & 
        \begin{tabular}{@{}c@{}}
            4 \\ 1.2\\ 
            0.7\\ 
            0.8 + 0.04\\
            3 \\ 
            0.04 + 3 + 0.04
        \end{tabular} \\ \hline
MM & \begin{tabular}{@{}c@{}}
            cover\\ air gap \\ ASU \\
            gas gap \\ cover
        \end{tabular} & 
        \begin{tabular}{@{}c@{}}
            steel \\ air \\ PCB \\
            Ar:CO$_2$ 93:7 \\ steel
        \end{tabular} & 
        \begin{tabular}{@{}c@{}}
            2\\ 0.5 \\ 1.2 \\
            3 \\ 2
        \end{tabular} \\
\hline
\end{tabular}
\end{table}

The digitization of the two technologies (RPWELL and MM) were parametrized in a similar way, thus modeled similarly. The parametric digitization was implemented in a few steps: 
\begin{itemize}
    \item Energy deposits outside the acceptance region (the sampling elements' active area) were ignored.
    \item Energy deposits inside the acceptance region were assigned the X-Y coordinates of the nearest pad. The position along the beam axis was given by the layer number. This defined a list of hits (fired pads). If more than a single energy deposit was assigned to the same pad, only one was considered.
    \item An underlying pad multiplicity distribution (mentioned in \ref{subsec:measureMIP}) was added to each hit, forming a cluster. The number of pads in the cluster was randomly sampled from the underlying pad multiplicity distribution, and neighboring pads were added to the list of hits accordingly.
    \item The HDE was applied to each hit. A random number was sampled from the uniform [0-1] distribution, and the hit was deleted if the value was greater than the quoted efficiency of the pad. A uniform efficiency was assumed across a sampling element.
\end{itemize}

The remaining list of hits represents the output of the simulation.

\subsubsection{Validation of the simulation}
\label{subsec:valid}

We modeled the DHCAL prototype described in Section \ref{subsec:expSetup} and simulated its response to low-energy (2--6 GeV) pions --- with $5\times 10^4$ pion events per energy. The agreement between the data and the simulation was validated in several steps. 

\paragraph{Closure test:}
Following the methodology discussed above, we measured the MIP detection efficiency and average pad multiplicity of each sampling element (MM and RPWELL) in the test-beam data and the simulation. These were expected to be in good agreement, the data being used as input in the simulation.

\begin{figure}[htbp]
\centering
    \begin{subfigure}[b]{0.4\textwidth}
        \includegraphics{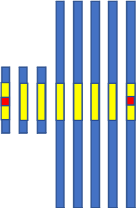}
        \centering
        \caption{MIP-like selection}
        \label{fig:miplike}
    \end{subfigure}
    \centering
    \begin{subfigure}[b]{0.4\textwidth}
        \includegraphics{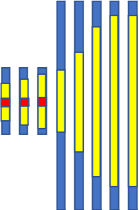}
        \centering
        \caption{Shower selection}
        \label{fig:generic}
    \end{subfigure}

\caption{\label{fig:accept} Schematic descriptions of (a) a MIP-like selection: events with a single hit in the first and last sampling elements and not more than three hits per sampling element; and (b) a shower selection region: events with exactly one hit in each of the three small sampling elements and no hits outside the expected shower-cone region, excluding MIP-like events. The red squares represent the single-hit requirements, and the yellow rectangles represent the acceptance regions.}
\end{figure}

\paragraph{Response to MIP-like events:}
A clean dataset of MIP-like events was defined by imposing strict requirements, selecting only events with a single hit in the first and last sampling elements and not more than three hits per sampling element (Figure \ref{fig:miplike}). Such events were selected in the data and the simulation. The resulting distributions of the number of hits per event were compared.

\paragraph{Response to showers:}
Shower events were selected out of those failing the MIP-like selection.
To reduce the probability of selecting events in which the shower started before the setup (Section \ref{subsec:expSetup}), events with exactly one hit in each of the first three (small) sampling elements were selected. In addition, based on typical shower topology, only events with hits confined within a cone (yellow markers in Figure \ref{fig:generic}) were selected. The resulting distributions of the number of hits per shower event in the data and the simulation were compared.

\subsection{Results}
\label{sec:results}

\subsubsection{Closure test}
\begin{figure}[ht]
\centering
    \begin{subfigure}[t]{0.48\textwidth}
        \centering
        \includegraphics[width=\textwidth]{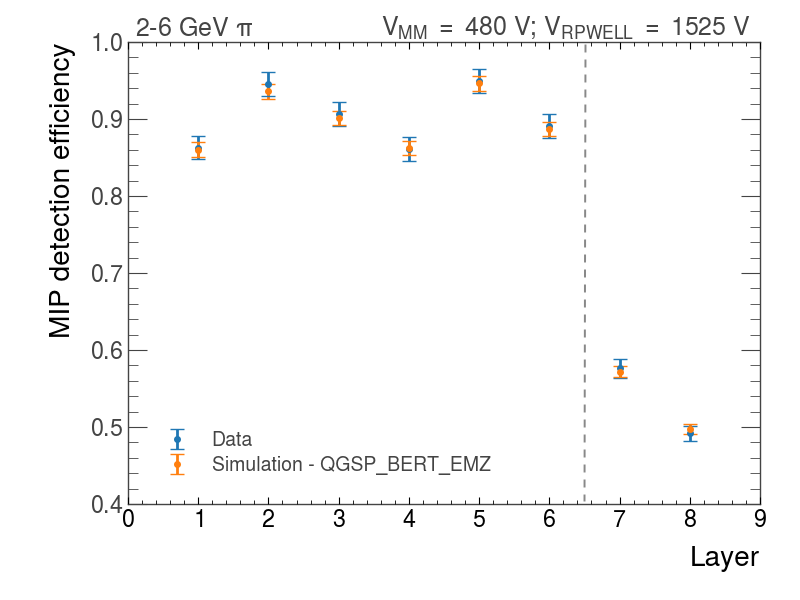}
        \caption{MIP detection efficiency}
        \label{fig:eff_Nov}
    \end{subfigure}
    \quad 
    \begin{subfigure}[t]{0.48\textwidth}
        \centering
        \includegraphics[width=\textwidth]{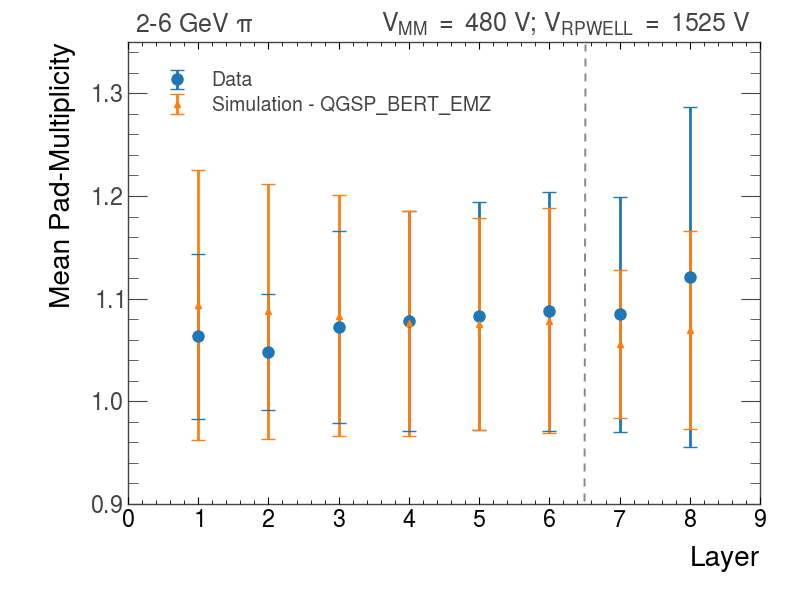}
        \caption{Average Pad Multiplicity}
        \label{fig:PM_Nov}
    \end{subfigure}

\caption{\label{fig:MIPandPM} A comparison of (a) the MIP detection efficiency and (b) average pad multiplicity of each layer as estimated from the data (blue dots) and simulation using QGSP\_BERT\_EMZ (orange triangles). The error bars represent the statistical uncertainty in (a) and the standard deviations of the pad multiplicity distributions in (b). The dashed-line separates between the MM and the RPWELL sampling elements (layers 1-6 and 7-8, respectively)}
\end{figure}

Figure \ref{fig:MIPandPM} compares the MIP-detection efficiency (left) and the average pad multiplicity (right) per sampling element measured with 2--6 GeV pions\footnote{in steps of one GeV} in the test-beam data and the simulation results. A good agreement (within statistical fluctuations) is observed, confirming that the simulated performance of an individual sampling element is consistent with the experimental values used in its parametrization.

\noindent

\subsubsection{Response to MIP-like events}
Table \ref{tab:mipLikeFrac} summarizes the number and fraction of the MIP-like events selected from the data and the simulations, for the same energies. The fractions in the data are lower than the expected ones from a calorimeter of this depth. In the simulation, we assume a pure pion beam and ignore interactions of pions with material before the calorimeter setup. Thus, imposing strict MIP selection criteria results in lower selection efficiency in the data than in the simulation.

\begin{figure}[ht]
  \centering
  \begin{tabular}[c]{ccc}
    \begin{subfigure}[c]{0.3\textwidth}
      \includegraphics[width=\textwidth]{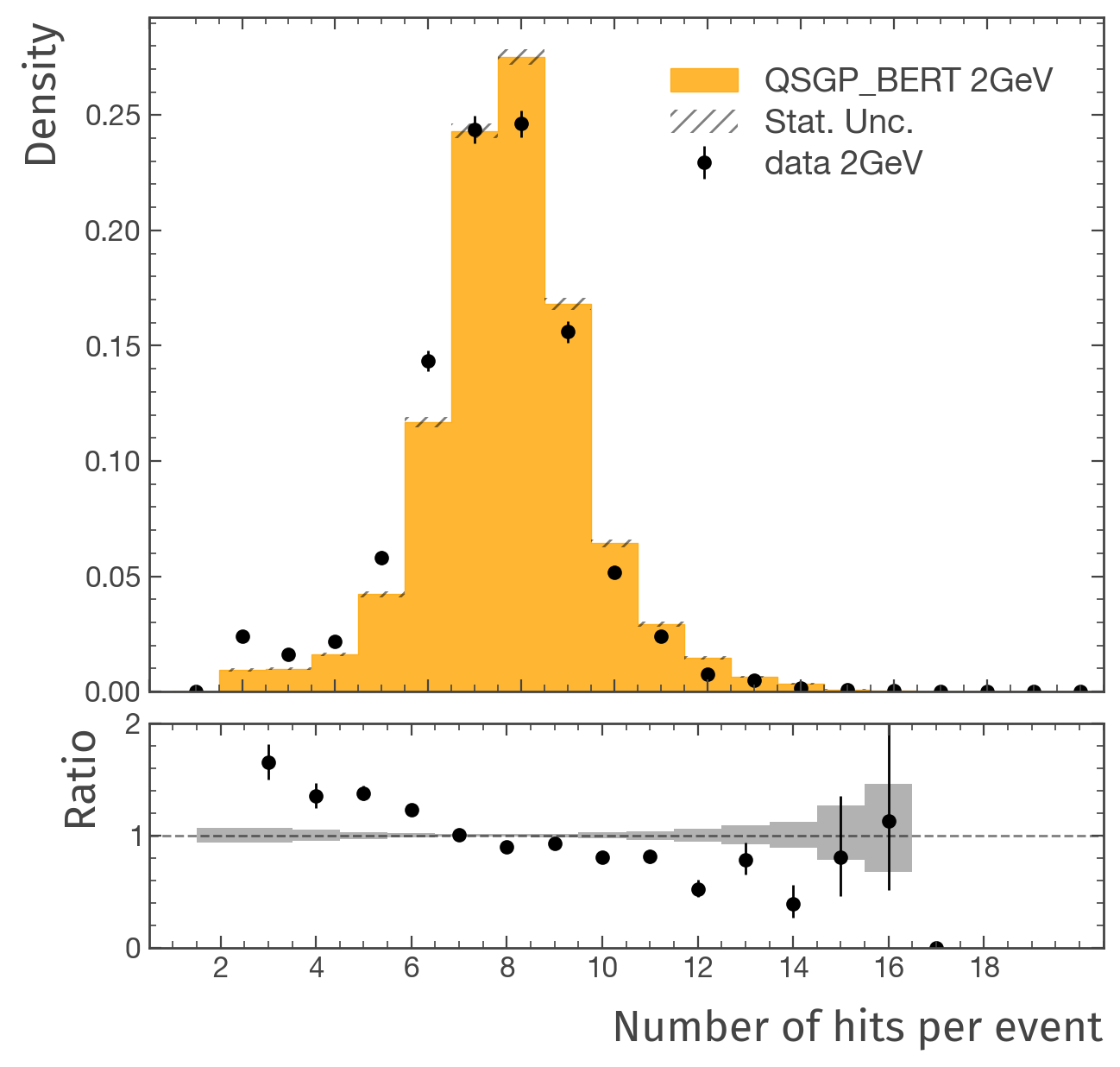}
    \end{subfigure}&
    \begin{subfigure}[c]{0.3\textwidth}
      \includegraphics[width=\textwidth]{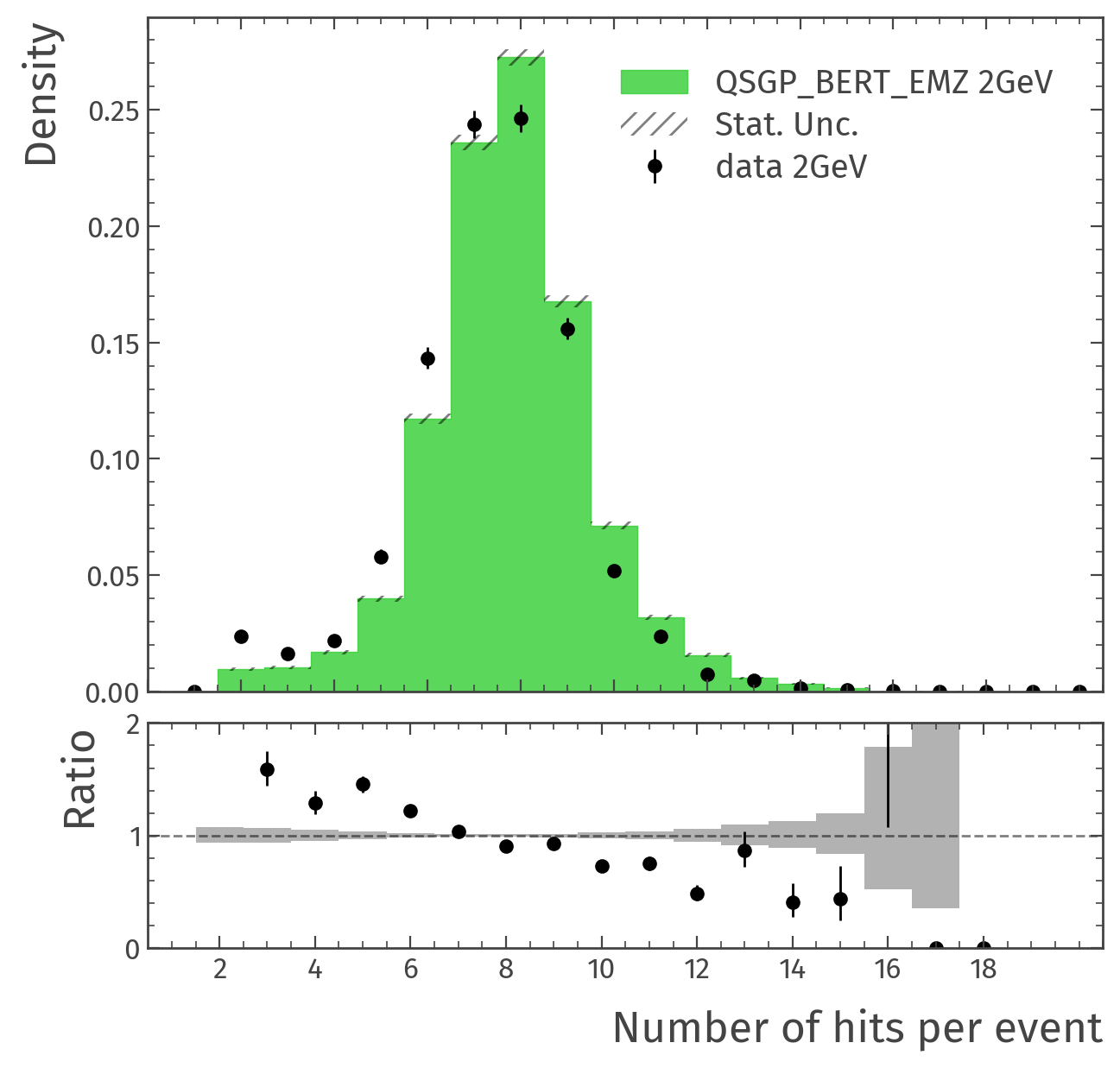}
    \end{subfigure}&
    \begin{subfigure}[c]{0.3\textwidth}
      \includegraphics[width=\textwidth]{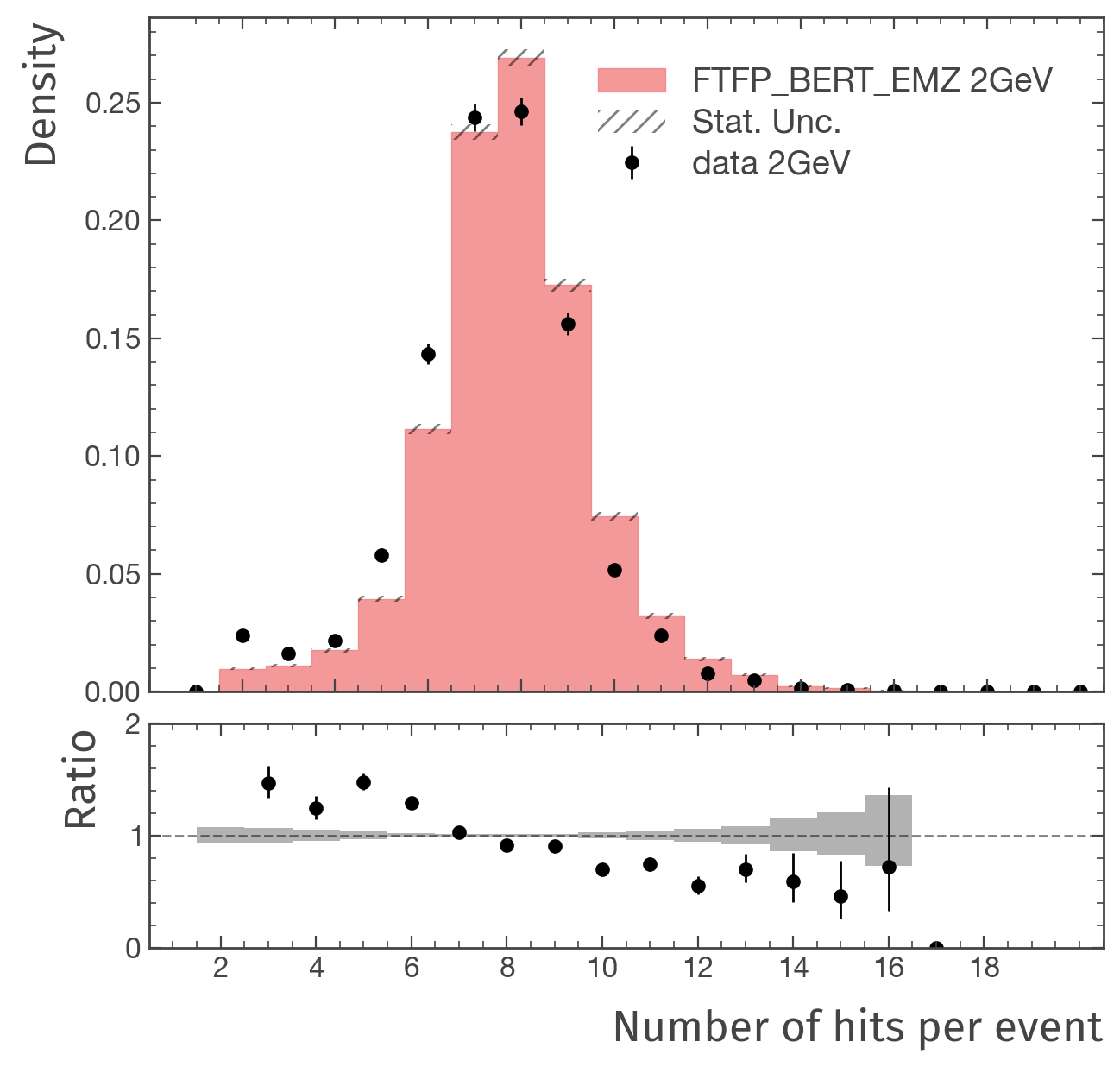}
    \end{subfigure}\\
   
    \begin{subfigure}[c]{0.3\textwidth}
      \includegraphics[width=\textwidth]{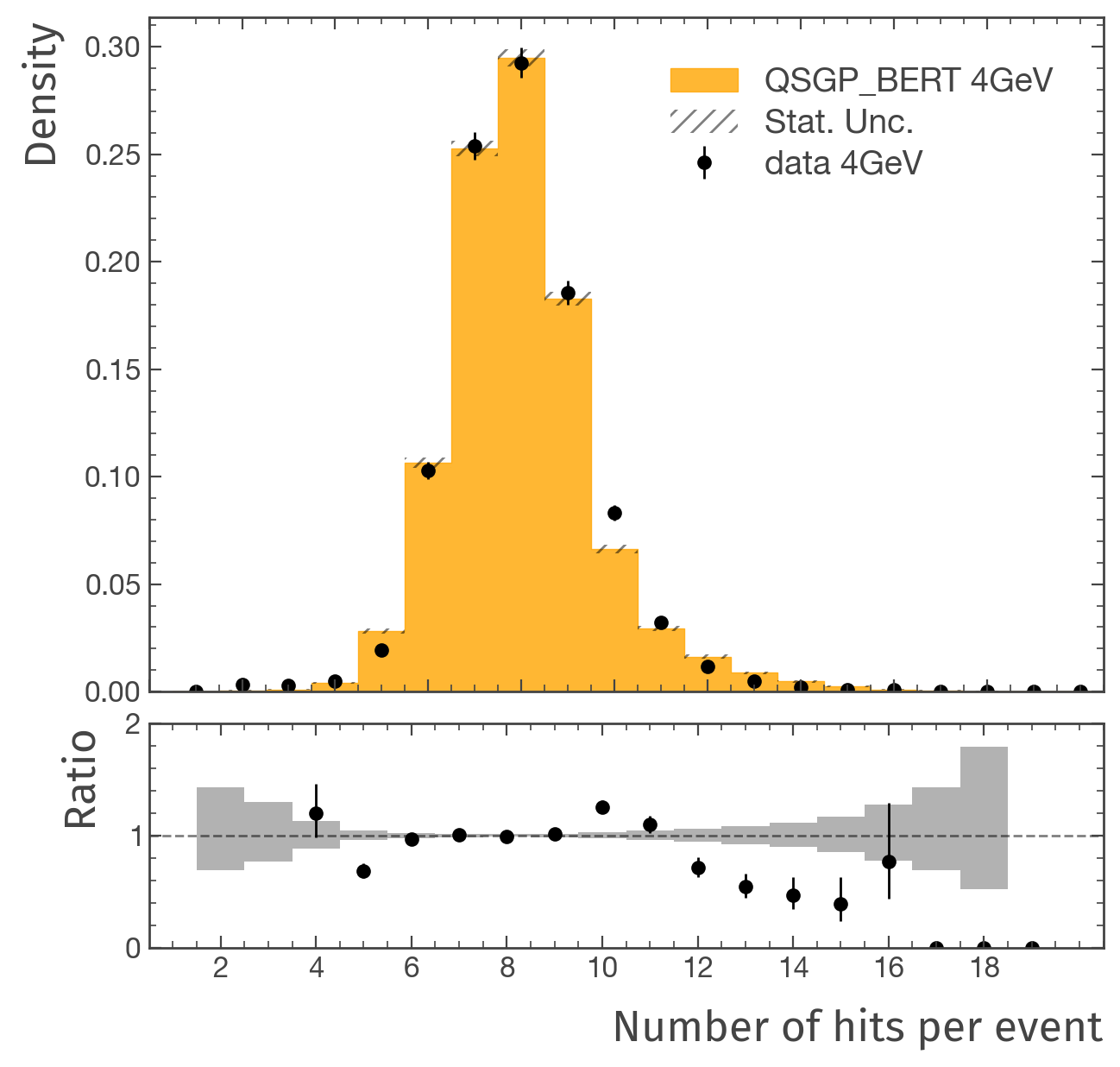}
    \end{subfigure}&
    \begin{subfigure}[c]{0.3\textwidth}
      \includegraphics[width=\textwidth]{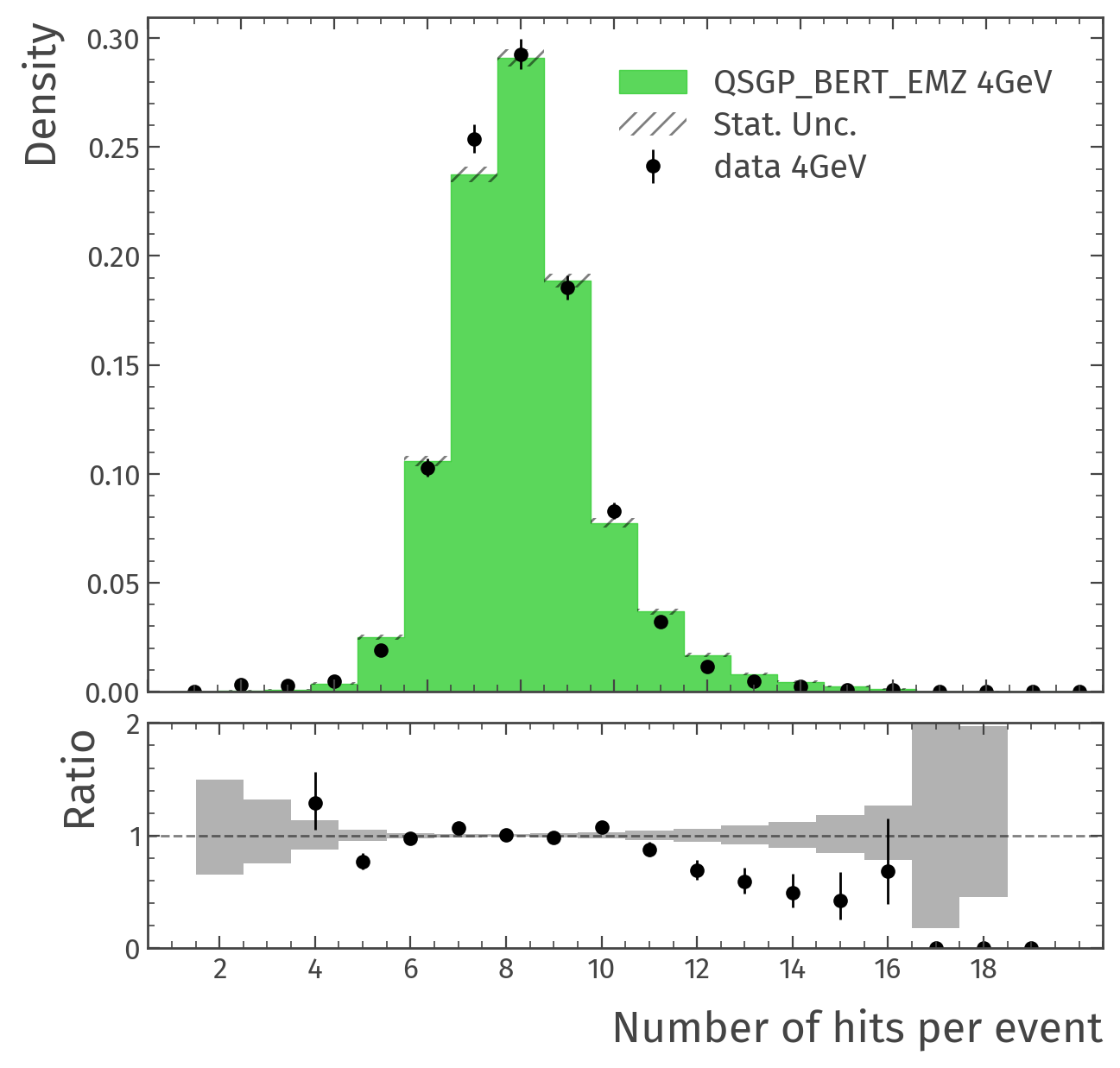}
    \end{subfigure}&
    \begin{subfigure}[c]{0.3\textwidth}
      \includegraphics[width=\textwidth]{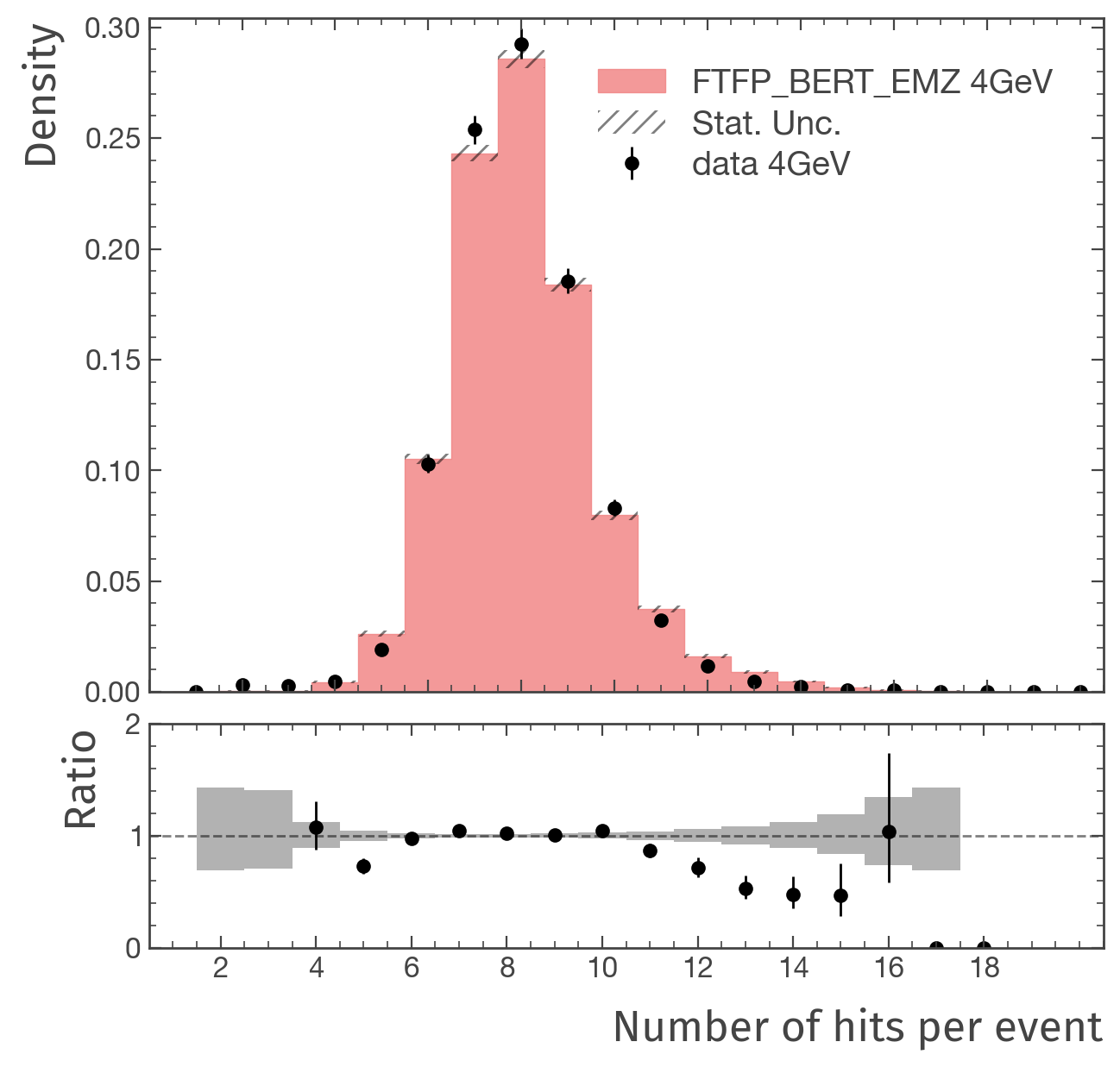}
    \end{subfigure}\\
   
    \begin{subfigure}[c]{0.3\textwidth}
      \includegraphics[width=\textwidth]{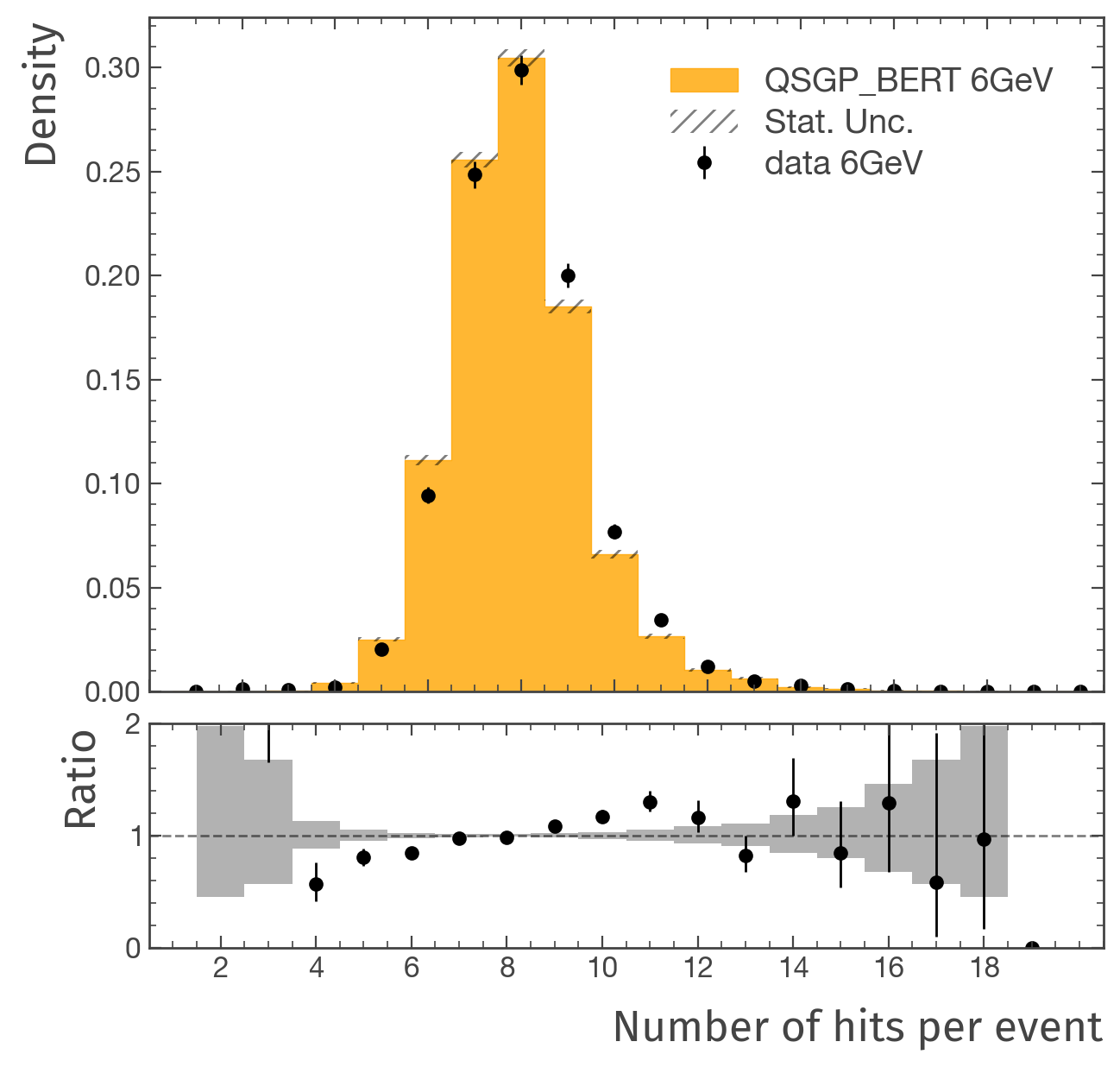}
    \end{subfigure}&
    \begin{subfigure}[c]{0.3\textwidth}
      \includegraphics[width=\textwidth]{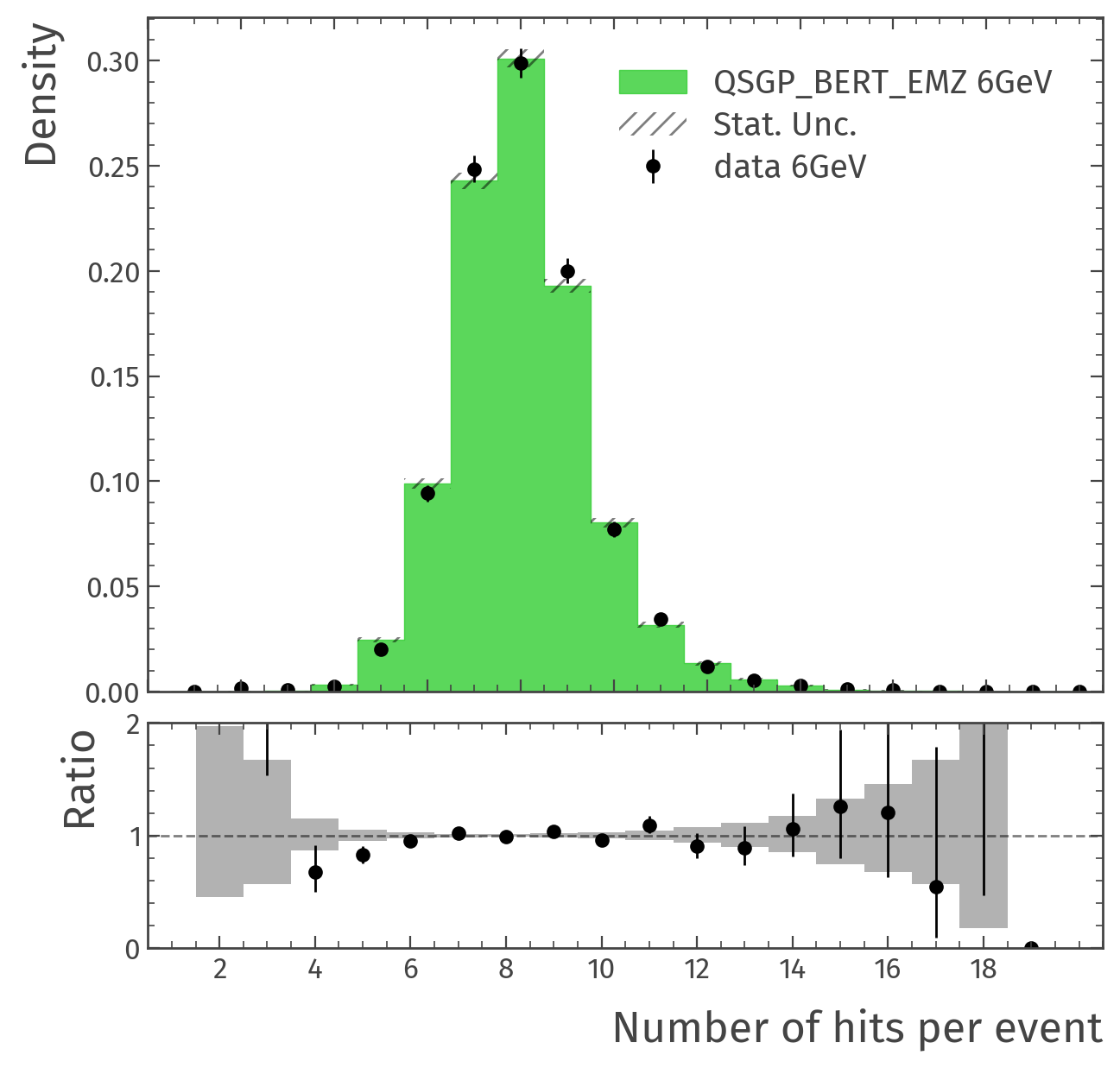}
    \end{subfigure}&
    \begin{subfigure}[c]{0.3\textwidth}
      \includegraphics[width=\textwidth]{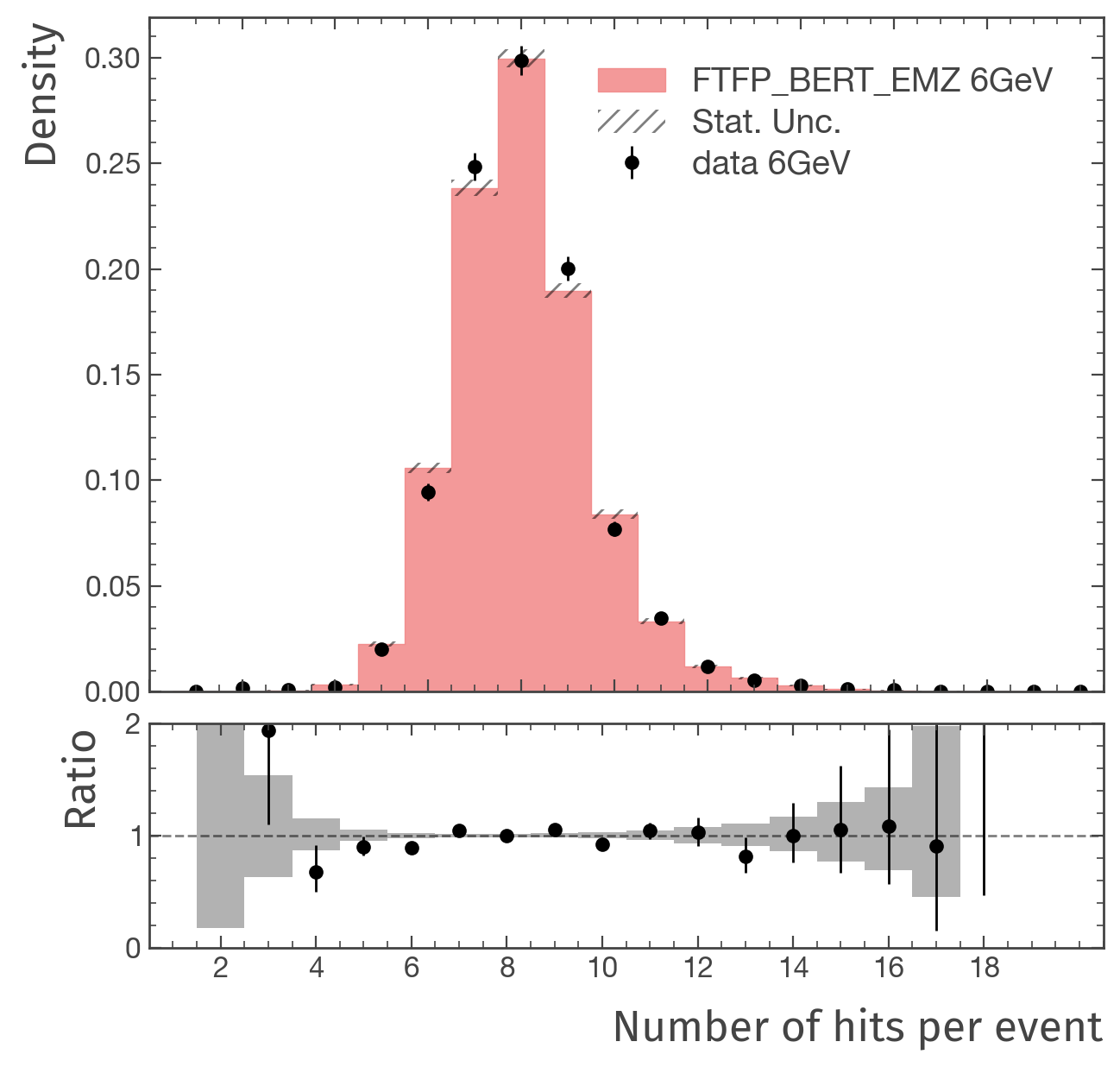}
    \end{subfigure}\\
  \end{tabular}    
  \caption{The distribution of the number of hits per MIP-like event (Figure \ref{fig:accept}-a) in the test-beam data (black dots) and the simulation using QGSP\_BERT (orange), QGSP\_BERT\_EMZ (green), and FTFP\_BERT\_EMZ (red) for different energy values of the pion beam (2, 4, and 6 GeV). The ratio of the data with respect to the simulation results is presented at the bottom of each plot. The error bars and the gray blocks represent the respective statistical uncertainty of the data and the simulation results.The top, middle, and bottom rows correspond to 2, 4 and 6 GeV, respectively.}
  \label{fig:miplike_comp}
\end{figure}

The distributions of the number of hits per event at different pion energies in MIP-like events are shown in Figure \ref{fig:miplike_comp} for pion energies of 2, 4, and 6 GeV. A similar trend was observed at 3 and 5 GeV. For each energy, the data is compared to the simulation results obtained with the three physics lists. The agreement between the data and the simulation results was improved for increasing beam energy. The distributions peak at eight hits per event corresponds to a single hit per sampling element. This is expected from detectors with an average pad multiplicity of 1.1 and such selection, containing mostly MIP events. 

\begin{table}[ht]
    \caption{\label{tab:mipLikeFrac}The number of MIP-like events selected from the data and the simulation conducted with the three physics lists. The fraction of these events from the total respective numbers of triggered and simulated ones is provided for both.}
    \centering
    \begin{tabular}{ccccc}
        \hline
        $\pi$ Energy & Data & QGSP\_BERT & QGSP\_BERT\_EMZ & FTFP\_BERT\_EMZ \\
        \hline
        2 GeV & 7250 (31.6\%) & 24600 (49.3\%) & 22771 (45.7\%) & 22764 (45.6\%) \\

        4 GeV & 6324 (32.4\%) & 19489 (39.0\%) & 18198 (36.4\%) & 18291 (36.6\%) \\
  
        6 GeV & 6144 (28.2\%) & 17831 (35.7\%) & 16619 (33.2\%) & 16653 (33.3\%) \\ 
        \hline
    \end{tabular}
    
\end{table}

\subsubsection{Response to showers:}
Table \ref{tab:showerFrac} summarizes the number and fraction of the shower events selected from the data and the simulations, for the same energies. Given the calorimeter's depth, the selection efficiency in the data and simulation is in agreement with the expectation. 
The fractions in the data agrees with those of the simulations, except in the 2 GeV. This discrepancy could originate from the beam contamination at this low energy, which was not simulated.

\begin{figure}[ht]
  \centering
  \begin{tabular}[c]{ccc}
    \begin{subfigure}[c]{0.3\textwidth}
      \includegraphics[width=\textwidth]{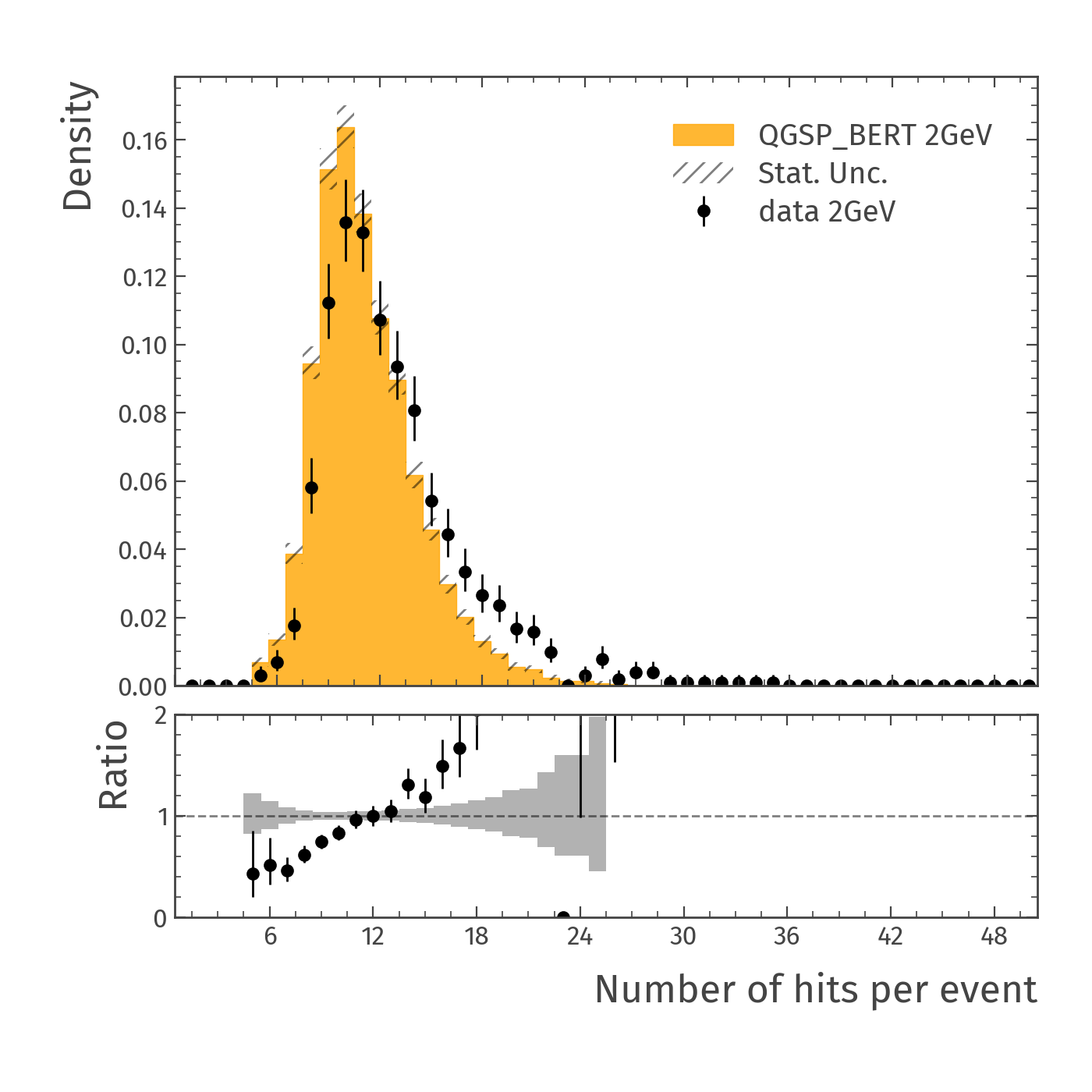}
    \end{subfigure}&
    \begin{subfigure}[c]{0.3\textwidth}
      \includegraphics[width=\textwidth]{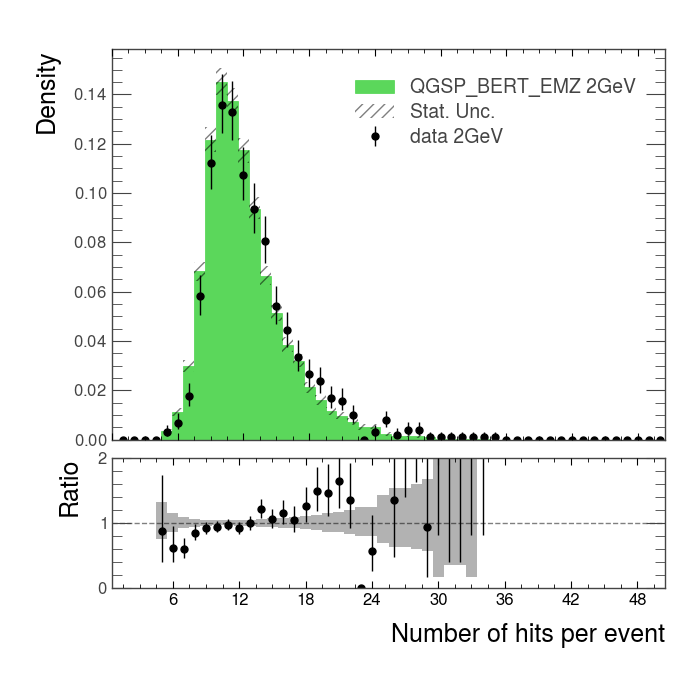}
    \end{subfigure}&
    \begin{subfigure}[c]{0.3\textwidth}
      \includegraphics[width=\textwidth]{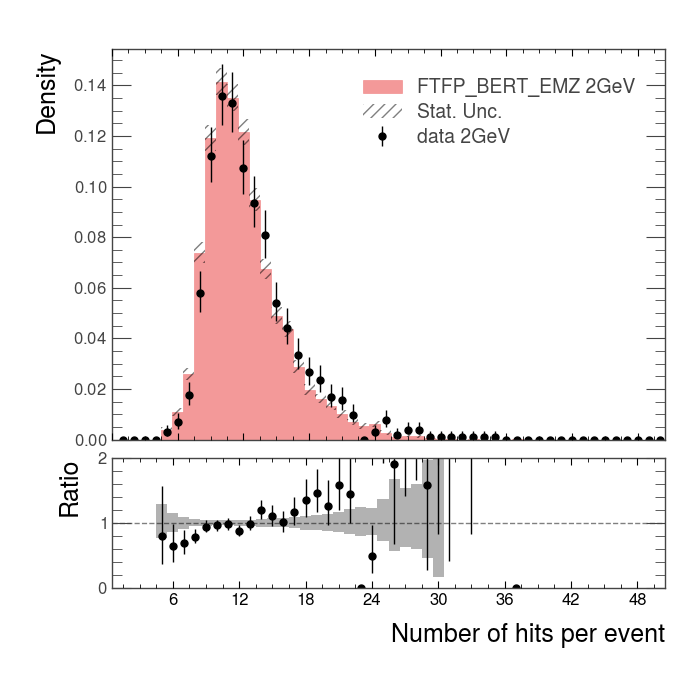}
    \end{subfigure}\\

    \begin{subfigure}[c]{0.3\textwidth}
      \includegraphics[width=\textwidth]{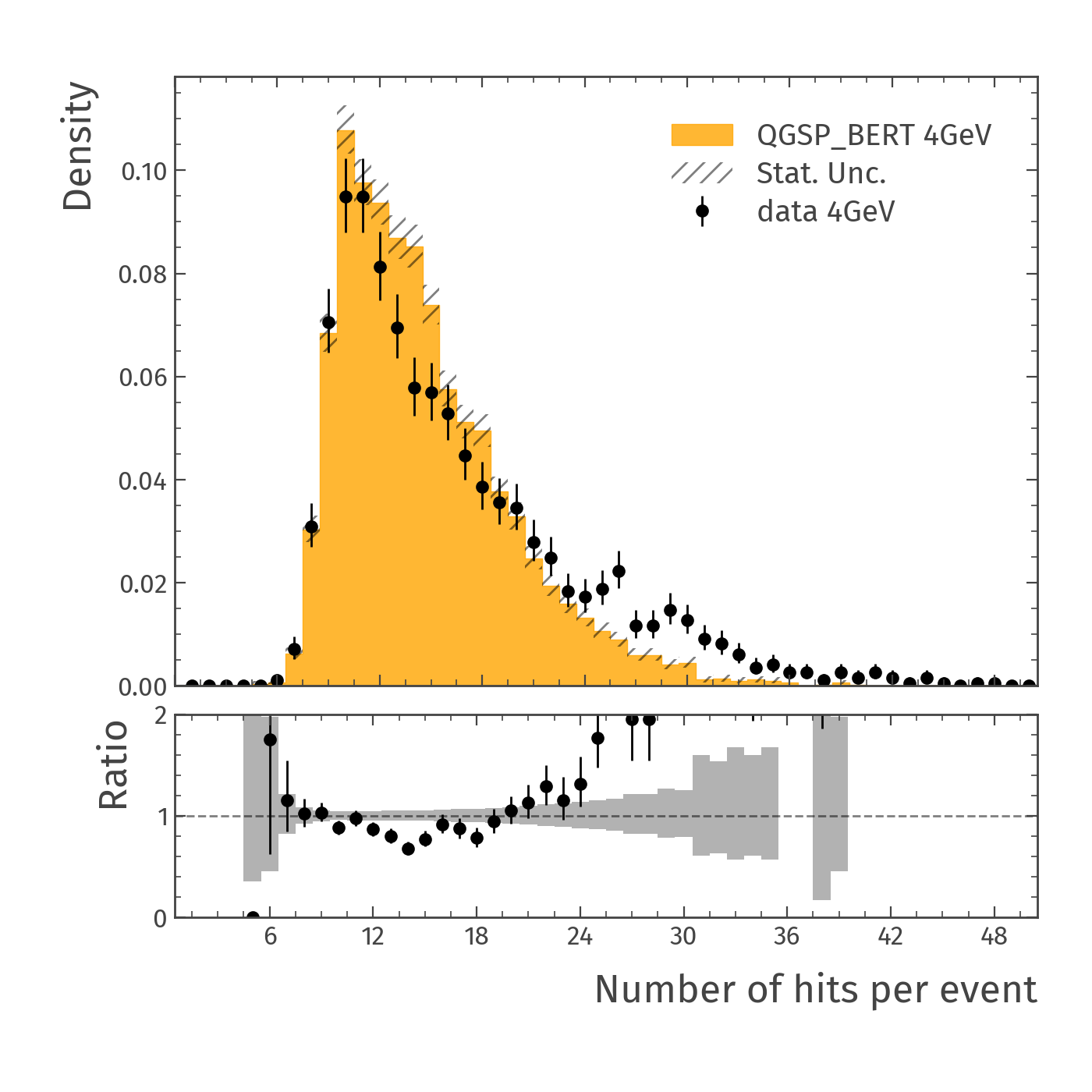}
    \end{subfigure}&
    \begin{subfigure}[c]{0.3\textwidth}
      \includegraphics[width=\textwidth]{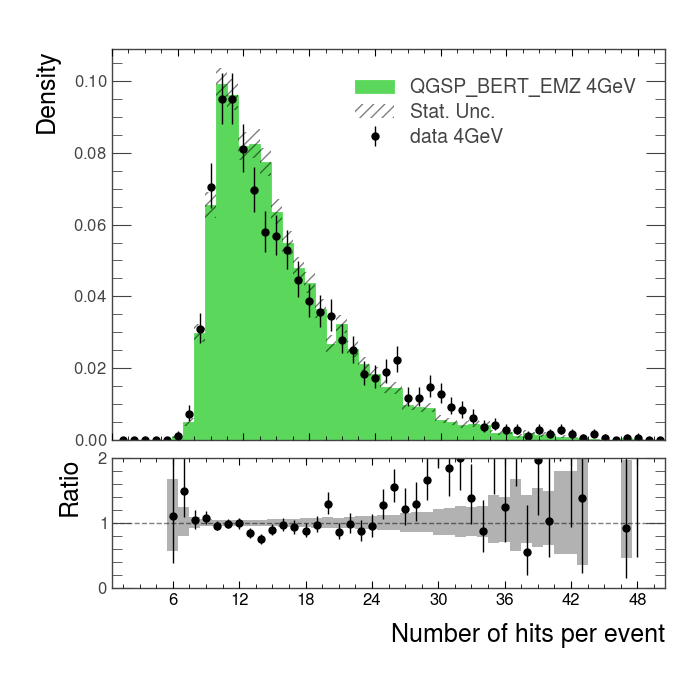}
    \end{subfigure}&
    \begin{subfigure}[c]{0.3\textwidth}
      \includegraphics[width=\textwidth]{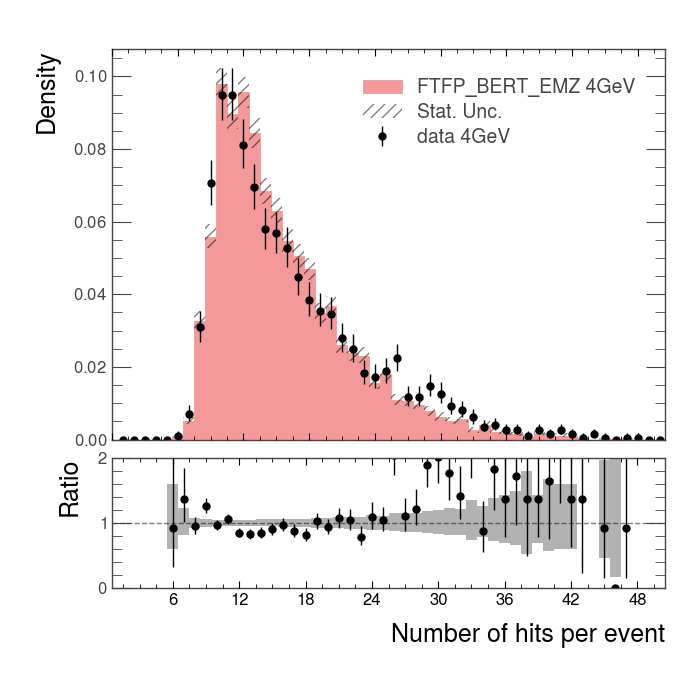}
    \end{subfigure}\\
    \begin{subfigure}[c]{0.3\textwidth}
      \includegraphics[width=\textwidth]{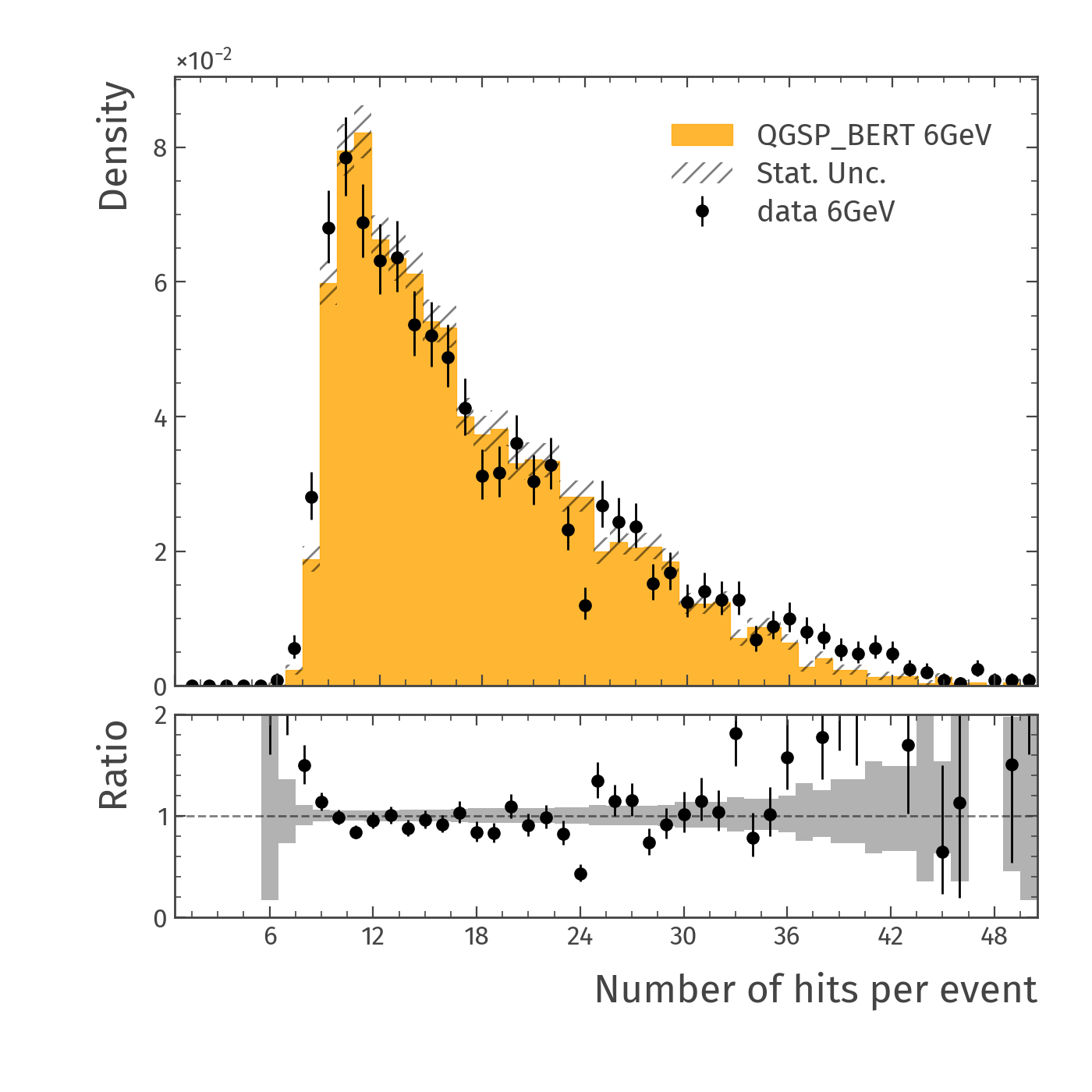}
    \end{subfigure}&
    \begin{subfigure}[c]{0.3\textwidth}
      \includegraphics[width=\textwidth]{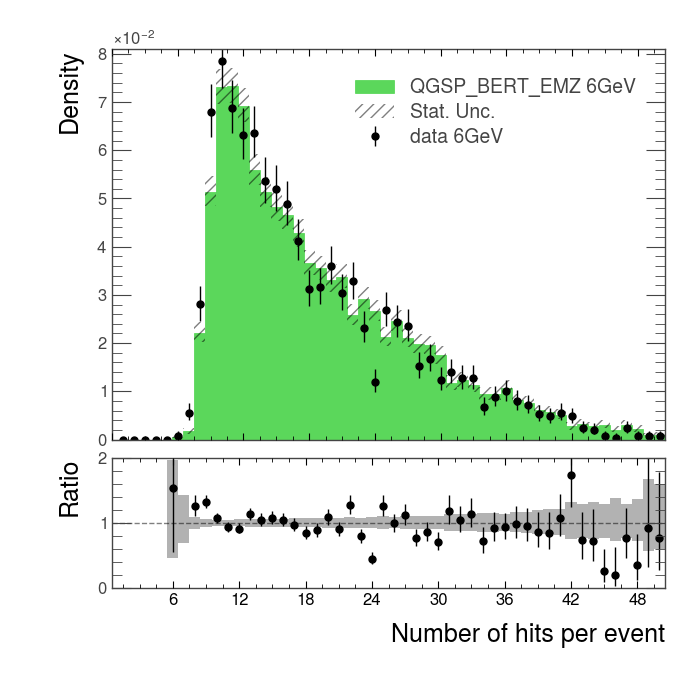}
    \end{subfigure}&
    \begin{subfigure}[c]{0.3\textwidth}
      \includegraphics[width=\textwidth]{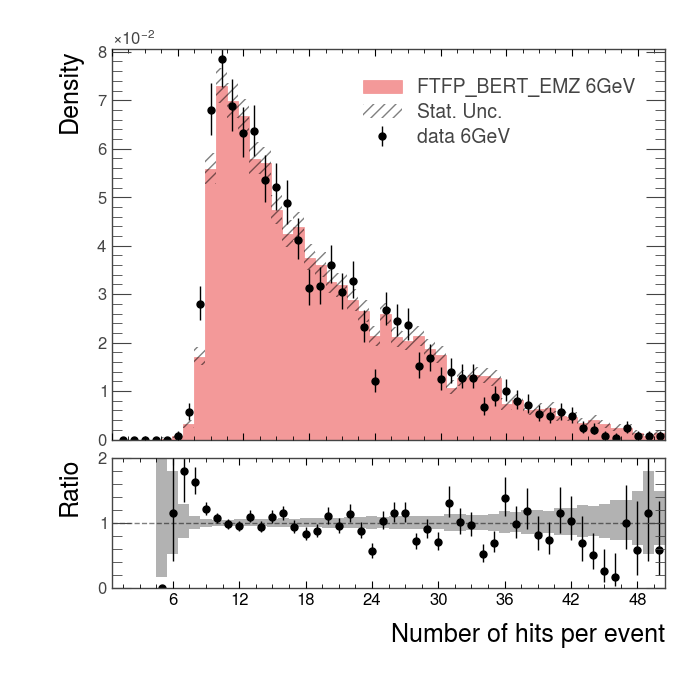}
    \end{subfigure}\\
  \end{tabular}    
  \caption{The distribution of the number of hits per shower event in the test-beam data (black dots) and the simulation using QGSP\_BERT (orange), QGSP\_BERT\_EMZ (green), and FTFP\_BERT\_EMZ (red) for different energy values of the pion beam  (2, 4, and 6 GeV). The ratio of the data with respect to the simulation results is presented at the bottom of each plot. The error bars and the gray blocks represent the respective statistical uncertainty of the data and the simulation results. The top, middle, and bottom rows correspond to 2, 4 and 6 GeV, respectively.}
  \label{fig:generic_comp}
\end{figure}

The distributions of the number of hits per shower at different pion energies are shown in Figure \ref{fig:generic_comp} for pion energies of 2, 4, and 6 GeV. A similar trend was observed at 3 and 5 GeV.For each energy, the data was compared to the simulation results obtained with the three physics lists. The agreement between the data and the simulation, also for showers, was improved as the beam energy increased. In events with larger number of hits, a better agreement between the data and the simulations is obtained with the physics lists containing the more precise EM interaction modeling (EMZ).

\begin{table}[ht]
    \caption{\label{tab:showerFrac}The number of shower events selected from the data and the simulation conducted with the three physics lists. The fraction of these events from the total respective numbers of triggered and simulated ones is provided for both.}
    \centering
    \begin{tabular}{ccccc}
        \hline
        $\pi$ Energy & Data & QGSP\_BERT & QGSP\_BERT\_EMZ & FTFP\_BERT\_EMZ \\
        \hline
        2 GeV & 1016 (4.4\%) & 4381 (8.8\%) & 4780 (9.6\%) & 4837 (9.7\%)\\
        4 GeV & 1970 (10.1\%) & 5175 (10.4\%) & 5426 (10.9\%)  & 5410 (10.8\%) \\
        6 GeV & 2515 (11.6\%) & 5659 (11.3\%) & 5806 (11.6\%) &  5804 (11.6\%) \\ 
        \hline
    \end{tabular}
\end{table}

\section{Expected Performance of a 50-layers RPWELL-based DHCAL}
\label{sec:50layers}

\begin{figure}[htbp]
    \centering
    \includegraphics[width=0.5\textwidth]{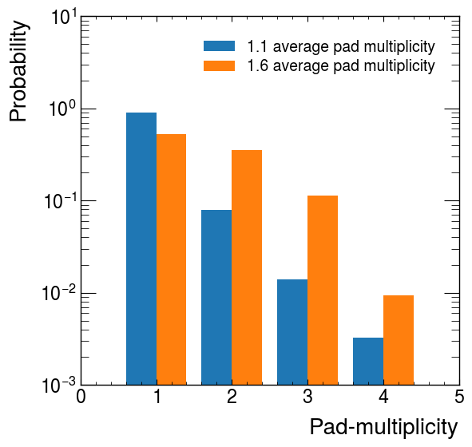}
    \caption{\label{fig:PMdist} The pad multiplicity distributions used for simulating average pad multiplicity of 1.1 and 1.6. The former was measured with a smaller RPWELL sampling element prototype using analog readout at a 150 GeV muon beam \cite{mediumSizeRPWELL}, and the latter was inspired by the average pad multiplicity value quoted for the RPC sampling elements \cite{9}}
\end{figure}

The simulation results with the QGSP\_BERT\_EMZ physics list showed a good agreement with the data for pions with energies above 4 GeV. Thus, we used the same framework to model a fully-equipped RPWELL-based DHCAL following the baseline design of calorimeters proposed for future collider experiments \cite{2,3}: 50 alternating layers of 2 cm thick steel absorbers and RPWELL sampling elements with 1$\times$1 cm$^2$ pads. We simulated the response of the DHCAL model to single pions over an energy range of 6–36 GeV, similar to the one tested with the RPC Fe-DHCAL prototype \cite{9}. $5\times 10^4$ single pion events were simulated for each energy. The expected performance of the DHCAL was evaluated for different MIP detection efficiencies and pad multiplicity distributions to study their effect on the pion energy resolution. MIP detection efficiency values in the range of 70–98\% were considered; the high value represents a realistic target --- based on the efficiency measured with smaller RPWELL prototypes \cite{mediumSizeRPWELL}; the lower value is intended to enhance possible MIP detection efficiency effect. Two pad multiplicity distributions, shown in Figure \ref{fig:PMdist}, with average values of 1.1 and 1.6, were investigated. The former was measured with a smaller RPWELL sampling element prototype using analog readout at a 150 GeV muon beam \cite{mediumSizeRPWELL}. The latter was inspired by the average pad multiplicity value quoted for the RPC sampling elements \cite{9}. 

\subsection{Methodology}\label{sec:methodSim}
\subsubsection{Event Selection}
To increase the probability for  pion energy deposition in the calorimeter (minimizing energy leakage), only events with the pion shower initiated in the first 10 layers were selected. Following \cite{eventSelection}, the selection relies on identifying \emph{the interaction layer} --- the first layer in a pion shower. It is based on the observation made in \cite{9}, that in a typical pion shower, the number of hits per layer increases in the first few layers. Each three consecutive layers are defined as a triplet. The average number of hits in the $i^{th}$ triplet is expressed as $N_{t}^i=(N_{hits}^i+N_{hits}^{i+1}+N_{hits}^{i+2})/3$, where $N_{hits}^i$ is the number of hits in the $i^{th}$ layer. The $i^{th}$ layer is defined as the interaction layer if it has more than two hits, and it is the first layer that fulfills:
\begin{equation}
    \frac{N_{t}^i}{N_{t}^{i-1}}>1.1\, \quad
\frac{N_{t}^{i+1}}{N_{t}^i}>1.1\, \quad
\frac{N_{t}^{i+2}}{N_{t}^{i+1}}>1.1
\end{equation}

Ten layers of the calorimeter are equivalent to about one $\lambda_\pi$. In agreement with the expectation, for pion energies of 6--36 GeV, the selection efficiency resulted to be $\sim$64\%.

\subsubsection{Energy Reconstruction} \label{subsec:erec}
The energy was reconstructed from the calorimeter response --- the relation between the number of hits per event ($N_{hits}$) and the beam energy ($E_{beam}$). This was parametrized using a power-law (eq. \ref{eq:pLaw} \cite{9}) and a logarithmic (eq. \ref{eq:log} \cite{12}) parametrizations (the two options suggested in \cite{9, 12}):
\begin{equation}\label{eq:pLaw}
  \langle N_{hits}\rangle= aE_{beam}^b \quad,\  b<1  
\end{equation}
\begin{equation}\label{eq:log}
\langle N_{hits} \rangle = \frac{a}{b}\log (b E_{beam}+1)
\end{equation}	    
where $\langle N_{hits}\rangle$ is the mean value of the Gaussian fit to the $N_{hits}$ distribution at $E_{beam}$, and $a$ and $b$ are free parameters. At this energy range, the $N_{hits}$ distribution is well described by a Gaussian distribution as shown in Appendix \ref{appendix:nhitsDist}.

For each shower, the reconstructed energy ($E_{rec}$) is estimated by inverting the response functions:
\begin{equation}\label{eq:pLawInv}
    E_{rec}=\sqrt[b]{\frac{N_{hits}}{a}}
\end{equation}

\begin{equation}\label{eq:logInv}
    E_{rec} = \frac{1}{b} \left[ \exp \left(\frac{b}{a} N_{hits} \right) - 1\right]
\end{equation}

For each beam energy, the average reconstructed energy ($\langle E_{rec} \rangle$ ), its associated error ($\varepsilon_{\langle E_{rec} \rangle}$), as well as the resolution ($\sigma_{rec}$) are extracted from a Gaussian fit to the reconstructed energy distribution. The agreement of the distribution with the Gaussian function is shown in Appendix \ref{appendix:erecDist}.

The bias on the energy reconstruction is defined by the relative difference between $\langle E_{rec} \rangle$ and $E_{beam}$ :
\begin{equation}
    bias =  \frac{\langle E_{rec} \rangle-E_{beam}}{E_{beam}}
\end{equation}

The energy resolution ($\frac{\sigma_{rec}}{\langle E_{rec}\rangle}$) is defined as the ratio between the width of the reconstructed energy distribution over its mean. Ignoring noise effects (electronics, detector, etc.), which were not simulated, it is parametrized as a function of the beam energy:
\begin{equation}\label{eq:res}
  \frac{\sigma_{rec}}{\langle E_{rec}\rangle}=\frac{S}{\sqrt{E_{beam}}}\oplus C
\end{equation}

\subsubsection{Uncertainties}

The uncertainty, $\varepsilon$, on the energy resolution, $\frac{\sigma_{rec}}{\langle E_{rec}\rangle}$, for each $E_{beam}$ is given by 
\begin{equation}
    \varepsilon=\varepsilon_{resp} \oplus \varepsilon_{rec}
\end{equation} 

The uncertainty of the reconstructed energy, $\varepsilon_{rec}$, is derived from the error propagation of $\varepsilon_{\langle E_{rec}\rangle}$ --- the error of the mean  value extracted from a Gaussian fit to the $E_{rec}$ distribution per beam energy. $\varepsilon_{resp}$ is the uncertainty associated with the response function. Its fit to the calorimeter response ($\langle N_{hits}\rangle$ vs. $E_{beam}$) takes into account the uncertainty on $\langle N_{hits}\rangle$. The resulting uncertainty of the fit parameters $a$, and $b$ were used to calculate the up and down variations of the response function. These were the combinations of 1$\sigma$ variations of the three parameters providing the largest up- and down-variation of the reconstructed energy, respectively. The up- and down-variations of  $\varepsilon_{resp}$ are given by the difference between the nominal energy resolution $\left(\frac{\sigma_{rec}}{\langle E_{rec} \rangle}\right)$ and its respective up- and down- variations.

\subsection{Results}\label{subsec:resultSim}
\subsubsection{Expected performance of an RPWELL-based DHCAL}
\begin{figure}[htbp]
\centering 
\includegraphics[width=0.7\textwidth]{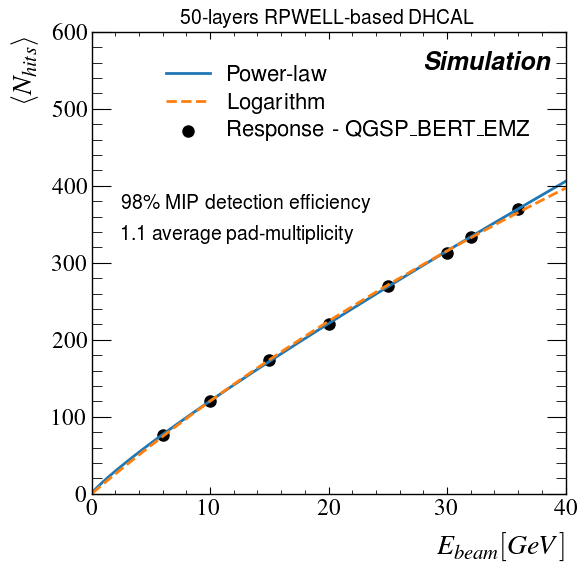}

\caption{\label{fig:res98eff1p1} The simulated response of an optimal 50-layers RPWELL-based DHCAL (black dots) and the fit of the power-law (blue line) and the logarithmic (orange dashed line) parametrizations.}
\end{figure}

Based on \cite{mediumSizeRPWELL}, we assumed an optimal RPWELL-based DHCAL consisting of sampling elements with 98\% MIP detection efficiency and 1.1 average pad multiplicity. Its response is shown in Figure \ref{fig:res98eff1p1} with the two parametrizations overlaid; the fit parameters are summarized in Table \ref{tab:response}. The $N_{hits}$ distributions of each beam energy and their corresponding Gaussian fits are presented in Appendix \ref{appendix:nhitsDist}.

\begin{figure}[htbp]
\centering 
\includegraphics[width=0.7\textwidth]{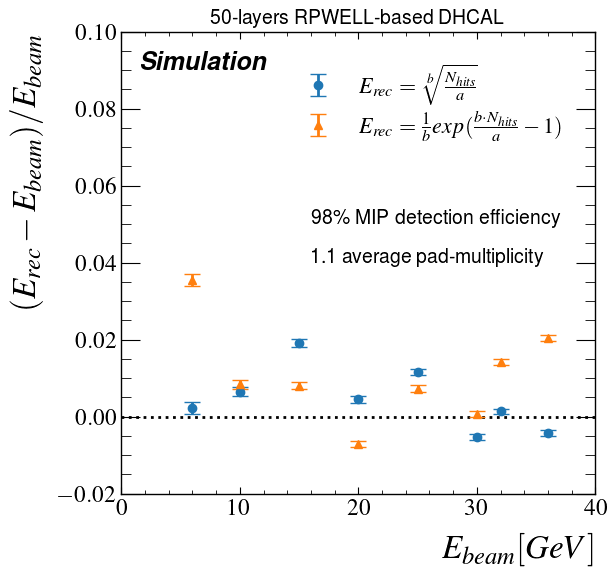}
\caption{\label{fig:bias98eff1p1} The bias on the energy reconstruction as a function of the beam energy. As calculated from the power-law (blue dots) and the logarithmic (orange triangles) response parametrizations of an optimal 50-layers RPWELL-based DHCAL.}
\end{figure}

The bias on the energy reconstruction as a function of the pion beam energy is shown for the two parametrizations in Figure \ref{fig:bias98eff1p1}. An average bias of 0.7\% (2\% at most) is obtained with the power-law parametrization, better than the average of 1.2\% (3.5\% at most) obtained with the logarithmic one. 

\begin{figure}[htbp]
\centering 
\includegraphics[width=0.7\textwidth]{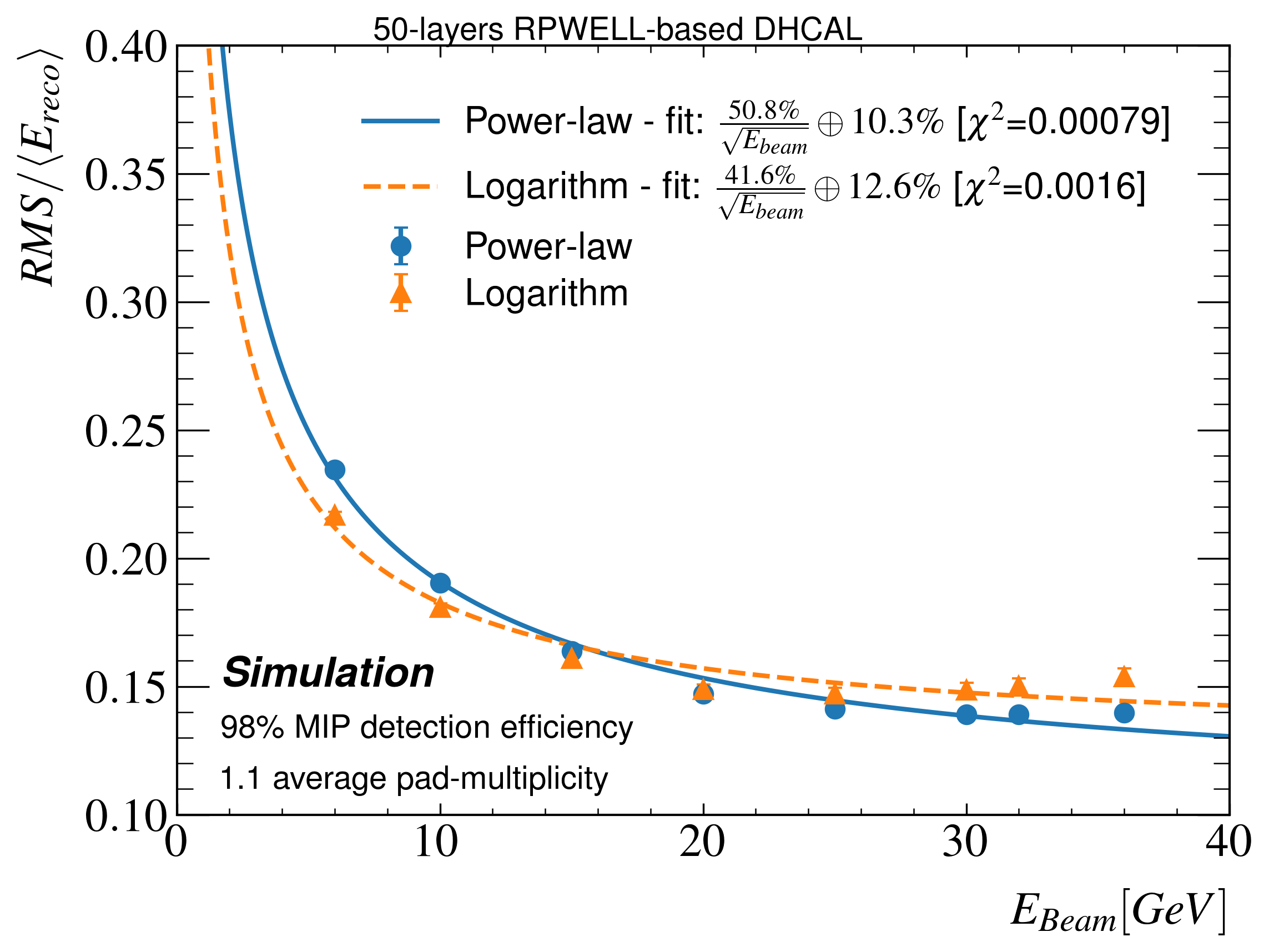}
\caption{\label{fig:eres98eff1p1} The simulated energy resolution as a function of the beam energy, as calculated from the power-law (blue dots) and the logarithmic (orange triangles) response parametrizations of an optimal 50-layers RPWELL-based DHCAL. Their corresponding fits are marked with a full line and dashed line, respectively.}
\end{figure}

The energy resolution as a function of the pion beam energy is shown for the two parametrizations in Figure \ref{fig:eres98eff1p1}. An energy resolution of $\frac{\sigma}{E[GeV]} =\frac{(50.8 \pm 0.3)\%}{\sqrt{E[GeV]}} \oplus(10.3 \pm 0.5)\%$ is expected when using the power-law parametrization, and $\frac{\sigma}{E[GeV]} =\frac{(41.6 \pm 0.4) \%}{\sqrt{E[GeV]}} \oplus (12.6 \pm 0.4)\%$ when using the logarithmic one. Thus, at pion energies below 15 GeV, better energy resolution is obtained using the logarithmic parametrization, and at higher energies, the power-law parametrization is preferable.

\subsubsection{The Effect of MIP Detection Efficiency}
\begin{figure}[htbp]

\centering
    \begin{subfigure}[b]{0.45\textwidth}
        \includegraphics[width=\textwidth]{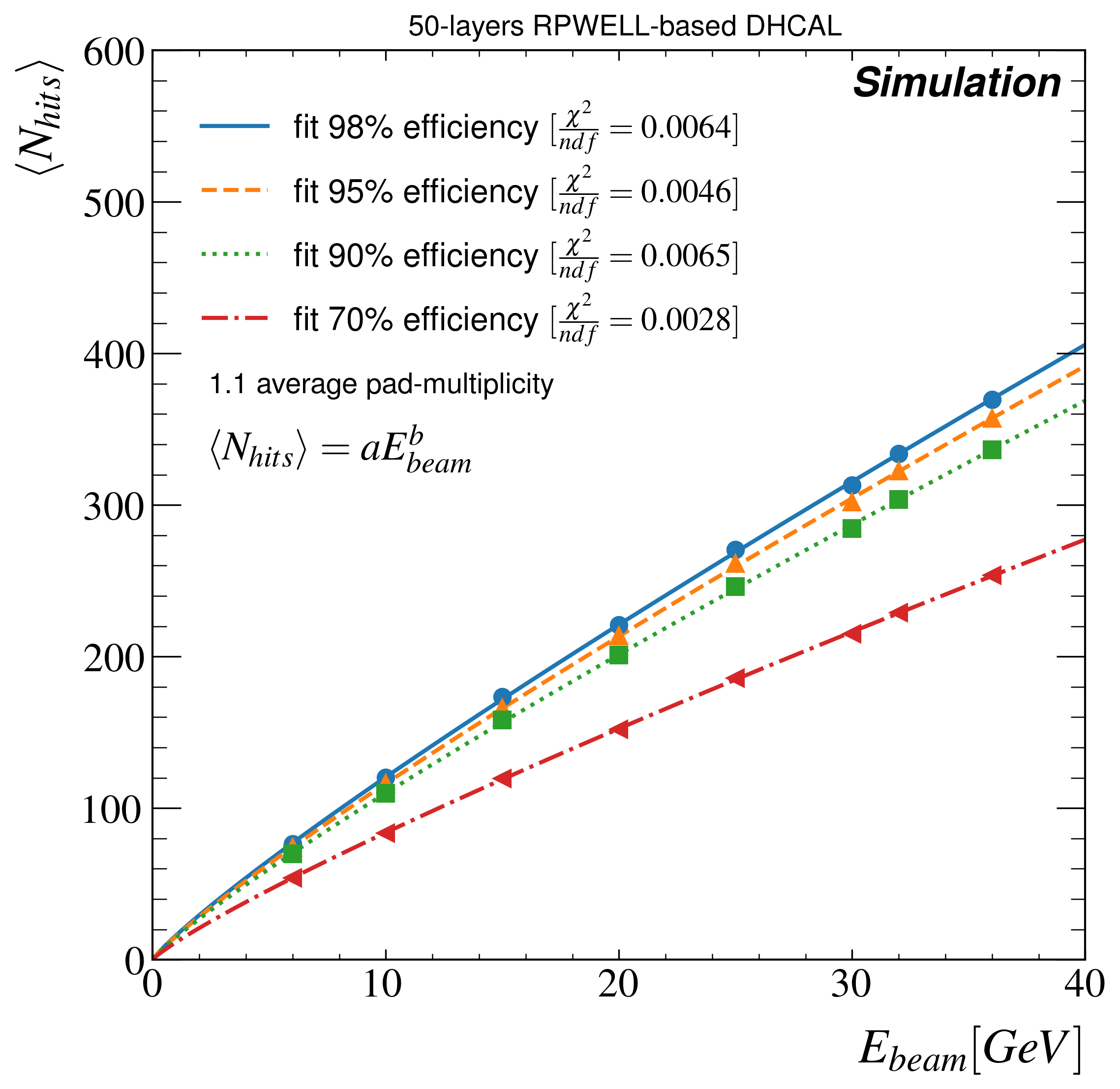}
        \centering
        \caption{Power-law}
        \label{fig:respPLaw}
    \end{subfigure}
    \qquad
    \centering
    \begin{subfigure}[b]{0.45\textwidth}
        \includegraphics[width=\textwidth]{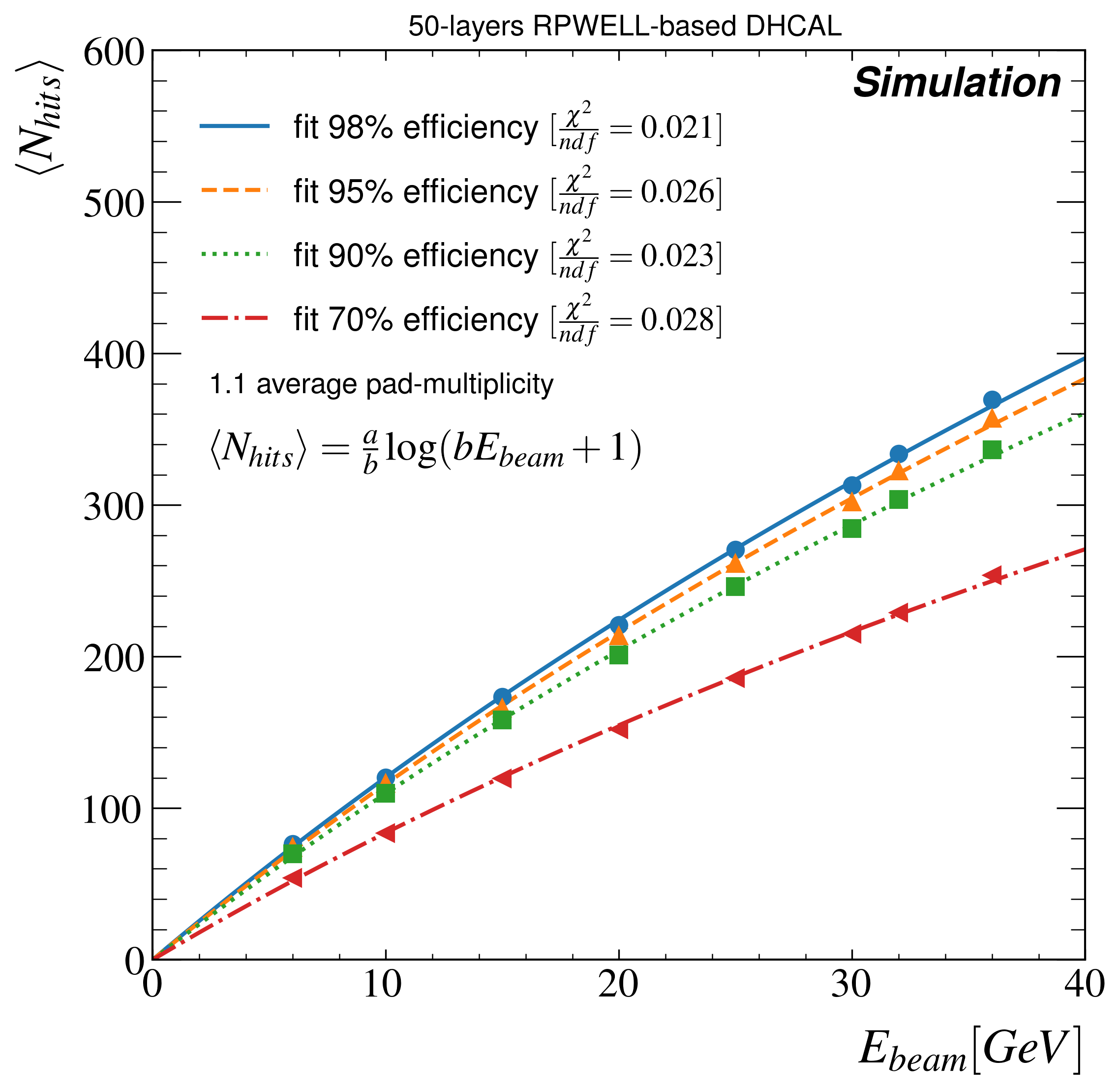}
        \centering
        \caption{Logarithm}
        \label{fig:respLog}
    \end{subfigure}

\caption{\label{fig:response} The simulated response (see Section \ref{subsec:erec}) of a 50-layers RPWELL-based DHCAL with an average pad multiplicity of 1.1 and MIP detection efficiency of 98\% (blue dots), 95\% (orange up-pointing triangles), 90\% (green squares), and 70\% (red left-pointing triangles). The response was obtained using (a) power-law and (b) logarithmic parametrizations.}
\end{figure}

We evaluated the effect of the MIP detection efficiency on the DHCAL performance at a fixed pad multiplicity distribution with an average of 1.1. For each efficiency value, the $N_{hits}$ distributions of each beam energy and their corresponding Gaussian fits are presented in Appendix \ref{appendix:nhitsDist}. The response for different MIP detection efficiency values is shown in Figure \ref{fig:response} for the power-law and logarithmic parametrizations. As indicated by the $\chi^2$ values, good fits were obtained. The fit parameters are summarized in Table \ref{tab:response}. 

\begin{table}[htbp]
\centering
\caption{\label{tab:response} The results of the response parametrization fit for different MIP detection efficiencies.}
\smallskip
\begin{tabular}{cccccc}
\hline
Parametrization	& \begin{tabular}{@{}c@{}} MIP-Detection \\ Efficiency\end{tabular}	& a	& b \\
\hline
    \begin{tabular}{@{}c@{}}
        Power-law \\$\langle N_{hits}\rangle= aE_{beam}^b$ \\(eq. \ref{eq:pLaw})
    \end{tabular} &
    \begin{tabular}{@{}c@{}}
        98\% \\ 95\%\\ 90\% \\ 70\%
     \end{tabular} &
     \begin{tabular}{@{}c@{}}
     16.04 $\pm$ 0.23 \\ 15.45 $\pm$ 0.19\\ 14.72 $\pm$ 0.22\\ 11.51 $\pm$ 0.13
     \end{tabular} &
     \begin{tabular}{@{}c@{}}
     0.876 $\pm$ 0.005 \\ 0.876 $\pm$ 0.004 \\ 0.873 $\pm$ 0.005 \\ 0.863 $\pm$ 0.004
     \end{tabular} \\
\hline
    \begin{tabular}{@{}c@{}}
        Logarithm \\ $\langle N_{hits} \rangle = \frac{a}{b}\log (b E_{beam}+1)$\\ (eq. \ref{eq:log})
    \end{tabular} &
    \begin{tabular}{@{}c@{}}
        98\% \\ 95\%\\ 90\% \\ 70\%
     \end{tabular} &
     \begin{tabular}{@{}c@{}}
     	13.04 $\pm$ 0.18 \\ 12.55 $\pm$ 0.20\\ 11.91 $\pm$ 0.18\\ 9.17 $\pm$ 0.18
     \end{tabular} &
     \begin{tabular}{@{}c@{}}
     	0.0172 $\pm$ 0.001 \\ 0.0169 $\pm$ 0.002 \\ 0.0176 $\pm$ 0.002 \\ 0.0196 $\pm$ 0.002
     \end{tabular} \\
\hline
\end{tabular}
\end{table}

\begin{figure}[htbp]
\centering
    \begin{subfigure}[b]{0.45\textwidth}
        \includegraphics[width=\textwidth]{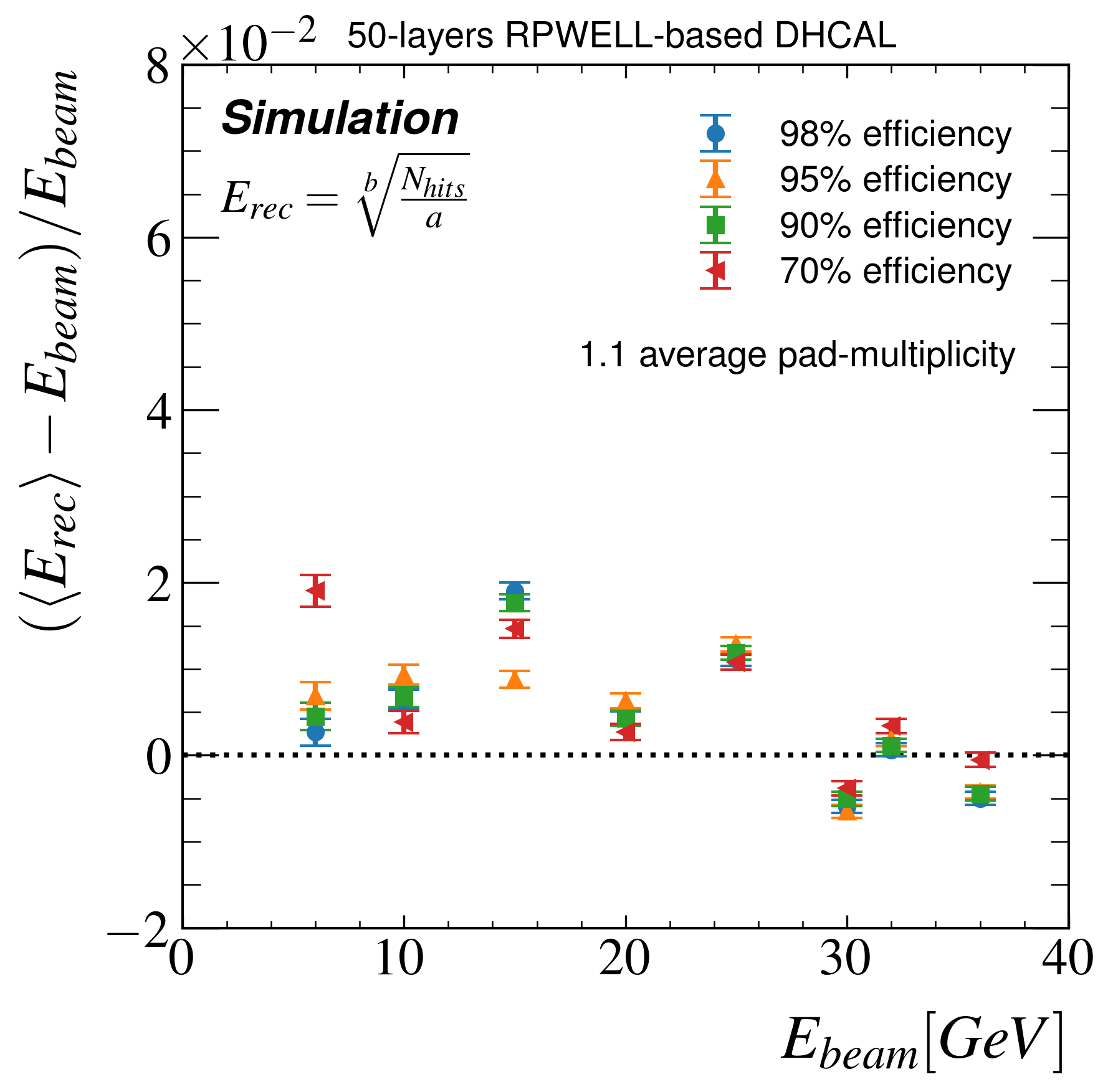}
        \centering
        \caption{Power-law}
        \label{fig:biasPLaw}
    \end{subfigure}
    \qquad
    \centering
    \begin{subfigure}[b]{0.45\textwidth}
        \includegraphics[width=\textwidth]{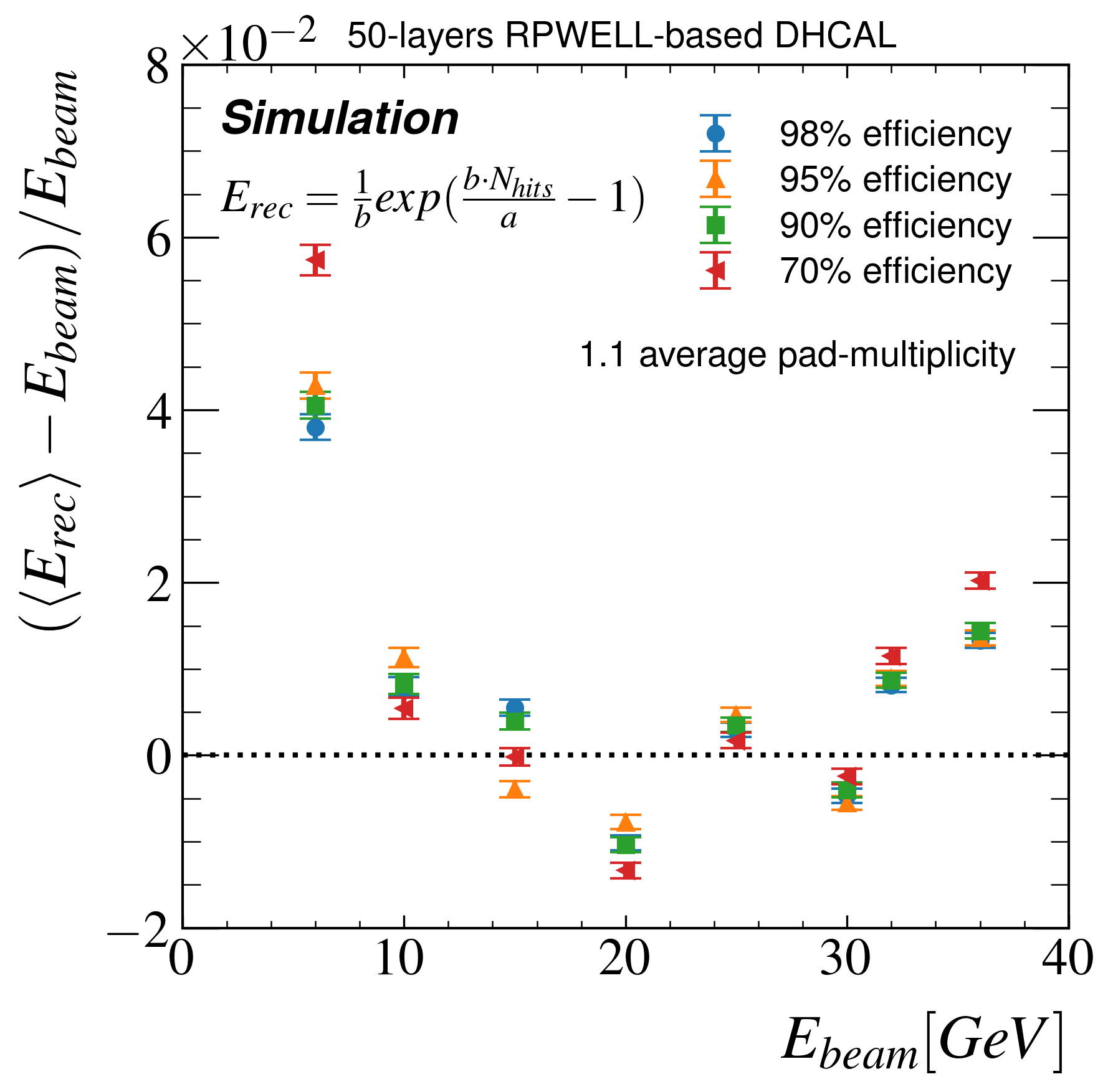}
        \centering
        \caption{Logarithm}
        \label{fig:biasLog}
    \end{subfigure}

\caption{\label{fig:bias} The pion energy reconstruction bias for the (a) power-law and (b) logarithmic parametrizations; it was obtained with a simulated 50-layers RPWELL-based DHCAL with an average pad multiplicity of 1.1 and MIP detection efficiency of 98\% (blue dots), 95\% (orange up-pointing triangles), 90\% (green squares), and 70\% (red left-pointing triangles).}
\end{figure}

Figure \ref{fig:bias} shows the bias on the pion energy reconstruction, for different MIP detection efficiencies, as a function of the beam energy for the power-law and logarithmic parametrizations. The bias measured with the former is smaller than the one measured with the latter, for which a large bias is measured at low and high energy values. In general, the change in the MIP detection efficiency has a small effect on the reconstruction bias.

\begin{figure}[htbp]

\centering
    \begin{subfigure}[b]{0.45\textwidth}
        \includegraphics[width=\textwidth]{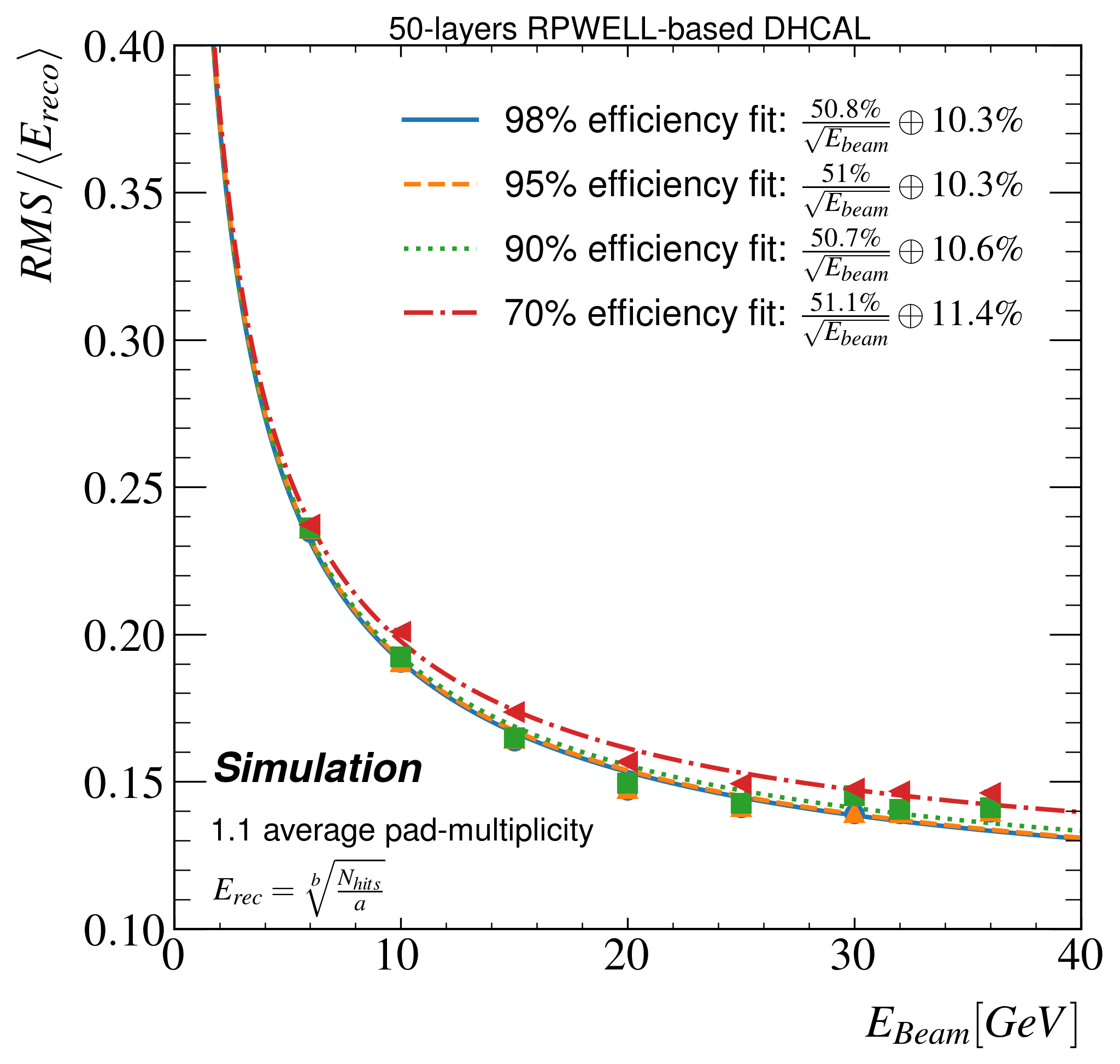}
        \centering
        \caption{Power-law}
        \label{fig:piResPLaw}
    \end{subfigure}
    \qquad
    \centering
    \begin{subfigure}[b]{0.45\textwidth}
        \includegraphics[width=\textwidth]{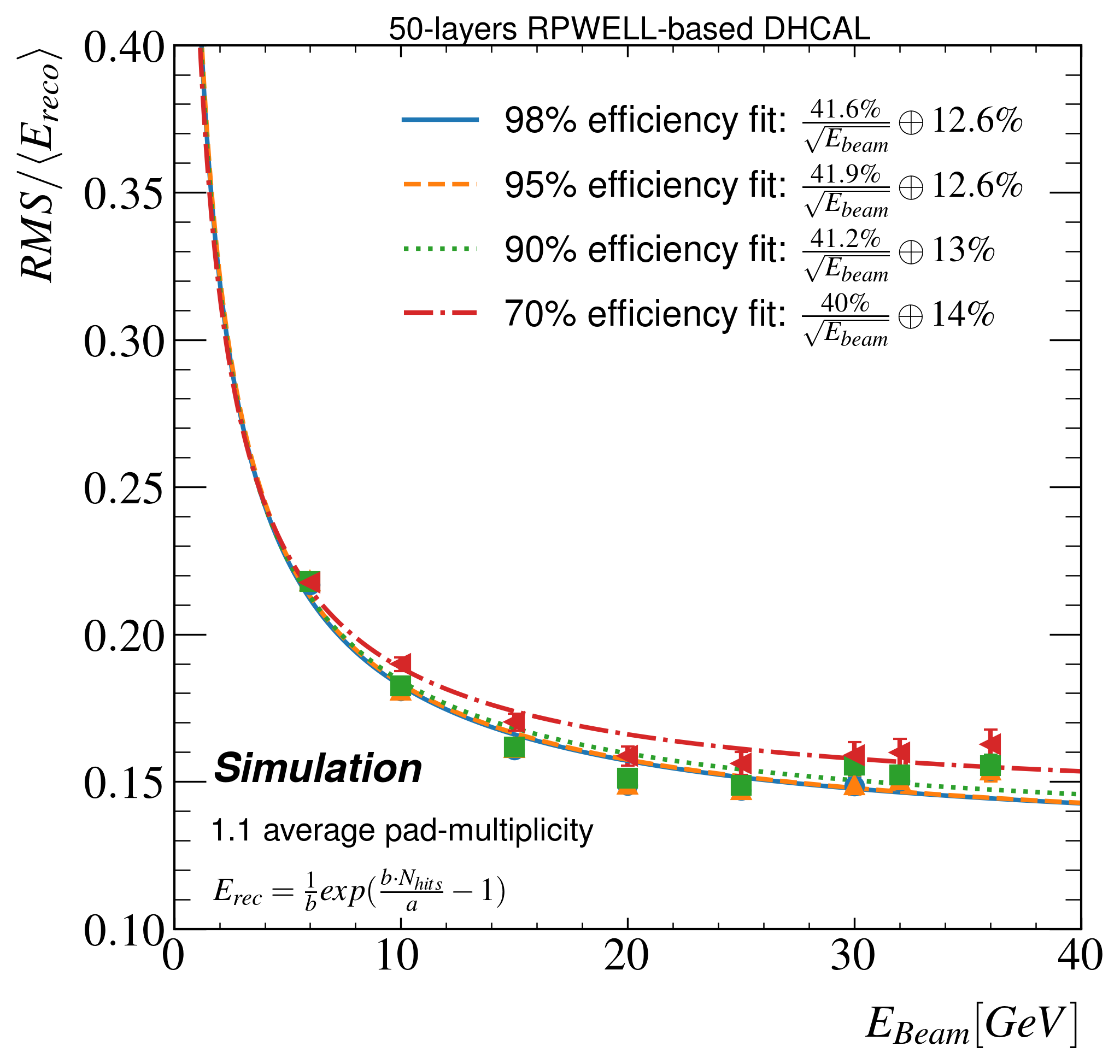}
        \centering
        \caption{Logarithm}
        \label{fig:piResLog}
    \end{subfigure}

\caption{\label{fig:piRes} The pion energy resolution of a simulated 50-layers RPWELL-based DHCAL with an average pad multiplicity of 1.1 and MIP detection efficiency of 98\% (blue dots), 95\% (orange up-pointing triangles), 90\% (green squares), and 70\% (red left-pointing triangles); using the (a) the power-law and (b) the logarithmic parametrizations.}
\end{figure}

Figure \ref{fig:piRes} presents the energy resolution for the different MIP detection efficiency values of the power-law and logarithmic parametrizations. The fit parameters and their corresponding uncertainties are listed in Table \ref{tab:eres}. A slightly better energy resolution is obtained at higher MIP detection efficiencies. In the power-law parametrization, lowering the MIP detection efficiency from 98\% to 70\% results in an insignificant difference in the stochastic term and a difference of 1.1\% in the constant term. In the logarithmic parametrization, respective differences of 1.4\% and 1.3\% are observed.
The effect on the energy resolution from a uniform change in the MIP detection efficiency is more pronounced with the logarithmic parametrization.

\begin{table}[htbp]
\centering
\caption{\label{tab:eres} A summary of the energy resolution values of a simulated 50-layers RPWELL-based DHCAL with an average pad multiplicity of 1.1 and MIP detection efficiency of 98\%, 95\% , 90\%, and 70\%, using the power-law and logarithmic parametrizations. The energy resolution is represented by the stochastic (S) and constant (C) terms (eq. \ref{eq:res})}
\smallskip
\begin{tabular}{ccccc}
\hline
\begin{tabular}{@{}c@{}} Response function\\ Parametrization\end{tabular}	& \begin{tabular}{@{}c@{}} MIP-Detection \\ Efficiency\end{tabular}	& \emph{S} [\% $\sqrt{\rm GeV}$] &	\emph{ C} [\%]\\
\hline
    \begin{tabular}{@{}c@{}}
        Power-law \\(eq. \ref{eq:pLaw})
    \end{tabular} &
    \begin{tabular}{@{}c@{}}
        98\% \\ 95\%\\ 90\% \\ 70\%
     \end{tabular} &
     \begin{tabular}{@{}c@{}}
        $50.8 \pm 0.3$ \\
     	$51.0 \pm 0.2$ \\ 
     	$50.7 \pm 0.3$ \\
     	$51.1\pm 0.2$
     \end{tabular} &
     \begin{tabular}{@{}c@{}}
        $10.29 \pm 0.05$ \\
     	$10.31 \pm 0.04$ \\ 
     	$10.64 \pm 0.06$ \\
     	$11.39 \substack{+0.05 \\ -0.04}$
     \end{tabular} \\
\hline
    \begin{tabular}{@{}c@{}}
        Logarithm \\ (eq. \ref{eq:log})
    \end{tabular}&
    \begin{tabular}{@{}c@{}}
        98\% \\ 95\%\\ 90\% \\ 70\%
     \end{tabular} &
     \begin{tabular}{@{}c@{}}
     	$41.6 \pm 0.4$ \\
     	$41.9 \substack{+0.4 \\ -0.5}$ \\ 
     	$41.2 \pm 0.5$ \\
     	$40.0 \substack{+0.7 \\ -0.8}$
     \end{tabular} &
     \begin{tabular}{@{}c@{}}
     	$12.6 \pm 0.4$ \\
     	$12.6 \pm 0.4$ \\ 
     	$13.0 \pm 0.4$ \\
     	$14.0 \pm 0.6$
     \end{tabular}  \\
\hline
\end{tabular}
\end{table}

\subsubsection{The Effect of Pad Multiplicity}
Motivated by the different average pad multiplicity values measured with the RPC \cite{9} and RPWELL \cite{mediumSizeRPWELL} sampling elements (Table \ref{tab:i}), we evaluated the DHCAL performance with two pad multiplicity distributions (average values of 1.1 and 1.6) at a fixed 98\% MIP detection efficiency. Following \cite{9}, the power-law parametrization was used for this evaluation. 

\begin{figure}[htbp]
    \centering
    \includegraphics[width=0.5\textwidth]{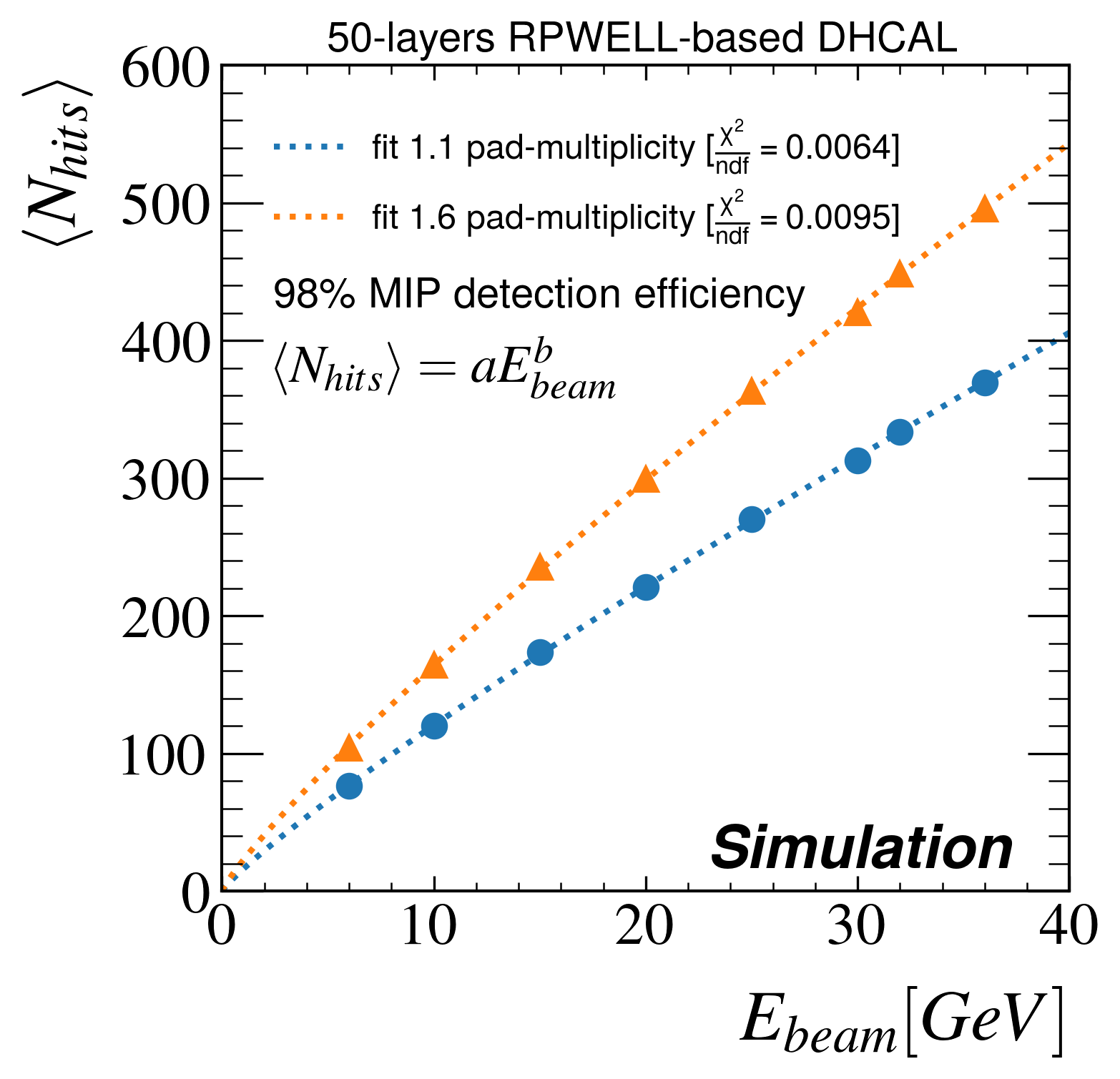}
    \caption{\label{fig:PMres} The response of an RPWELL-based DHCAL with  pad multiplicity averages of 1.1 (blue circles) and 1.6 (orange triangles) and 98\% MIP detection efficiency.}
\end{figure}

\begin{figure}[htbp]
\centering
    \begin{subfigure}[b]{0.43\textwidth}
        \includegraphics[width=\textwidth]{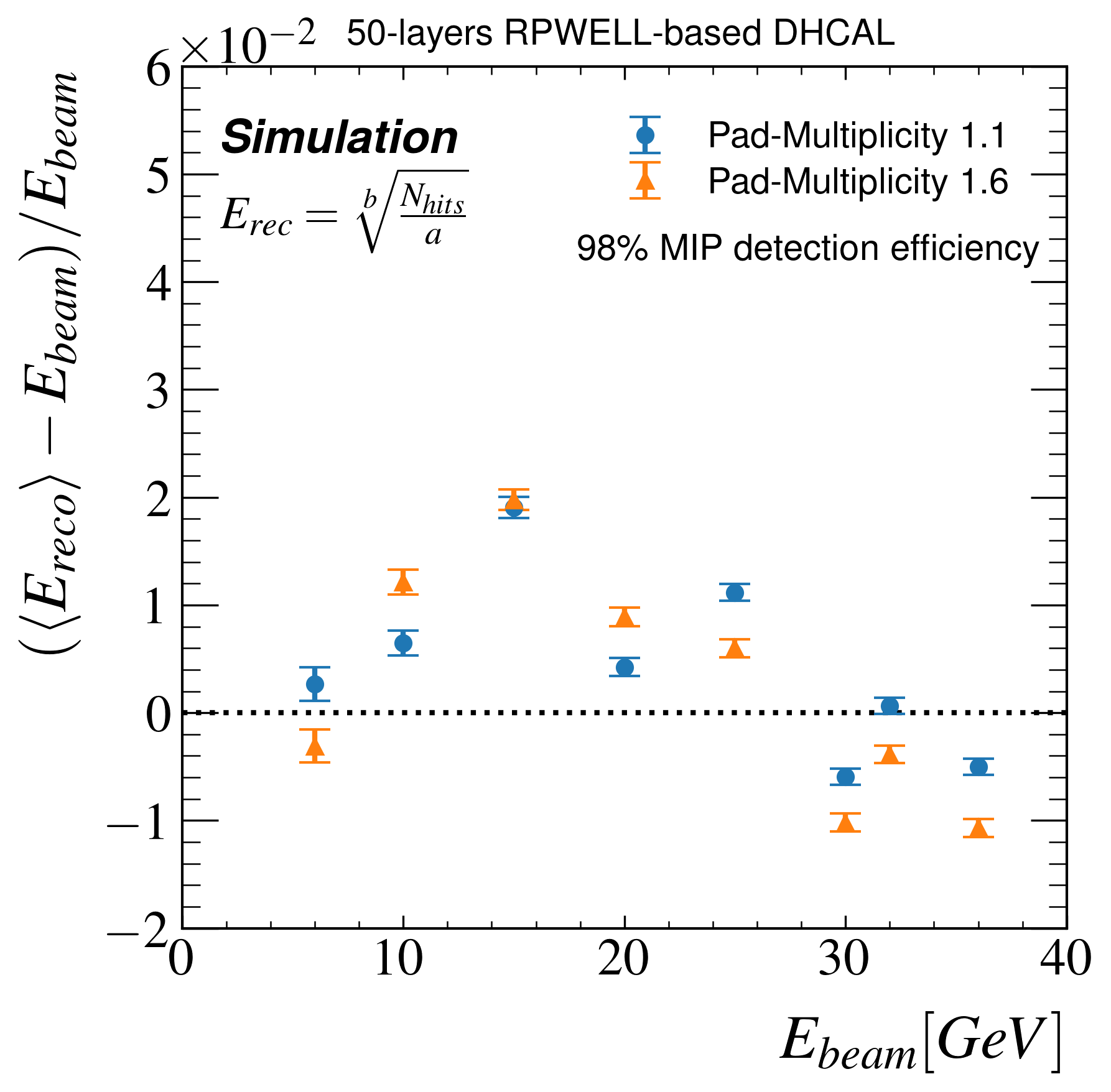}
        \centering
        \caption{Reconstruction bias}
        \label{fig:biasPM}
    \end{subfigure}
    \qquad
    \centering
    \begin{subfigure}[b]{0.43\textwidth}
        \includegraphics[width=\textwidth]{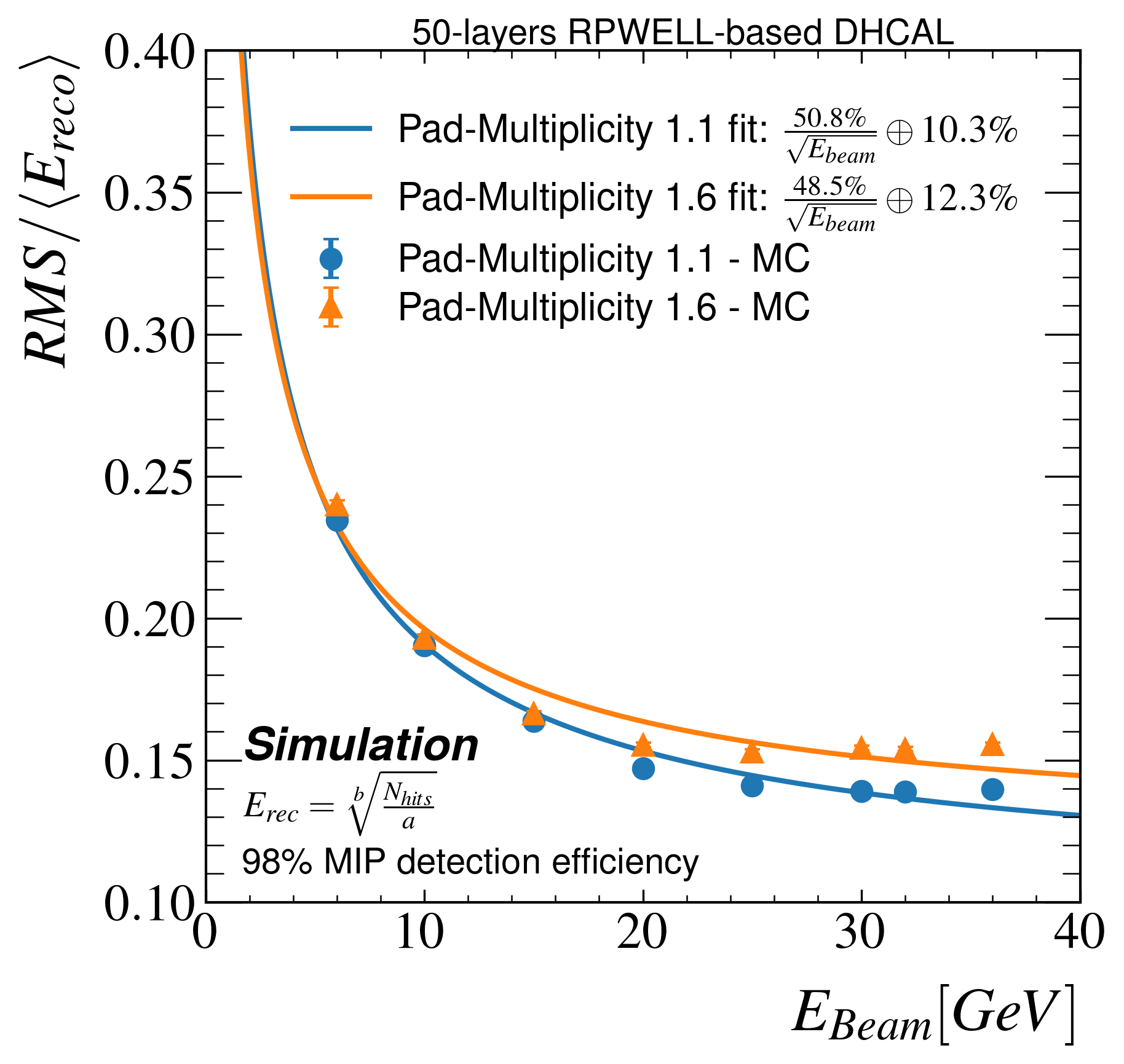}
        \centering
        \caption{Energy resolution}
        \label{fig:eresPM}
    \end{subfigure}

\caption{\label{fig:PM} The (a) energy reconstruction bias and (b) energy resolution of an RPWELL-based DHCAL with pad multiplicity averages of 1.1 (blue circles) and 1.6 (orange triangles) and 98\% MIP detection efficiency.}
\end{figure}

Figure \ref{fig:PMres} shows the calorimeter response to pions, using the two distributions of pad multiplicity, and the resulting fit parameters are presented in Table \ref{tab:PMres}. The ratio between the average number of hits per event measured with the two distributions is compatible with the ratio of their average values (1.6:1.1), although at larger energies some deviations occur. This is attributed to the larger deviation from linearity of the response at higher pad multiplicity. Figure \ref{fig:PM} shows their energy reconstruction bias and energy resolution. While the bias on the energy reconstruction is similar in both pad multiplicity distributions, the energy resolution degrades for larger pad multiplicity.

\begin{table}[htbp]
\centering
\caption{\label{tab:PMres} A summary of the fit parameters of the response using the two average pad multiplicity values of Figure \ref{fig:PMres}.}
\smallskip
\begin{tabular}{cccc}
\hline
\begin{tabular}{@{}c@{}} Average \\ Pad Multiplicity\end{tabular}	& a	& b\\
\hline
   1.1 & 16.04 $\pm$ 0.23 & 0.876 $\pm$ 0.005 \\
   1.6 & 13.04 $\pm$ 0.18 & 0.017 $\pm$ 0.002 \\
\hline
\end{tabular}
\end{table}

\section{Summary and Discussion}\label{subsec:simDis}
We presented a study towards the potential design of an RPWELL-based DHCAL. Data collected with an eight-layers MPGD-DHCAL in a low-energy (2--6 GeV) pion beam was used to validate a GEANT4-based framework for simulating the response of DHCAL prototypes. A good agreement was found between test-beam data and simulation results for pions with $E > 4$ GeV. The best agreement was obtained with the physics lists that include the EMZ model, as pointed out also in \cite{9}. This indicates that EM interactions play an important role in the modeling of hadronic showers, as the EMZ model offers the a high precision modeling of the former.

Once validated, the simulation framework was used to study the expected performance of a 50-layer RPWELL-based Fe-DHCAL (equivalent to $\sim 4.9\lambda_\pi$) with 6--36 GeV pions. Assuming optimal RPWELL sampling elements with a uniform 98\% MIP detection efficiency and 1.1 average pad multiplicity, a pion energy resolution of $\frac{\sigma}{E[GeV]} = \frac{(50.8 \pm 0.3) \%}{\sqrt{E[GeV]}}\oplus(10.3 \pm 0.05)\%$ is expected. Such performance suggests that within the framework of particle-flow, an RPWELL-based DHCAL with a close-to-optimal performance could reach the targeted jet-energy resolution \cite{1}.

Following \cite{9,12}, two optional response parametrizations were investigated: a power-law (eq. \ref{eq:pLaw}) and a logarithmic (eq. \ref{eq:log}) one\footnote{Following \cite{offset}, a power-law parametrization with an offset was also tested. It could provide additional flexibility in incorporating an energy threshold for obtaining at least one hit in the DHCAL. However, repeating the same procedure with this parametrization showed inconsistent results, particularly in the dependency of the energy threshold on the MIP detection efficiency.}. The power-law parametrization is somewhat superior to the logarithmic one; it has a smaller energy reconstruction bias; a better expected energy resolution at energies between 15--36 GeV, and a lower sensitivity to uniform reduction of the MIP detection efficiency.

Using the power-law parametrization, Table \ref{tab:sumEres} summarizes the expected pion energy resolution of an RPWELL-based DHCAL. Various MIP detection efficiencies and pad multiplicity distributions are considered. The measured performance of an RPC Fe-DHCAL \cite{9} is shown for comparison. 

\begin{table}[htbp]
\centering
\caption{\label{tab:sumEres} A summary of the energy resolution of an RPWELL-based DHCAL with different uniform MIP detection efficiency and average pad multiplicity values. The energy reconstruction is based on the power-law parametrization. For comparison, the last row includes the experimental results of the CALICE RPC Fe-DHCAL, which includes offline software compensation.}
\smallskip
\begin{tabular}{ccccc}
\hline
&\begin{tabular}{c}
     Average  \\
     Pad Multiplicity 
\end{tabular} & \begin{tabular}{c} MIP Detection \\ Efficiency \end{tabular} & S[\% GeV] & C [\%]\\
\hline
\begin{tabular}{c}
     Simulation  \\
     {\footnotesize (uniform efficiency)} 
\end{tabular}
 &
\begin{tabular}{c}
     1.1  \\ 1.1 \\ 1.6
\end{tabular} &
\begin{tabular}{c}
     98\%  \\ 70\% \\ 98\%
\end{tabular} &
\begin{tabular}{c}
     $50.8 \pm 0.3$  \\
     $51.1 \pm 0.2$ \\
     $48.5 \pm 0.3$ \\
\end{tabular} &
\begin{tabular}{c}
     $10.29 \pm 0.06$  \\
     $11.39 \pm 0.05$ \\
     $12.26 \pm 0.07$ \\
\end{tabular}\\

\hline
\begin{tabular}{c}
     CALICE DHCAL \cite{9}  \\
     {\footnotesize (software compensation)} 
\end{tabular}
  & 1.69	& 97\% & $48.4 \pm 1.5 $ & 10.6 $\pm$ 0.5\\
\hline

\end{tabular}
\end{table}

To get a visible effect, the effects of lower MIP detection efficiency values, down to 70\% were tested. While the stochastic terms remained the same (fitted values within uncertainties), the constant term was lower by $\sim$1\% at 98\% efficiency, indicating better energy resolution at all investigated energies. 

The effect of the pad multiplicity distribution is more pronounced; relative to an average pad multiplicity of 1.1, at an average of 1.6, the stochastic term is lower by $\sim$2\% and the constant term is higher by $\sim$2\%.  This can be explained by the larger probability of two nearby shower fragments to occupy the same readout pad; int this case, fewer hits would be recorded, so that the variance of their distribution (Poisson error), and thus, the stochastic term would also be smaller. The loss of hits is equivalent to the loss of information. Thus, the constant term would be increased. 

Relative to the RPC-based Fe-DHCAL \cite{9}, an RPWELL-based one with 98\% MIP detection efficiency and 1.1 average pad multiplicity is expected to yield similar pion energy resolution. This comparison should be taken with a grain of salt. On the one hand, the energy resolution reported in \cite{9} was measured with test-beam data and was subject to uncertainties not considered in our simulation. On the other hand, the reconstruction algorithm used was more advanced. It exploited information from individual layers and calibrated the number of hits based on the hit-density across adjacent layers. Nevertheless, given its significantly lower average pad multiplicity and the fact that it can be operated using environment-friendly gas mixtures, it is worth considering the RPWELL as a potential candidate for future DHCALs.

Based on \cite{12}, the RPWELL- and MM-DHCAL are compared in terms of their simulated response in Table \ref{tab:MMcomp}. The operation principle of the two technologies is similar, and they have minor differences in the material budget (Table \ref{tab:model}). As expected, their response is similar. Choosing one technology over the other could depend on considerations such as cost, stability, and robustness.

\begin{table}[htbp]
\centering
\caption{\label{tab:MMcomp} A comparison of the response to pions with simulated results of RPWELL-based DHCAL and measured results of virtual MM DHCAL  \cite{12}. The response is presented using the logarithmic parametrization ($\langle N_{hits} \rangle= \frac{a}{b}\log(1+bE_{beam}) $).}
\smallskip
\begin{tabular}{cccc}
\hline
Technology	& MIP-Detection Efficiency	& a	& b\\
\hline
    RPWELL &\begin{tabular}{@{}c@{}}
     	98\% \\ 95\% 
     \end{tabular} &
     \begin{tabular}{@{}c@{}}
     	13.04 $\pm$ 0.18 \\ 12.55 $\pm$ 0.20 
     \end{tabular} &
     \begin{tabular}{@{}c@{}}
     	0.017 $\pm$ 0.002 \\ 0.017 $\pm$ 0.002 
     \end{tabular}\\
     \hline
    MM \cite{12} & 96.6\% & 12.31 & 0.009\\
\hline

\end{tabular}
\end{table}

The simulation framework used the Hit Detection Efficiency (HDE) and pad multiplicity distribution as an input. Thus, it can easily be extended to other technologies, provided that the HDE and the pad multiplicity are known. So far, an assumption of uniform HDE and pad multiplicity distribution has been made. A realistic simulation should take into account also non-uniformity in the response. For each pion energy, these are expected to smear the distribution of $N_{hits}$ and degrade the energy resolution. Finally, the optimization of future DHCAL designs would require an operation in a complete particle-flow experiment framework \cite{cvpf, MLPF}. E.g., the MIP detection efficiency and average pad multiplicity are expected to affect the DHCAL performance also in its contribution to the confusion term, the uncertainty of energy deposit assignment to a specific particle, and thus the jet energy resolution.

\acknowledgments
We thank Prof. Wang Yi from Tsinghua university in China for kindly providing us with the silicate glass tiles. This research was supported in part by the I-CORE Program of the Planning and Budgeting Committee, the Nella and Leon Benoziyo Center for High Energy Physics at the Weizmann Institute of Science, the sir charles clore prize, the common fund of the RD51 collaboration at CERN (the Sampling Calorimetry with Resistive Anode MPGDs, SCREAM, project), Grant No 712482 from the Israeli Science Foundation (ISF), Grant No 5029538 from the Structural Funds, ERDF and ESF, Greece, and the Chateaubriand fellowship from the French Embassy in Israel.

\begin{appendices}
\section{Pad Multiplicity and Hit Detection Efficiency}\label{appendix:HDE}

We distinguish between the hit detection efficiency (HDE) and the measured MIP detection efficiency as well as between 
the underlying pad multiplicity and the measured one. The underlying pad multiplicity is defined as the number of pads with signals induced on them, prior to any efficiency losses. The HDE is the probability that a hit will be measured from a pad with an induced signal. The measured MIP detection efficiency was defined as the ratio of the number of efficient events over the total number of MIP tracks, where an event was considered efficient if a cluster was measured at the tested layer less that 1 cm away from the point of intersection with the MIP track. The measured pad multiplicity was defined by the number of hits in the cluster (Section \ref{sec:method}). 

The HDE and underlying pad multiplicity are related to the measured MIP detection efficiency and pad multiplicity by binomial statistics.

\begin{equation}\label{eq:a1}
    N_m^*=\sum_{n\leq m}\binom{n}{m} N_n \varepsilon_h^m (1-\varepsilon_h )^{(n-m)}
\end{equation}
\begin{equation}
    N_0^*=N_{events}(1-\varepsilon_M)
\end{equation}
where $\varepsilon_M$ and $N_m^*$ denote the MIP detection efficiency and the measured number of events with pad multiplicity $m$, respectively; and $\varepsilon_h$ and $N_m$ are the HDE and the underlying number of events with pad multiplicity $m$. Under the assumption that the maximum underlying pad multiplicity is 4, we get the following set of equations:
\begin{subequations}\label{eq:sys}
\begin{align}
    \label{eq:sys:1}
    N_4^* & = N_4 \varepsilon_h^4
    \\
    \label{eq:sys:2}
    N_3^* & = [N_3+4N_4 (1-\varepsilon_h )] \varepsilon_h^3
    \\
    \label{eq:sys:3}
    N_2^* & =[N_2+3N_3 (1-\varepsilon_h )+6N_4 (1-\varepsilon_h )^2 ] \varepsilon_h^2
    \\
    \label{eq:sys:4}
    N_1^* & =[N_1+2N_2 (1-\varepsilon_h )+3N_3 (1-\varepsilon_h )^2+4N_4 (1-\varepsilon_h )^3 ] \varepsilon_h
    \\
    \label{eq:sys:5}
    N_{events} &=\sum_{m=1}^4 N_m
\end{align}
\end{subequations}

This set of equations result in a linear equation:
\begin{equation} \label{eq:lin}
    N_{events} \varepsilon_h^4-C_1 \varepsilon_h^3+C_2 \varepsilon_h^2-C_3 \varepsilon_h^1+C_4=0
\end{equation}
 where:
 
\begin{subequations}\label{eq:C_measure}
    \begin{align}
        \label{eq:C_measure:1}
        C_4 & =N_4^* =N_4 \varepsilon_h^4
        \\
        \label{eq:C_measure:2}
        C_3 & =N_3^*+4N_4^* =(N_3+4N_4 ) \varepsilon_h^3
        \\
        \label{eq:C_measure:3}
        C_2 & =N_2^*+3N_3^*+6N_4^* =(N_2+3N_3+6N_4 ) \varepsilon_h^2
        \\
        \label{eq:C_measure:4}
        C_1 & =N_1^*+2N_2^*+3N_3^*+4N_4^* =(N_1+2N_2+3N_3+4N_4 ) \varepsilon_h
    \end{align}
\end{subequations}

Thus, the solution of eq. \ref{eq:lin} yields the values for $\varepsilon_h, N_1,N_2,N_3$, and $N_4$ --- the HDE and the underlying pad multiplicity distribution.
\newpage
\FloatBarrier

\section{$N_{hits}$ distribution in RPWELL-DHCAL}\label{appendix:nhitsDist}

\begin{figure}[htbp]
    \centering
    \includegraphics[width=0.95\textwidth]{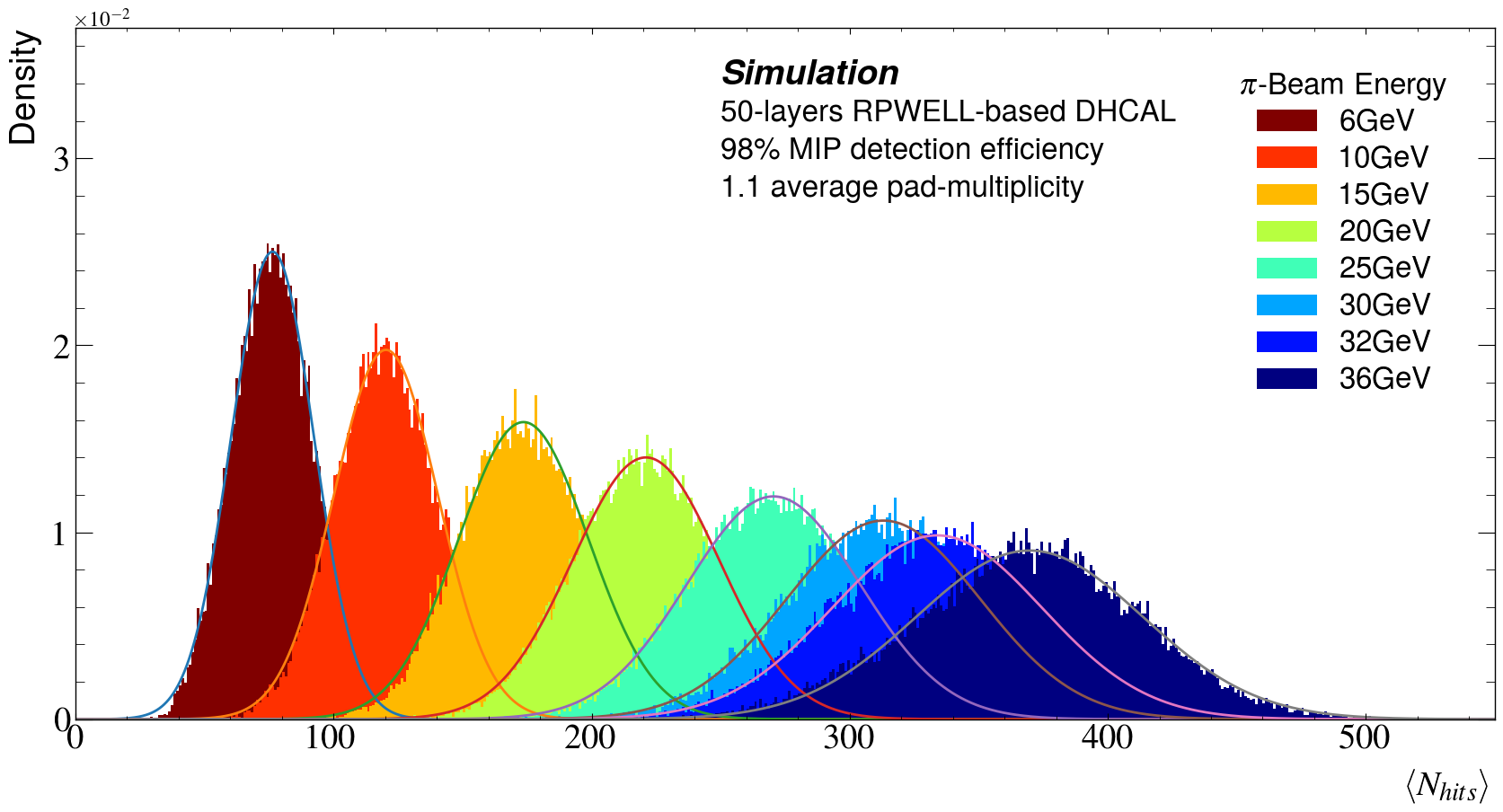}
    \caption{\label{fig:nhitsDist98} The number of hits ($N_{hits}$) distributions per pion beam energy, as simulated using 50-layers RPWELL-based DHCAL with pad multiplicity of 1.1 and 98\% MIP detection efficiency. The Gaussian fits used for extracting the DHCAL response are presented by solid lines.}
\end{figure}

\begin{figure}[htbp]
    \centering
    \includegraphics[width=0.95\textwidth]{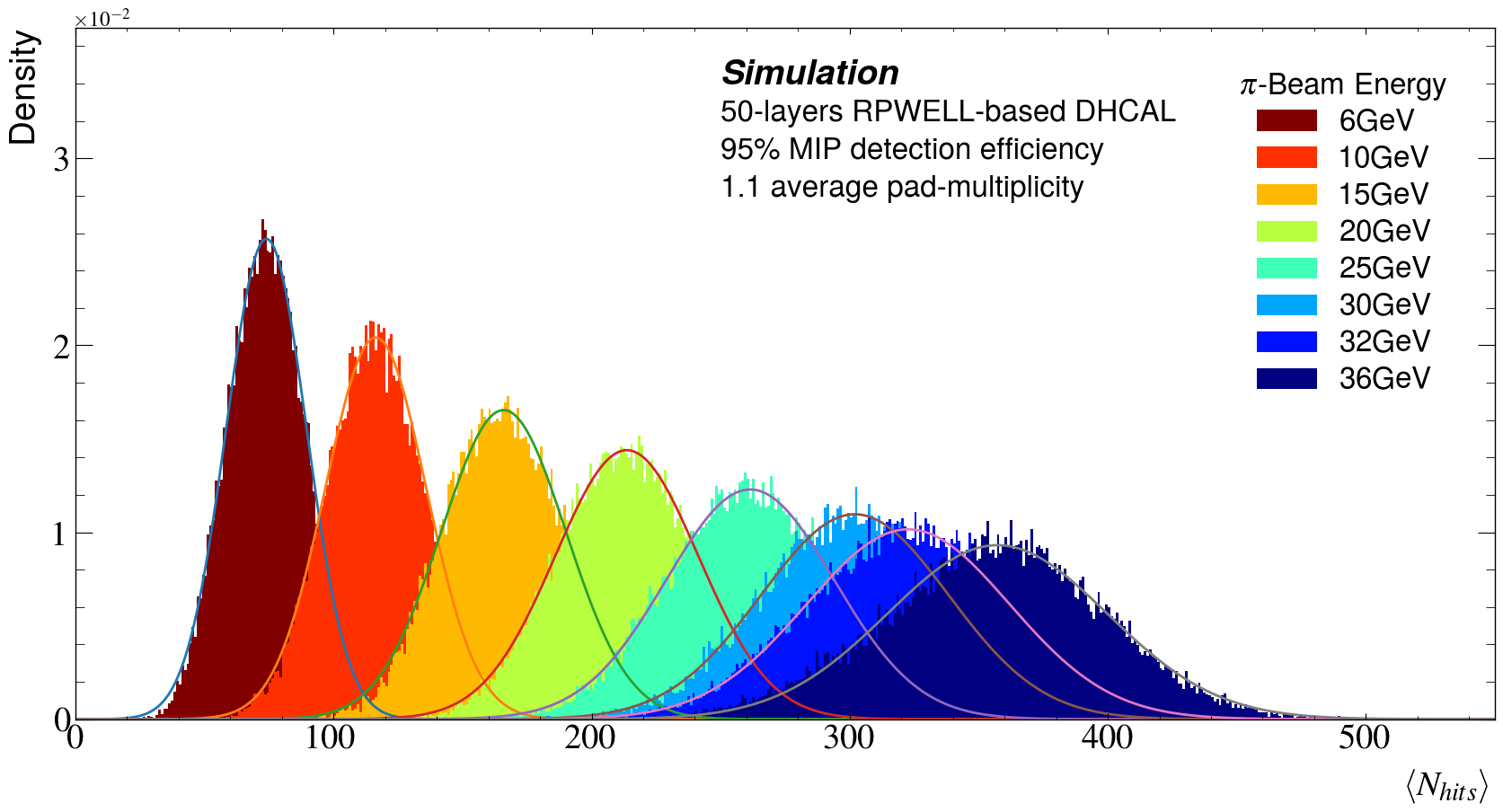}
    \caption{\label{fig:nhitsDist95} The number of hits ($N_{hits}$) distributions per pion beam energy, as simulated using 50-layers RPWELL-based DHCAL with pad multiplicity of 1.1 and 95\% MIP detection efficiency. The Gaussian fits used for extracting the DHCAL response are presented by solid lines.}
\end{figure}

\begin{figure}[htbp]
    \centering
    \includegraphics[width=0.95\textwidth]{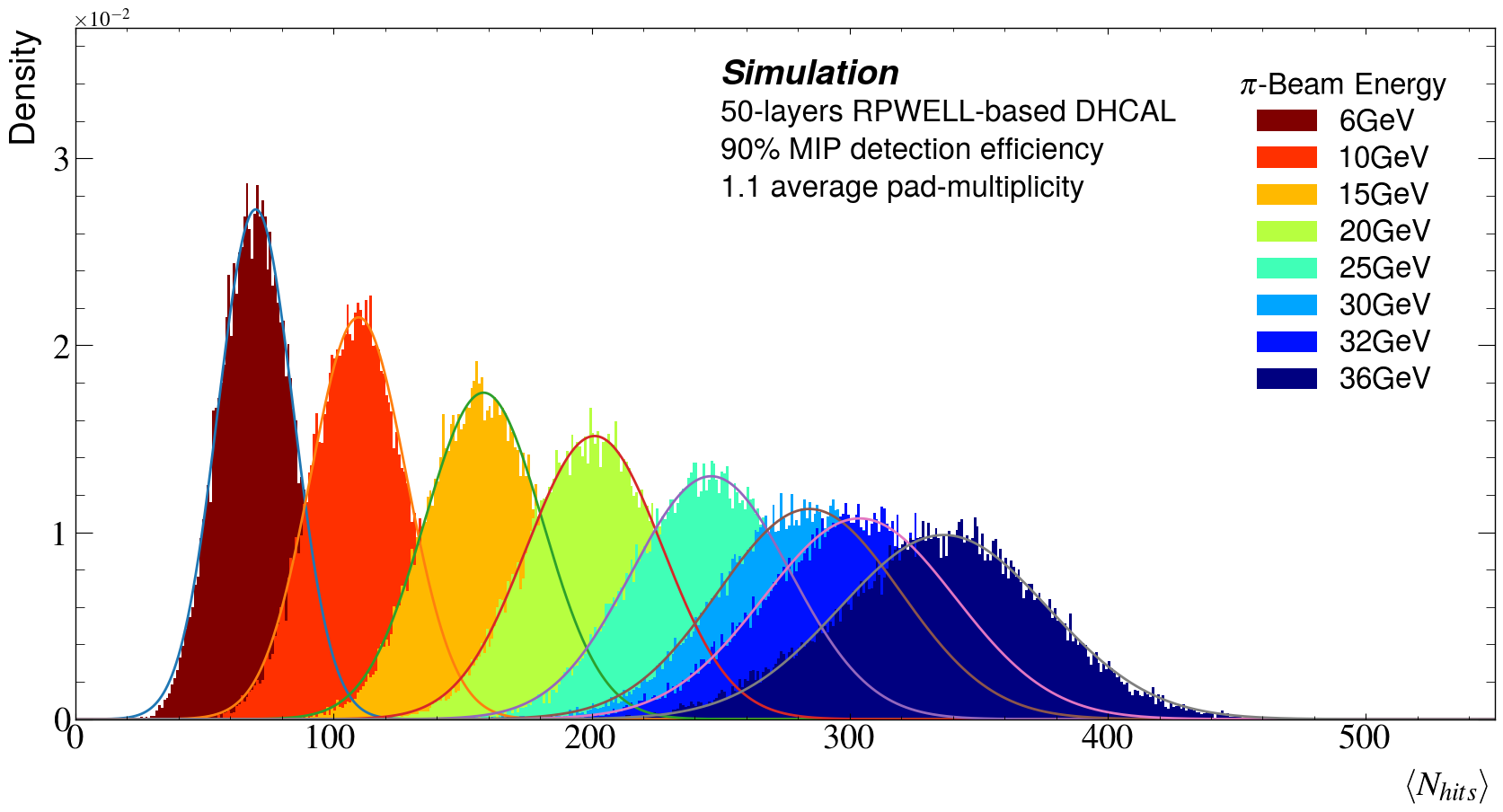}
    \caption{\label{fig:nhitsDist90} The number of hits ($N_{hits}$) distributions per pion beam energy, as simulated using 50-layers RPWELL-based DHCAL with pad multiplicity of 1.1 and 90\% MIP detection efficiency. The Gaussian fits used for extracting the DHCAL response are presented by solid lines.}
\end{figure}

\begin{figure}[htbp]
    \centering
    \includegraphics[width=0.95\textwidth]{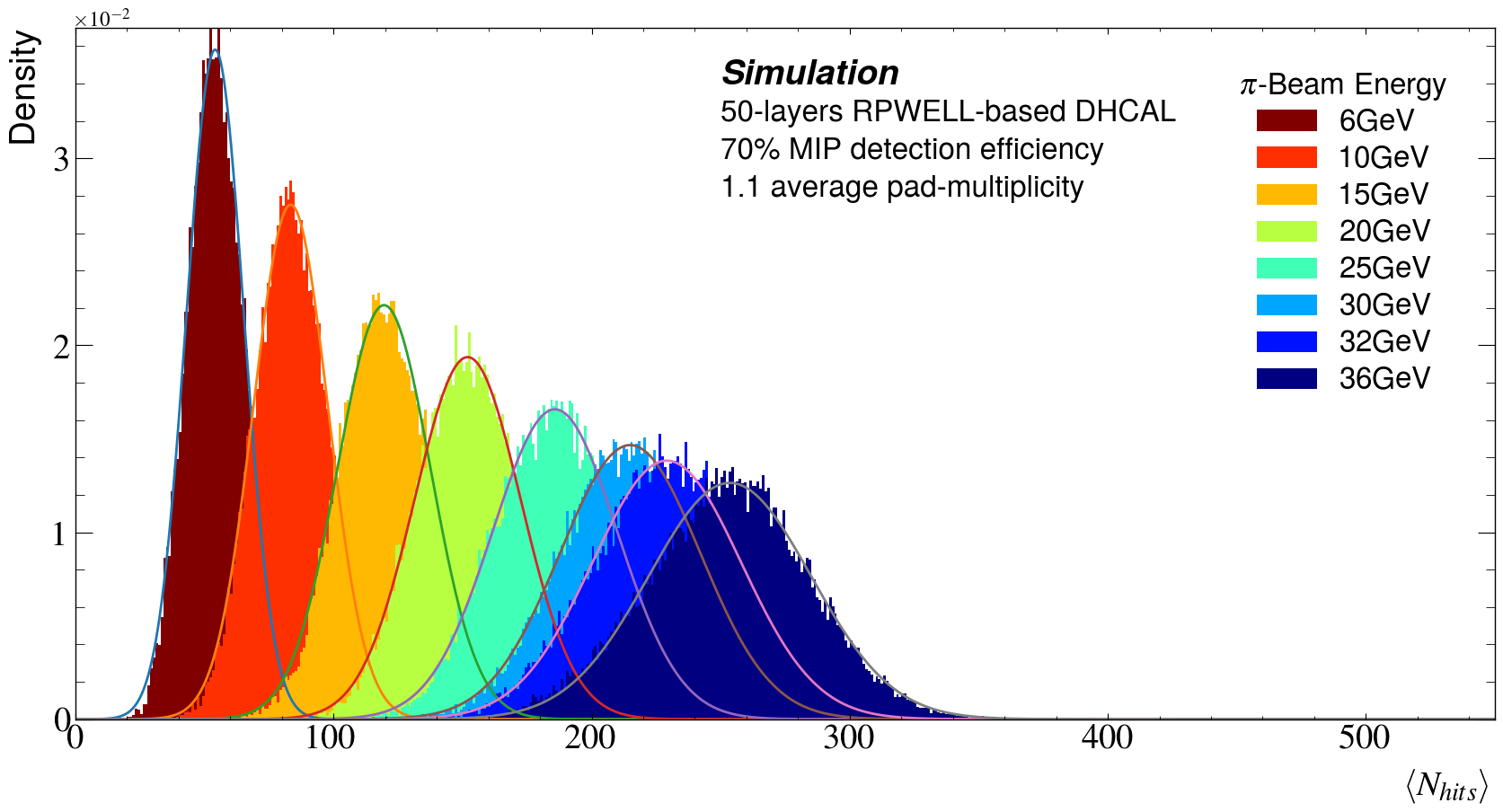}
    \caption{\label{fig:nhitsDist70} The number of hits ($N_{hits}$) distributions per pion beam energy, as simulated using 50-layers RPWELL-based DHCAL with pad multiplicity of 1.1 and 70\% MIP detection efficiency. The Gaussian fits used for extracting the DHCAL response are presented by solid lines.}
\end{figure}

\begin{figure}[htbp]
    \centering
    \includegraphics[width=0.95\textwidth]{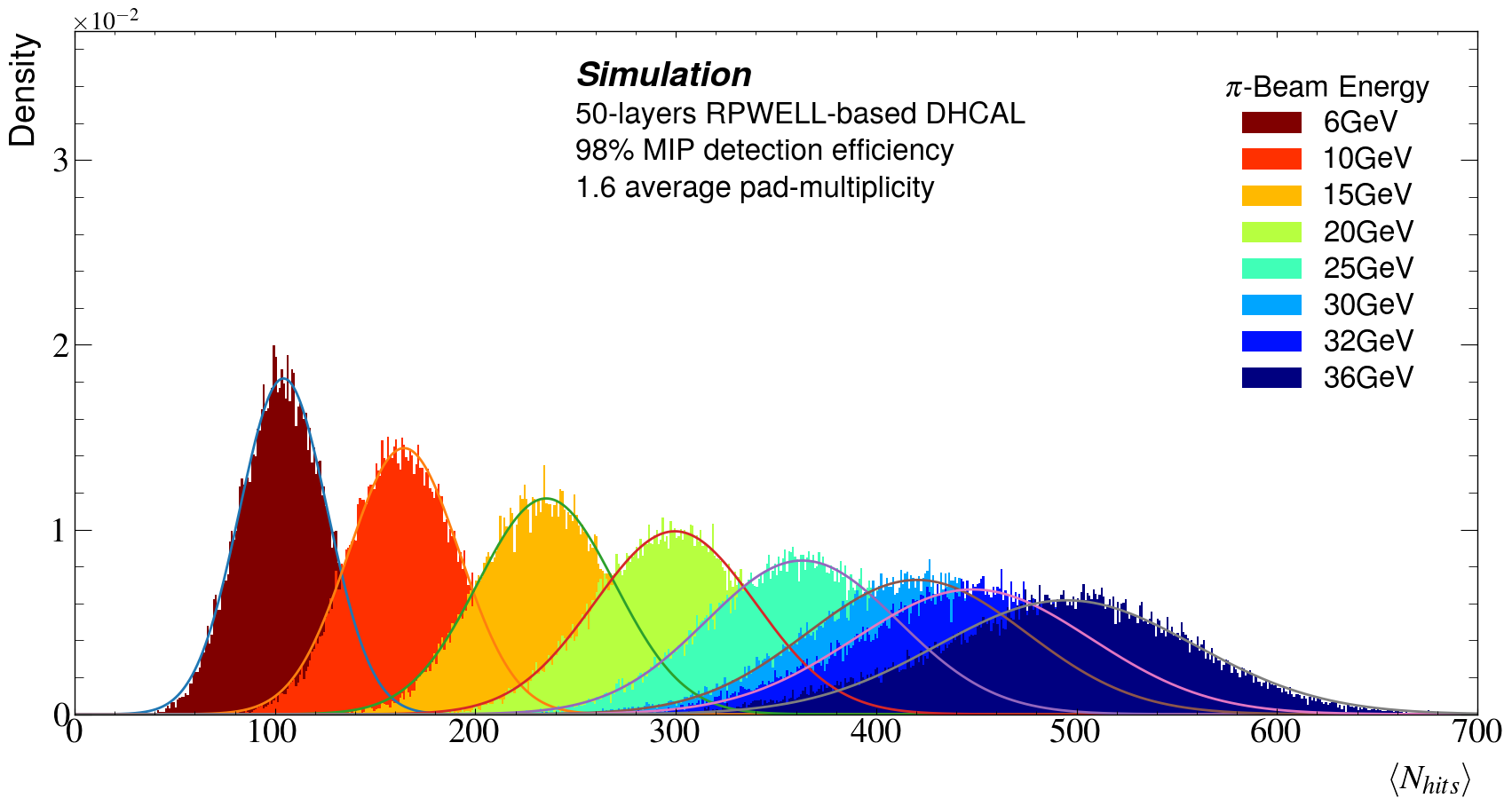}
    \caption{\label{fig:nhitsDist98_1p6} The number of hits ($N_{hits}$) distributions per pion beam energy, as simulated using 50-layers RPWELL-based DHCAL with pad multiplicity of 1.6 and 98\% MIP detection efficiency. The Gaussian fits used for extracting the DHCAL response are presented by solid lines.}
\end{figure}

\FloatBarrier

\section{$E_{rec}$ distribution in RPWELL-DHCAL}\label{appendix:erecDist}

\begin{figure}[htbp]
    \centering
    \includegraphics[width=0.95\textwidth]{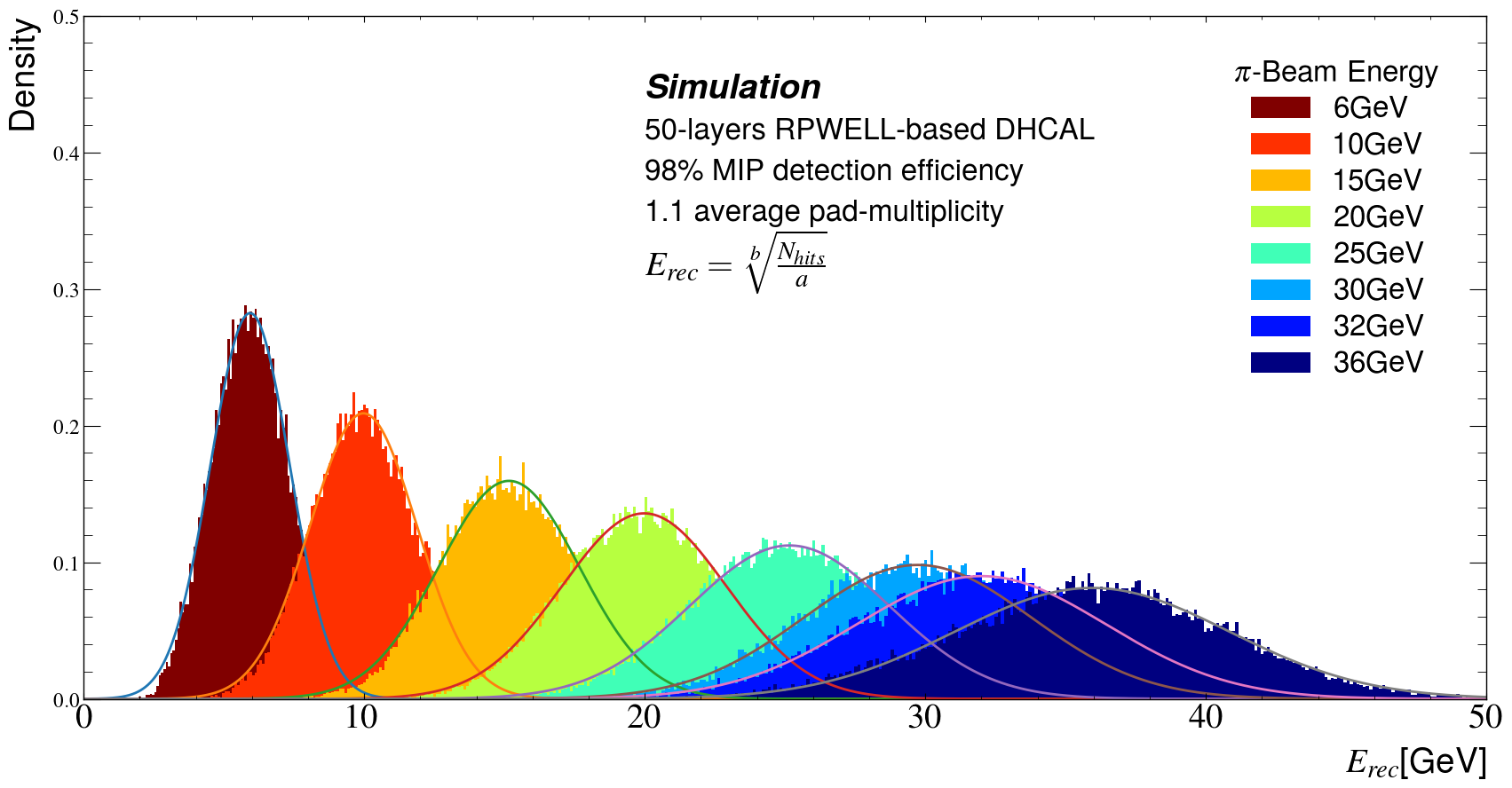}
    \caption{\label{fig:erecDist98power} The reconstructed energy ($E_{rec}$) distributions per pion beam energy, as simulated using 50-layers RPWELL-based DHCAL with pad multiplicity of 1.1 and 98\% MIP detection efficiency. The energy was reconstructed using the power-law parametrization. The Gaussian fits used for extracting the DHCAL response are presented by solid lines.}
\end{figure}

\begin{figure}[htbp]
    \centering
    \includegraphics[width=0.95\textwidth]{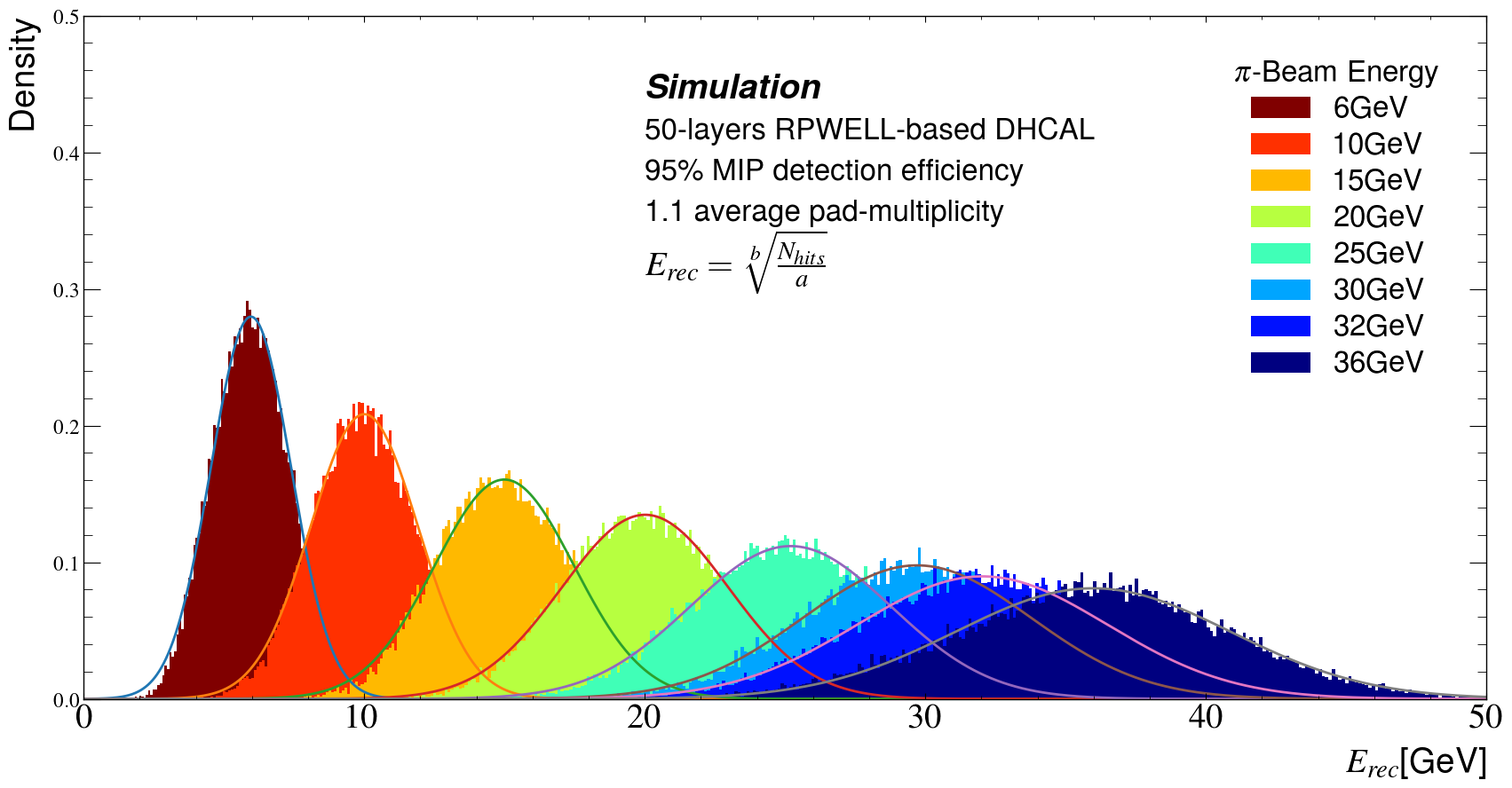}
    \caption{\label{fig:erecDist95power} The reconstructed energy ($E_{rec}$) distributions per pion beam energy, as simulated using 50-layers RPWELL-based DHCAL with pad multiplicity of 1.1 and 95\% MIP detection efficiency. The energy was reconstructed using the power-law parametrization. The Gaussian fits used for extracting the DHCAL response are presented by solid lines.}
\end{figure}

\begin{figure}[htbp]
    \centering
    \includegraphics[width=0.95\textwidth]{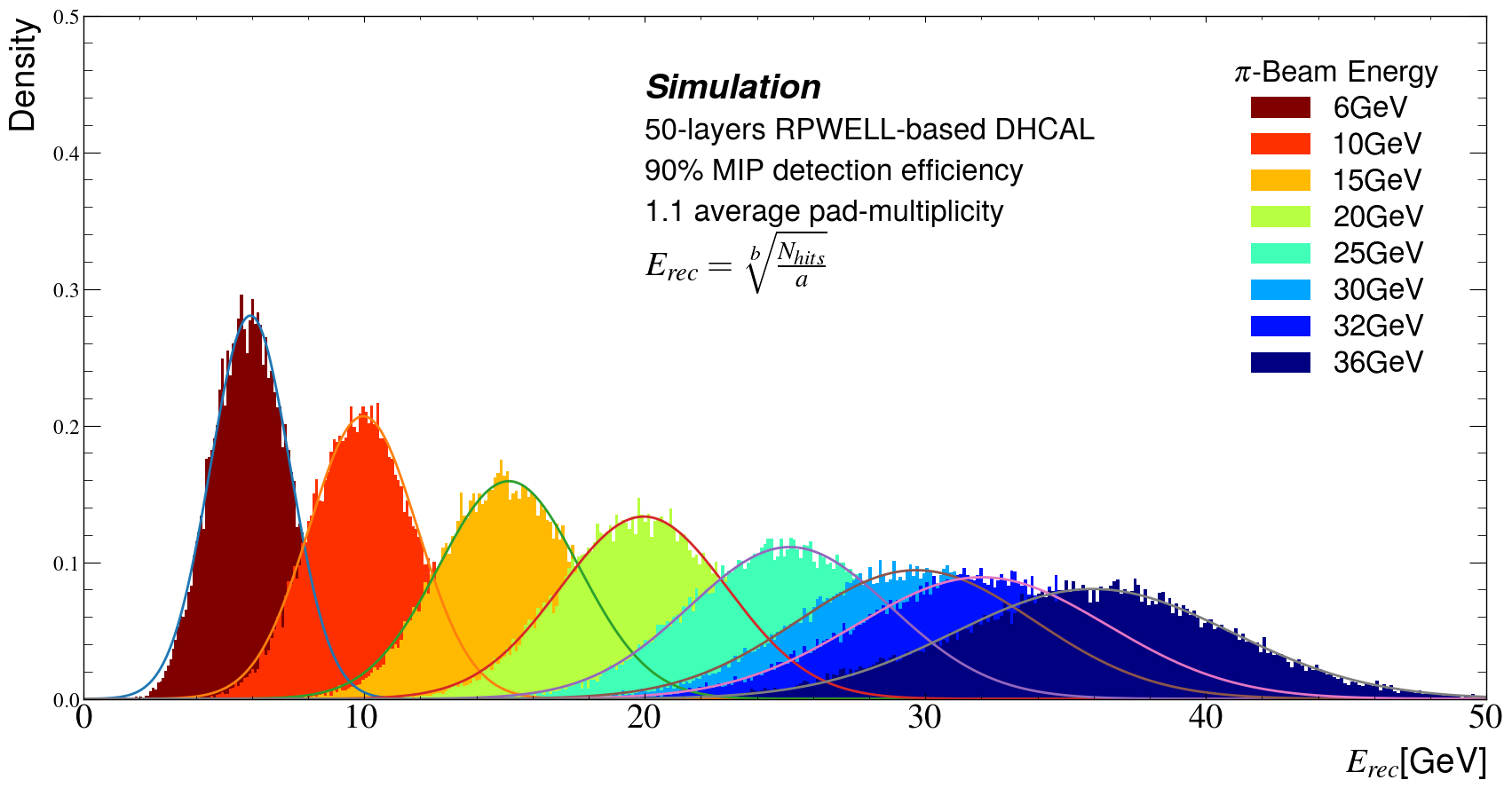}
    \caption{\label{fig:erecDist90power} The reconstructed energy ($E_{rec}$) distributions per pion beam energy, as simulated using 50-layers RPWELL-based DHCAL with pad multiplicity of 1.1 and 90\% MIP detection efficiency. The energy was reconstructed using the power-law parametrization. The Gaussian fits used for extracting the DHCAL response are presented by solid lines.}
\end{figure}

\begin{figure}[htbp]
    \centering
    \includegraphics[width=0.95\textwidth]{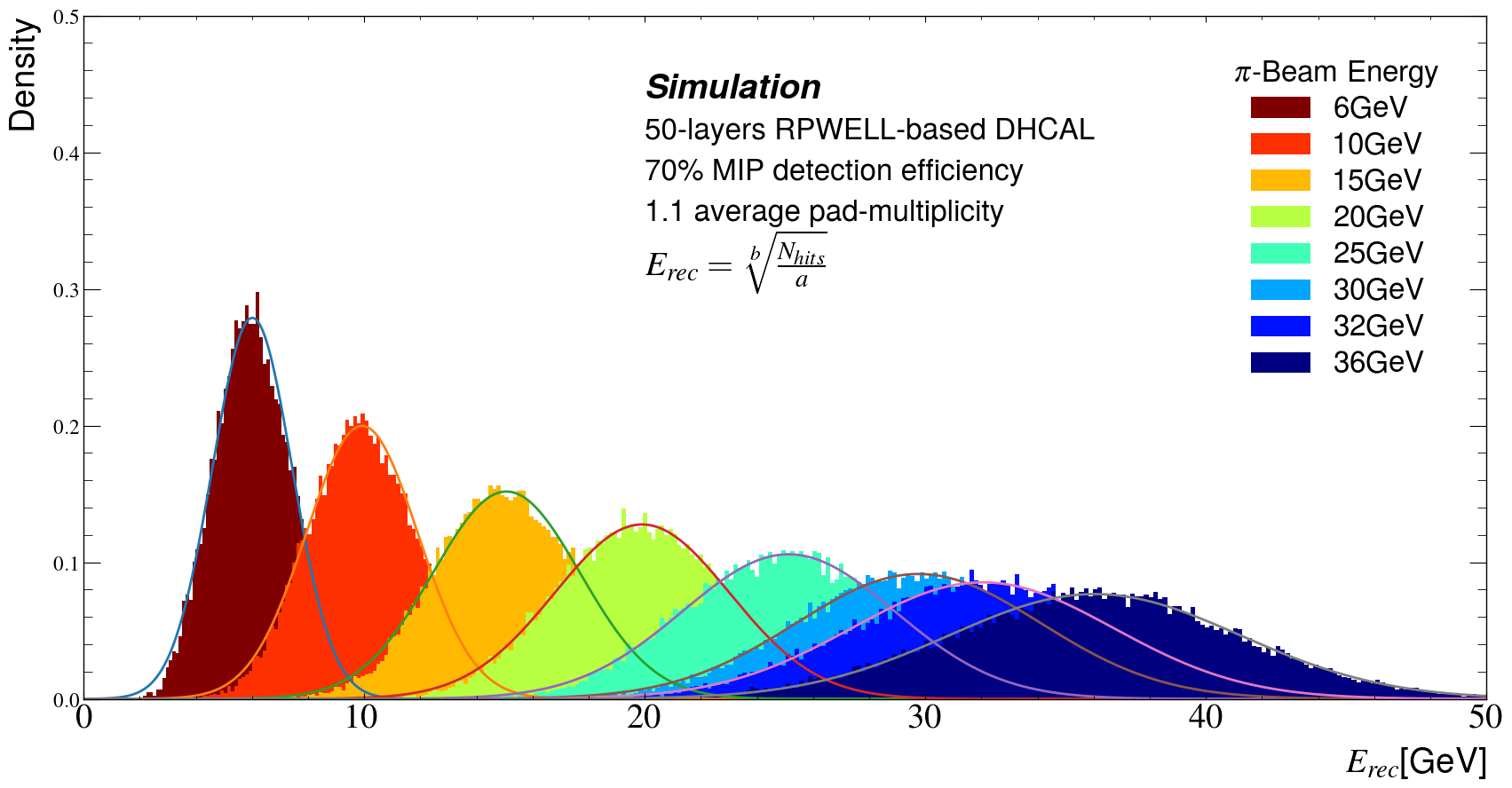}
    \caption{\label{fig:erecDist70power} The reconstructed energy ($E_{rec}$) distributions per pion beam energy, as simulated using 50-layers RPWELL-based DHCAL with pad multiplicity of 1.1 and 70\% MIP detection efficiency. The energy was reconstructed using the power-law parametrization. The Gaussian fits used for extracting the DHCAL response are presented by solid lines.}
\end{figure}

\begin{figure}[htbp]
    \centering
    \includegraphics[width=0.95\textwidth]{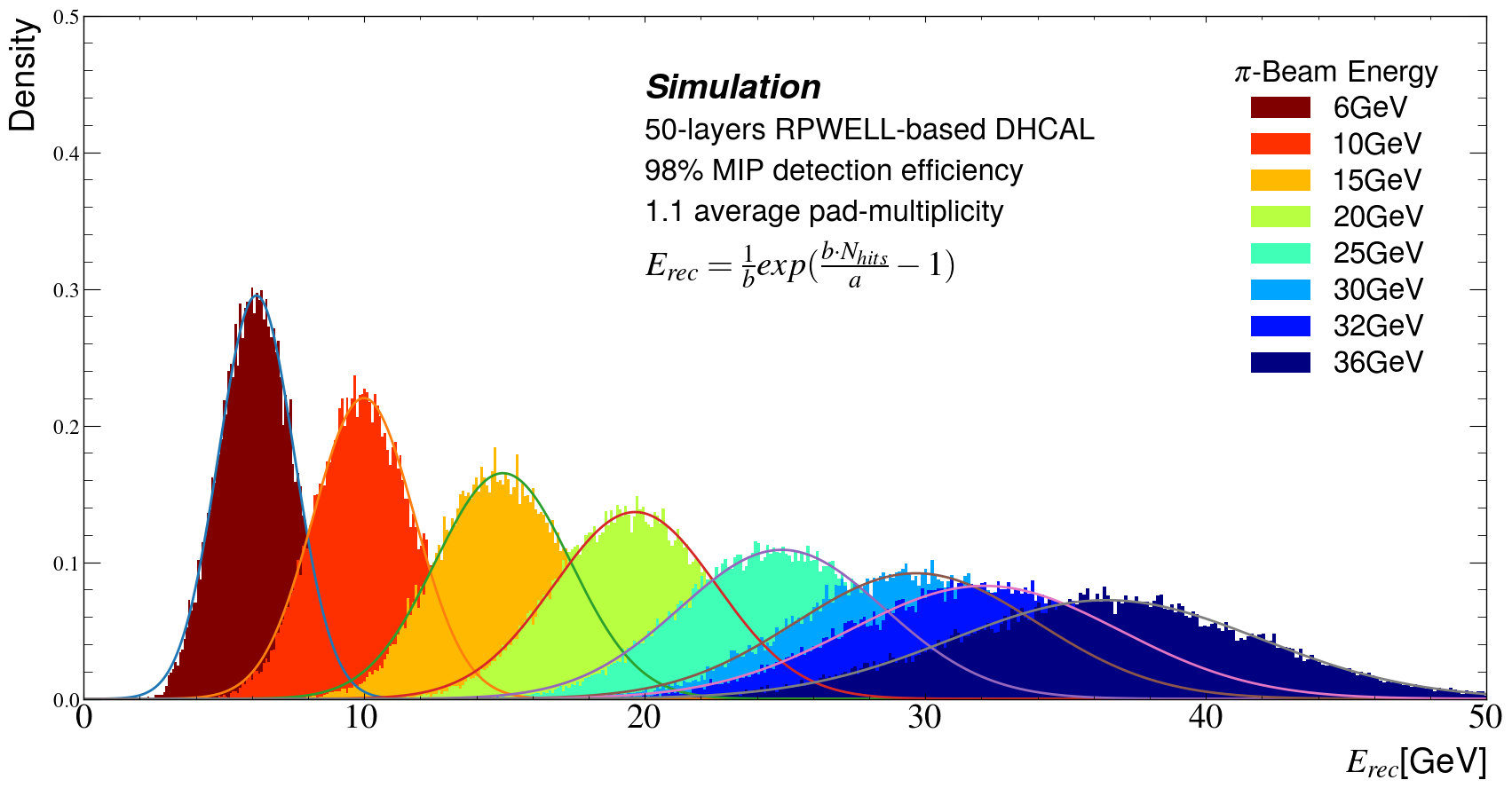}
    \caption{\label{fig:erecDist98log} The reconstructed energy ($E_{rec}$) distributions per pion beam energy, as simulated using 50-layers RPWELL-based DHCAL with pad multiplicity of 1.1 and 98\% MIP detection efficiency. The energy was reconstructed using the logarithmic parametrization. The Gaussian fits used for extracting the DHCAL response are presented by solid lines.}
\end{figure}

\begin{figure}[htbp]
    \centering
    \includegraphics[width=0.95\textwidth]{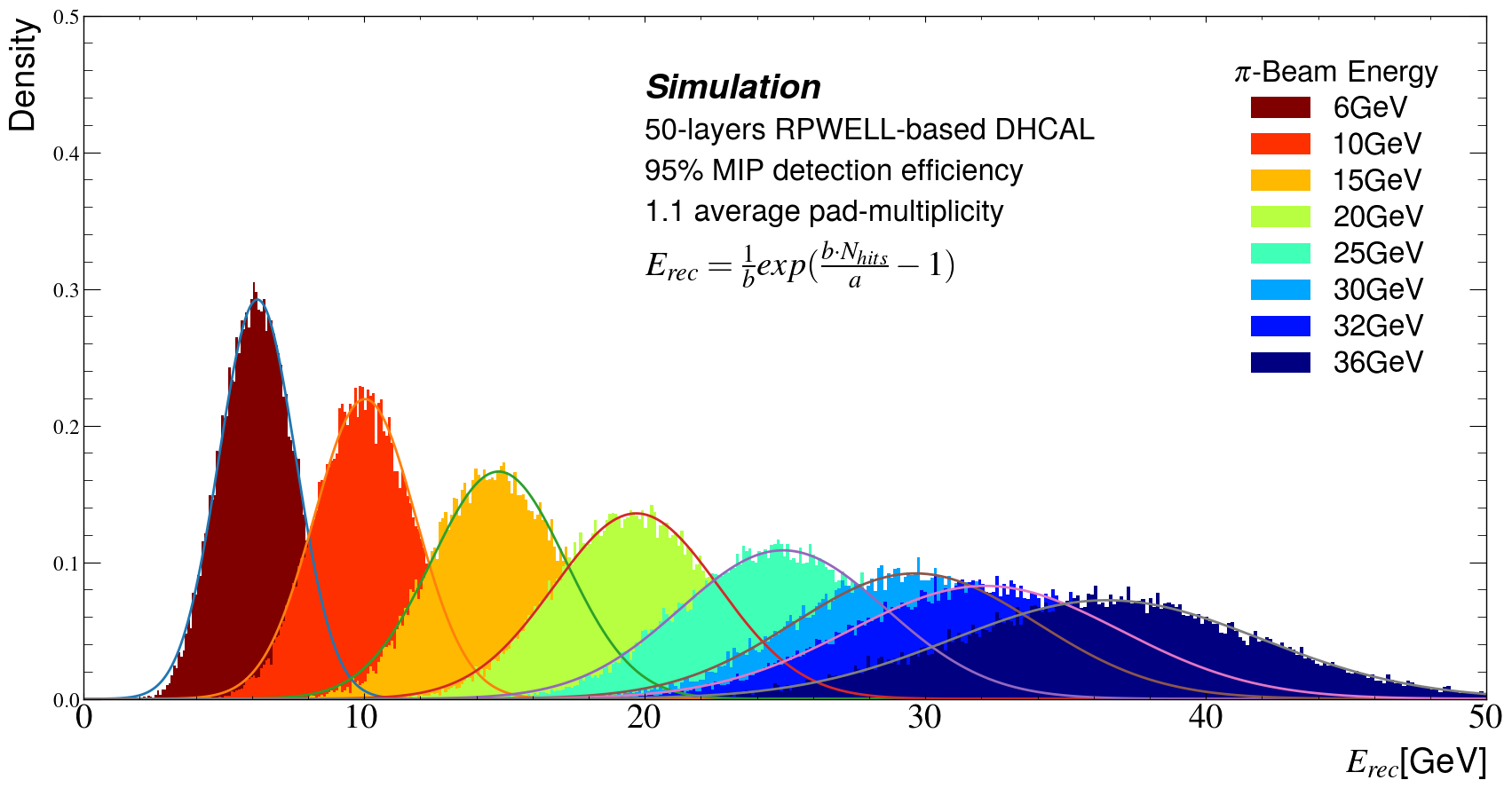}
    \caption{\label{fig:erecDist95log} The reconstructed energy ($E_{rec}$) distributions per pion beam energy, as simulated using 50-layers RPWELL-based DHCAL with pad multiplicity of 1.1 and 95\% MIP detection efficiency. The energy was reconstructed using the logarithmic parametrization. The Gaussian fits used for extracting the DHCAL response are presented by solid lines.}
\end{figure}

\begin{figure}[htbp]
    \centering
    \includegraphics[width=0.95\textwidth]{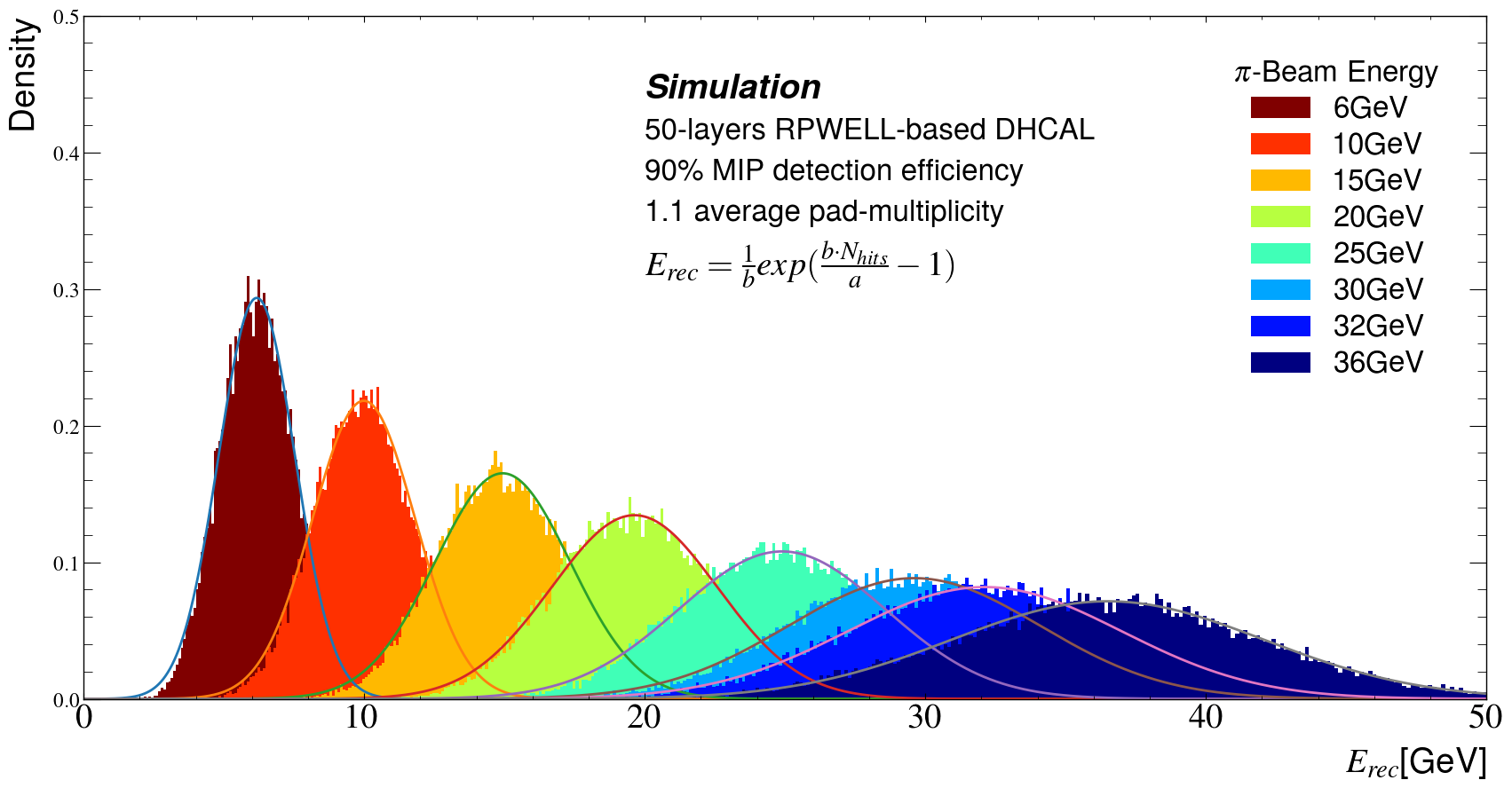}
    \caption{\label{fig:erecDist90log} The reconstructed energy ($E_{rec}$) distributions per pion beam energy, as simulated using 50-layers RPWELL-based DHCAL with pad multiplicity of 1.1 and 90\% MIP detection efficiency. The energy was reconstructed using the logarithmic parametrization. The Gaussian fits used for extracting the DHCAL response are presented by solid lines.}
\end{figure}

\begin{figure}[htbp]
    \centering
    \includegraphics[width=0.95\textwidth]{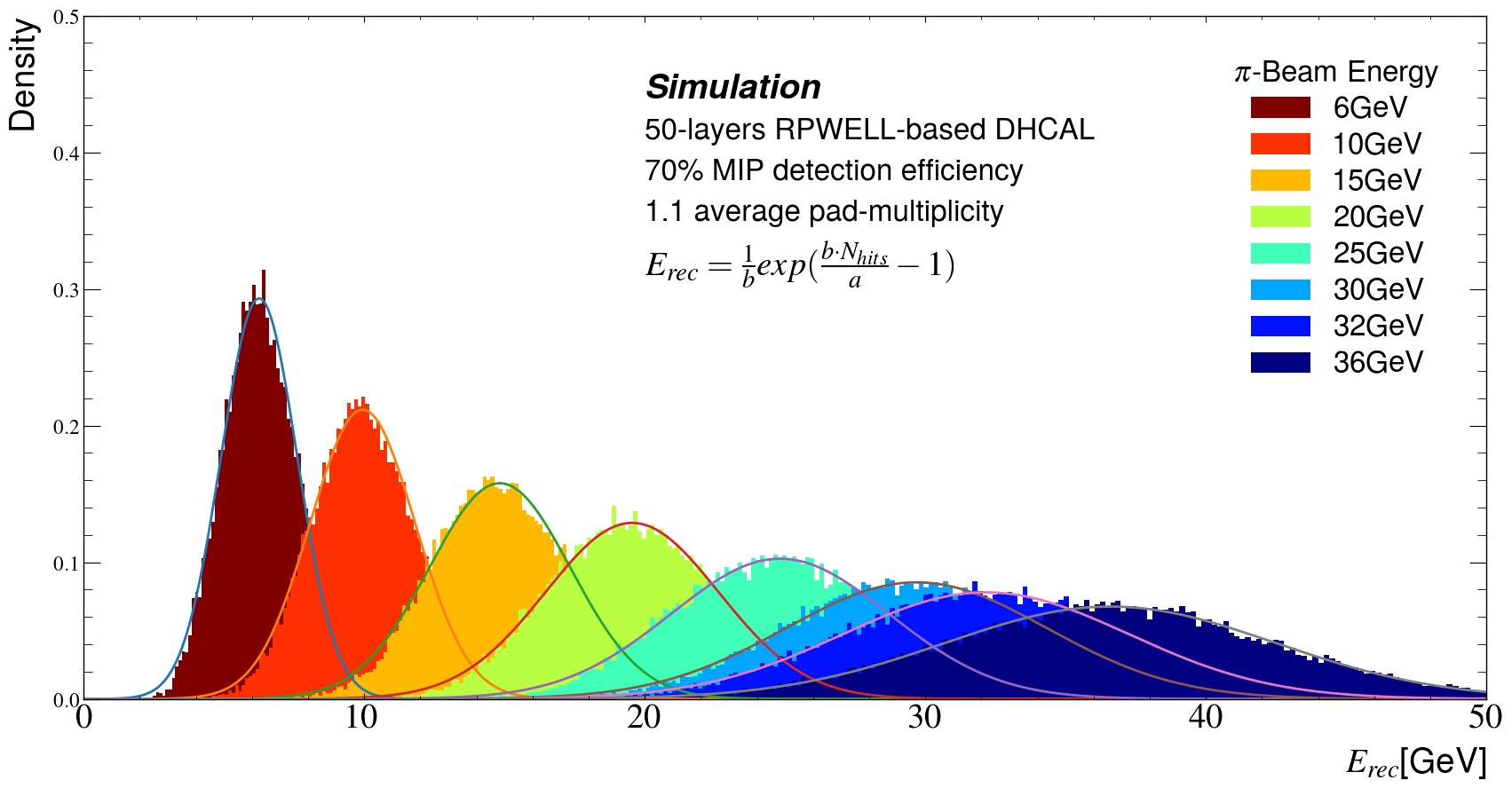}
    \caption{\label{fig:erecDist70log} The reconstructed energy ($E_{rec}$) distributions per pion beam energy, as simulated using 50-layers RPWELL-based DHCAL with pad multiplicity of 1.1 and 70\% MIP detection efficiency. The energy was reconstructed using the logarithmic parametrization. The Gaussian fits used for extracting the DHCAL response are presented by solid lines.}
\end{figure}

\begin{figure}[htbp]
    \centering
    \includegraphics[width=0.95\textwidth]{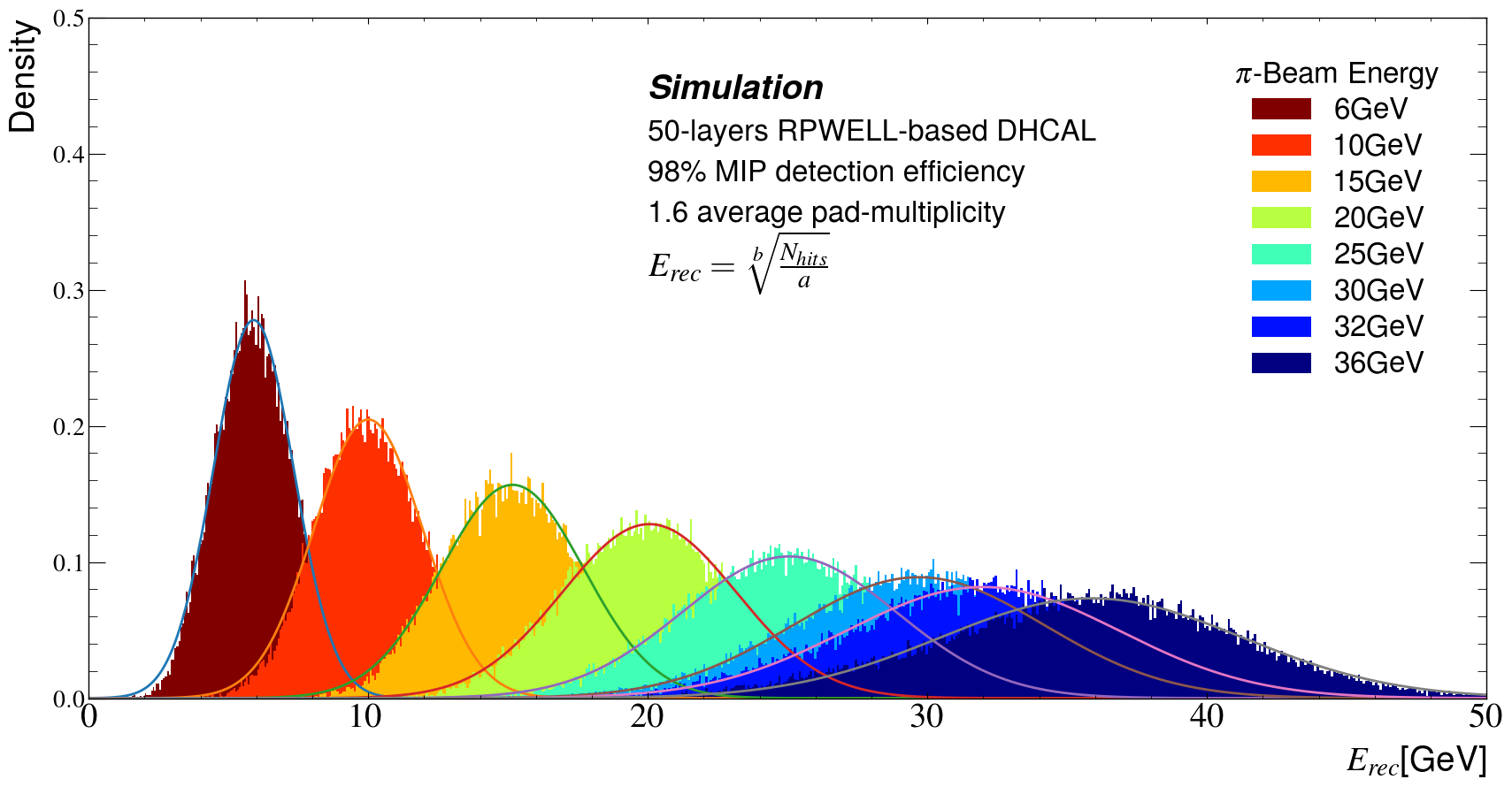}
    \caption{\label{fig:erecDist98_1p6} The reconstructed energy ($E_{rec}$) distributions per pion beam energy, as simulated using 50-layers RPWELL-based DHCAL with pad multiplicity of 1.6 and 98\% MIP detection efficiency. The energy was reconstructed using the power-law parametrization. The Gaussian fits used for extracting the DHCAL response are presented by solid lines.}
\end{figure}

\end{appendices}

\FloatBarrier

\end{document}